\documentclass[usenatbib]{mn2e}
\usepackage{epsf}
\usepackage{graphicx}
\usepackage{times}

\def\jnl@style#1{{\rmfamily#1}}%
\def\jref@jnl#1{{\jnl@style#1}}%

\newcommand\aj{\jref@jnl{AJ}}%
\newcommand\araa{\jref@jnl{ARA\&A}}%
\newcommand\apj{\jref@jnl{ApJ}}%
\newcommand\apjl{\jref@jnl{ApJ}}%
\newcommand\apjs{\jref@jnl{ApJS}}%
\newcommand\ao{\jref@jnl{Appl.~Opt.}}%
\newcommand\apss{\jref@jnl{Ap\&SS}}%
\newcommand\aap{\jref@jnl{A\&A}}%
\newcommand\aapr{\jref@jnl{A\&A~Rev.}}%
\newcommand\aaps{\jref@jnl{A\&AS}}%
\newcommand\azh{\jref@jnl{AZh}}%
\newcommand\baas{\jref@jnl{BAAS}}%
\newcommand\jrasc{\jref@jnl{JRASC}}%
\newcommand\memras{\jref@jnl{MmRAS}}%
\newcommand\mnras{\jref@jnl{MNRAS}}%
\newcommand\pra{\jref@jnl{Phys.~Rev.~A}}%
\newcommand\prb{\jref@jnl{Phys.~Rev.~B}}%
\newcommand\prc{\jref@jnl{Phys.~Rev.~C}}%
\newcommand\prd{\jref@jnl{Phys.~Rev.~D}}%
\newcommand\pre{\jref@jnl{Phys.~Rev.~E}}%
\newcommand\prl{\jref@jnl{Phys.~Rev.~Lett.}}%
\newcommand\pasp{\jref@jnl{PASP}}%
\newcommand\pasj{\jref@jnl{PASJ}}%
\newcommand\qjras{\jref@jnl{QJRAS}}%
\newcommand\skytel{\jref@jnl{S\&T}}%
\newcommand\solphys{\jref@jnl{Sol.~Phys.}}%
\newcommand\sovast{\jref@jnl{Soviet~Ast.}}%
\newcommand\ssr{\jref@jnl{Space~Sci.~Rev.}}%
\newcommand\zap{\jref@jnl{ZAp}}%
\newcommand\nat{\jref@jnl{Nature}}%
\newcommand\iaucirc{\jref@jnl{IAU~Circ.}}%
\newcommand\aplett{\jref@jnl{Astrophys.~Lett.}}%
\newcommand\apspr{\jref@jnl{Astrophys.~Space~Phys.~Res.}}%
\newcommand\bain{\jref@jnl{Bull.~Astron.~Inst.~Netherlands}}%
\newcommand\fcp{\jref@jnl{Fund.~Cosmic~Phys.}}%
\newcommand\gca{\jref@jnl{Geochim.~Cosmochim.~Acta}}%
\newcommand\grl{\jref@jnl{Geophys.~Res.~Lett.}}%
\newcommand\jcp{\jref@jnl{J.~Chem.~Phys.}}%
\newcommand\jgr{\jref@jnl{J.~Geophys.~Res.}}%
\newcommand\jqsrt{\jref@jnl{J.~Quant.~Spec.~Radiat.~Transf.}}%
\newcommand\memsai{\jref@jnl{Mem.~Soc.~Astron.~Italiana}}%
\newcommand\nphysa{\jref@jnl{Nucl.~Phys.~A}}%
\newcommand\physrep{\jref@jnl{Phys.~Rep.}}%
\newcommand\physscr{\jref@jnl{Phys.~Scr}}%
\newcommand\planss{\jref@jnl{Planet.~Space~Sci.}}%
\newcommand\procspie{\jref@jnl{Proc.~SPIE}}%
\newcommand{\comment}[1]{}

\newcommand{\sm}[1]{\mbox{{\scriptsize #1}}}

\newcommand{\be}{\begin{equation}}
\newcommand{\ee}{\end{equation}}
\newcommand{\bea}{\begin{eqnarray}}
\newcommand{\eea}{\end{eqnarray}}
\newcommand{\bdm}{\begin{displaymath}}
\newcommand{\edm}{\end{displaymath}}
\newcommand{\bef}{\begin{figure}}
\newcommand{\eef}{\end{figure}}
\newcommand{\befone}{
  \begin{figure*}
  \centering
  \begin{minipage}{\textwidth}
}
\newcommand{\eefone}{\end{minipage}\end{figure*}}

\newcommand{\cm}{\mbox{cm}}
\newcommand{\km}{\mbox{km}}
\newcommand{\AU}{\mbox{AU}}
\renewcommand{\sec}{\mbox{s}}
\newcommand{\g}{\mbox{g}}
\newcommand{\Msol}{\mbox{$M_{\sun}$}}
\newcommand{\pc}{\mbox{pc}}

\newcommand{\K}{\mbox{K}}
\newcommand{\yr}{\mbox{year}}
\newcommand{\ys}{\mbox{years}}

\newcommand{\Ma}{{\cal M}}

\newcommand{\di}{\mbox{d}}

\newcommand{\vx}{\bf x}

\def\eps@scaling{0.95}

\def\showone#1{
  \centering
  \leavevmode
  \epsfxsize=\eps@scaling\linewidth
  \epsfbox{#1.eps}
}

\def\showtwo#1#2{
  \centering
  \leavevmode
  \epsfxsize=\eps@scaling\linewidth
  \epsfbox{#1.eps} 
  \\
  \epsfxsize=\eps@scaling\linewidth
  \epsfbox{#2.eps}
}

\def\showthree#1#2#3{
  \centering
  \leavevmode
  \epsfxsize=\eps@scaling\linewidth
  \epsfbox{#1.eps} 
  \\
  \epsfxsize=\eps@scaling\linewidth
  \epsfbox{#2.eps}
  \epsfxsize=\eps@scaling\linewidth
  \epsfbox{#3.eps}
}

\def\epstwo@scaling{0.44}

\def\showsix#1#2#3#4#5#6{
  \centering
  \leavevmode
  \epsfxsize=\epstwo@scaling\linewidth
  \epsfbox{#1.eps} \hfil
  \epsfxsize=\epstwo@scaling\linewidth
  \epsfbox{#2.eps} \hfil
  \epsfxsize=\epstwo@scaling\linewidth
  \epsfbox{#3.eps} \hfil
  \epsfxsize=\epstwo@scaling\linewidth
  \epsfbox{#4.eps} \hfil
  \epsfxsize=\epstwo@scaling\linewidth
  \epsfbox{#5.eps} \hfil
  \epsfxsize=\epstwo@scaling\linewidth
  \epsfbox{#6.eps}
}

\title[Formation and Evolution of Protostellar Disks]
      {The Formation and Evolution of Protostellar Disks; \\ 
	3D AMR Hydro-Simulations of Collapsing, Rotating Bonnor-Ebert-Spheres}
\author[R. Banerjee, R. E. Pudritz \& L. Holmes]{Robi Banerjee$^{1}$,
  Ralph E. Pudritz and Lindsay Holmes\\ 
\\
$^{1}$Sharcnet Postdoctoral Fellow \\
Department of Physics and Astronomy, McMaster University, Hamilton,
      Ontario L8S 4M1, Canada}

\begin{document}

\maketitle

\begin{abstract}

We present a detailed study of the collapse of molecular cloud cores
using high resolution 3D adaptive mesh refinement (AMR) numerical
simulations. In this first in a series of investigations our initial
conditions consists of spherical molecular core obeying the
hydrostatic Bonnor-Ebert-Profile with varying degrees of initial
rotation. Our simulations cover both the formation of massive disks in
which massive stars form as well as low mass disks. We use a
customized version of the FLASH code whose AMR technique allows us to
follow the formation of a protstellar disk and protostellar core(s)
through more than ten orders in density increase while continuously
resolving the local Jeans length (i.e. obeying the Truelove
criterion,~\citet{Truelove97}). Our numerical simulations also
incorporate the energy loss due to molecular line emission in order to
obtain a more realistic picture of protostellar core and disk
formation. Out initial states model system of mass $168\,\Msol$ and
$2.1\,\Msol$ that will form high and low mass stars, respectively. We
follow many features such as the development complex shock structures,
and the possible fragmentation of the disk. We find that slowly
rotating cores ($\Omega\,t_{\sm{ff}} = 0.1$) produce disks in which a
strong bar develops but which does not fragment. Faster initial
rotation rates ($\Omega\,t_{\sm{ff}} = 0.2$) result in the formation
of a ring which may fragment into two star-forming cores. The size of
the rings found in our simulated disks agree with the observations of
similar systems.

\end{abstract}

\begin{keywords}
accretion, accretion discs, hydrodynamics, ISM: clouds, evolution,
methods: numerical
\end{keywords}

\section{Introduction}

The formation of protostars and their surrounding disks from
collapsing, dense, molecular cores within molecular clouds is still an
actively debated topic. Early pioneering works
by~\citet{Bodenheimer68}, \citet{Larson69}, and \citet{Penston69}
addressed this problem using numerical simulations for various initial
conditions. \citeauthor{Bodenheimer68} studied collapses of isothermal
spherical symmetric gas clouds which are initially out of equilibrium.
They pointed out that the solutions depend only very weakly on the
choice of the initial density profile (homogeneous or
Chandrasekhar-type). \citet{Larson69} and \citet{Penston69} included
the effect of energy dissipation by radiation in their 1D simulations
and found that cooling during the initial collapse phase is efficient
enough to keep the gas at its initial temperature. Although both
studies start with different initial conditions (\citeauthor{Larson69}
used a homogeneous density distribution whereas \citeauthor{Penston69}
used a Bonnor-Ebert-Profile) they obtained similar results during the
isothermal collapse. Starting with a highly unstable, initial singular
isothermal sphere \citet{Shu77} obtained a different evolution
scenario: a self-similar inside-out collapse and claimed that the
former solutions are physically artificial. \citet{Hunter77} in his
study of unstable isothermal spheres pointed out that both
investigations address the same problem but refer to two different
stages of the isothermal collapse. The
\citeauthor{Larson69}-\citeauthor{Penston69} studies investigate the
collapse of molecular clouds prior to the formation of protostellar
core whereas \citeauthor{Shu77}'s study apply to the accretion phase
of an non-rotating cloud with a central protostar.

In reality, molecular cores have some inital angular momentum so that
most of the collapsing material ends up in a rotating accretion disk
(due to their non-vanishing initial angular momentum). This
significantly alters the dynamics of the protostellar core's
evolution.  More recently, this issue was addressed with the help of
2D and 3D numerical simulations by several groups \citep[for a review
see][]{Bodenheimer00}. For instance \citet{Burkert93} studied the
fragmentation of a rotating sphere of uniform density with a $m =2$
perturbation on a fixed nested grid. Their simulations showed that a
bar formed which then fragmented into two low mass binaries and several
lighter secondary fragments. \citet{Truelove98} used a 3D AMR
technique that resolves the local Jeans length throughout the
simulation to reinvestigate the fragmentation problem. These authors
found that a bar like structure collapses instead to a single filament
without further fragmentation. These differences can be shown to
depend upon the ability to resolve the local Jeans length during the
collapse \citep[e.g. review][]{Bodenheimer00}. An equivalent
requirement for SPH simulations was found by~\citet{Bate97} who showed
that the minimal resolvable mass must be smaller than the Jeans mass
to avoid numerical fragmentation. But these authors also noted that
the inclusion of artificial viscosity and heating, which increases
the Jeans length and slows the collapse, lead to physical
fragmentation of bars which is predicted in the work
of~\citet{Inutsuka92}.  Clouds with an initial Gaussian density profile
were studied by \citet{Boss93} who concluded that the most sensitive
parameter for clouds to fragment is $\alpha$, the ratio of the thermal
to gravitational energy.

\citet{Matsumoto03} did an extensive parameter study of rotating
Bonnor-Ebert-Spheres with different rotation profiles and rotation
speeds. Their investigations are based on a three dimensional nested
grid technique which resolves the local Jeans length and the use of
two different equation of states depending on the central density (an
isothermal EOS in the low density regime and a barotropic EOS in the
high density regime). They showed that the final structure of the
collapsed cloud (bar, ring, or binary system) depend sensitively on
the parameter $t_{\sm{ff}} \, \Omega$, where $t_{\sm{ff}}$ and
$\Omega$ are the initial free fall time and the angular velocity,
respectively.

Another important physical process that alters the dynamics of the
core formation and disk evolution is the cooling process during the
collapsing phase. An often used simplification to account for the
decreasing efficiency of the cooling mechanism is the previously
mentioned sudden switch of the EOS, depending on the local density. We
do not follow this approach but rather include the loss of energy due
to molecular line emission by collisional excitations. In this work we
show that the effective equation is a complex function of time,
density, and space which influences the structure of the
protoplanetary disk and its possible fragmentation.

The study of Bonnor-Ebert-Spheres as initial states for star formation
are of particular interest because they resemble marginal stable
molecular cores in pressure equilibrium and have flat topped rather
than singular density profiles. Observed evidence for
Bonnor-Ebert-type cores are found in a variety of molecular clouds
\citep[e.g.][]{Racca02, Harvey01, Alves01} and in numerical
simulations of star cluster formation in molecular
clouds~\citep{Tilley04}. The oberservation of massive disks undergoing
high mass star formation \citep{Chini04} suggests that a similar
scenario pertains to massive star formation \citep[e.g.][]{Yorke02}.

In this paper we investigate low and high mass star formation from
rotating molecular cores through AMR simulations.  This paper is
organized as follows: in Sec.~\ref{sec:numerics} we explain our
numerical scheme including the cooling procedure, and summarize the
properties of the static Bonnor-Ebert-Sphere. Our results of the
collapse in the isothermal and non-isothermal regime are given in
Sec.~\ref{sec:core_formation}. This sections includes also a review of
the pure isothermal collapse of a overcritial Bonnor-Ebert-Sphere. The
formation of disks in low and high mass rotating cores are discussed
in Sec.~\ref{sec:disk_formation} and in Sec.~\ref{sec:fragmentation}
we present our results on the formation of rings and bars in the disk
and their possible fragmentation. Finally, we summarize and discuss
our findings in Sec.~\ref{sec:summary}.

\section{Numerical scheme and initial conditions}
\label{sec:numerics}

For our studies of collapsing gas clouds we used the FLASH
code~\citep{FLASH00} which solves the coupled gravito-hydrodynamic
equations on an adaptive mesh.  FLASH is based on a block structured
adaptive mesh refinement (AMR) technique which is implemented in the
PARAMESH library~\citep{PARAMESH99}. This AMR technique allows us to
follow the core formation over more the ten orders of magnitude in
density increase without violating the Truelove
criterion~\citep{Truelove97}. For this purpose we implemented a new
refinement criterion to the FLASH code which assures that the local
Jeans length
\be 
\lambda_J = \left(\frac{\pi\,c^2}{G\,\rho}\right)^{1/2}
\label{eq:jeans_lenth}
\ee 
is at least resolved by $N > 4$ grid points, where, $c$, $G$, and
$\rho$ are the isothermal sound speed, the gravitational constant, and
the local mass density, respectively. The runs we present in this work
are performed with an even smaller Jeans number $J = \Delta
x/\lambda_J$ of $1/8$ or $1/12$ ($\Delta x$ is the grid spacing in one
dimension at the point $x$).

\subsection{Properties of Bonnor-Ebert-Spheres}

We choose as an initial setup a non magnetized spherical gas cloud
in a marginal stable hydrostatic equilibrium, i.e. a
Bonnor-Ebert-Sphere~\citep{Ebert55, Bonnor56}. With the definitions 
\bea
\rho(\xi) & \equiv & \rho_0\,e^{-\Phi(\xi)}
\label{eq:BE_def_density}
\\
\xi & \equiv & \frac{r}{r_0} \quad ; \quad
 r_0 = \frac{c}{\sqrt{4\pi\,G\,\rho_0}}
\label{eq:BE_def_radius}
\eea
the radial density profile is given by the solution of the Lane-Emden
equation \citep[cf.][]{Chandra67}:
\be
\frac{1}{\xi^2}\,
  \frac{\di}{\di\xi}\left(\xi^2\,\frac{\di\Phi}{\di\xi}\right) 
  =  e^{-\Phi}
\label{eq:BE_diff}
\ee
\be
\Phi(0) = \Phi^{\prime}(0) = 0
\label{eq:BE_diff_boundcond}
\ee
where $\rho_0$ is the initial central density, $c$ is the
thermal sound speed, $r_0$ is the characteristic radius of the gas
cloud, and a prime ($^{\prime}$) denotes the differentiation with
respect to $\xi$.  
The solution is characterized by a density profile with a flat core
of density $\rho_0$ and an envelope which falls off with radius
roughly as $r^{-2}$ until truncated by the pressure of the surrounding
medium. 
Note that the function $\phi(\xi)$ is
 -- up to an integration constant -- equal to the gravitational
 potential $\psi$ (cf. Eq.~(\ref{eq:BE_pot_solution})).

We use a fourth order Runge-Kutta numerical scheme to solve
Eq.~(\ref{eq:BE_diff}) for the potential $\phi(\xi)$, and the
acceleration $\phi^{\prime}(\xi)$ which give the density profile
$\rho(\xi)$ and the mass parameter $q(\xi)$. 
Fig.~\ref{fig:BE_parms} summarizes the radial dependences of the
dimensionless parameters which determine the Bonnor-Ebert-Sphere with
the cut-off radius of $\xi_c = 6.5$.

\bef
\showone{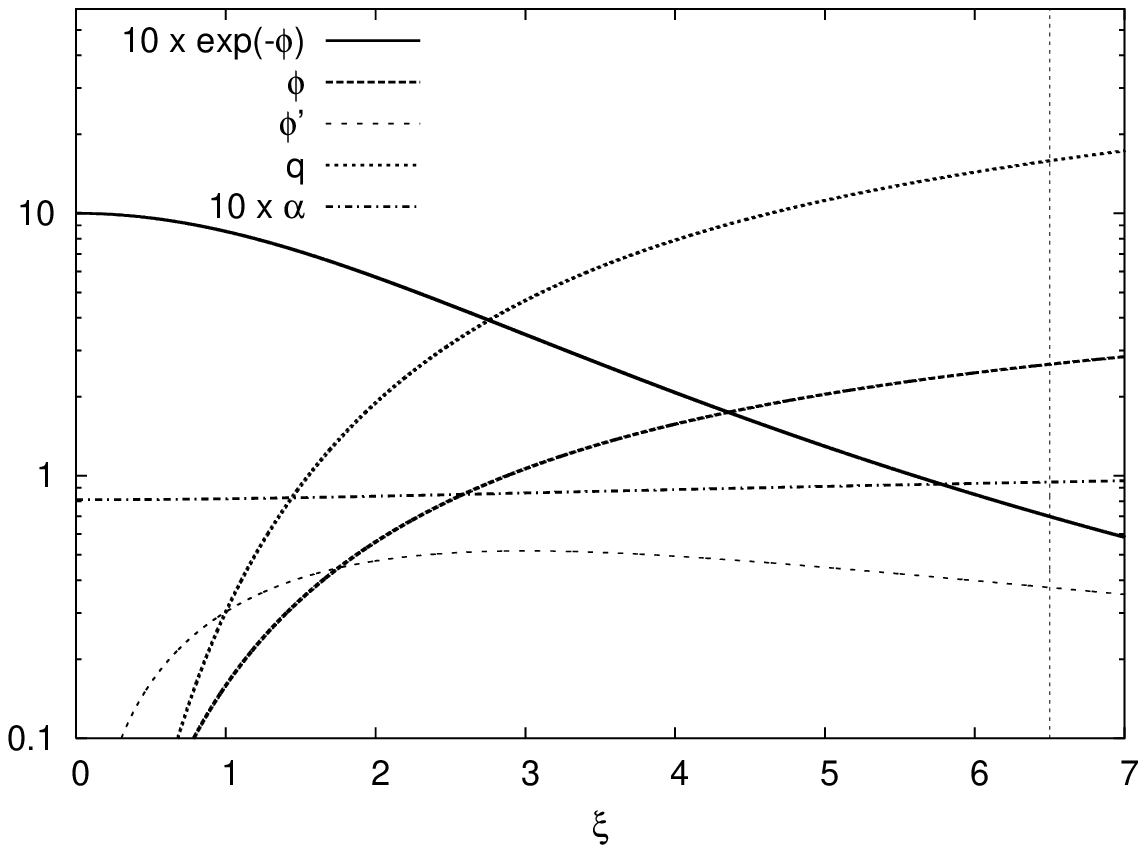}
\caption{Shows the dimensionless parameters of a Bonnor-Ebert-Sphere:
  the density profile (solid line) multiplied by a factor of $10$, the
  potential $\phi$ (dashed line), the gravitational acceleration
  $\phi^{\prime}$ (thin dashed line), the mass parameter $q$ (dotted line),
  and the ratio of the thermal energy to gravitational energy $\alpha$
  (dash-dotted line) multiplied by a factor of 10 for a sphere with
  $\xi_c = 6.5$. The vertical line indicates this cut-off radius.} 
\label{fig:BE_parms}
\eef

The total mass within a Bonnor-Ebert sphere at the radius $\xi$ is
given by
\bea
M(r) & = & \frac{c^3}{\sqrt{4\pi\,G^3\,\rho_0}}\,q \\
   & \approx & 1\, \Msol \, \left(\frac{T}{20\,\K}\right)^{3 \over 2} \,
     \left(\frac{\rho_0}{8\times 10^{-17}\,\g\,\cm^{-3}}\right)^ 
	  {-{1 \over 2}} 
\nonumber \\
   & & \times \left(\frac{q}{15.85}\right) \quad , \nonumber
\label{eq:BE_sphere_mass}
\eea
where $q$ is the integrated dimensionless mass factor \citep[see
  also][]{McLaughlin96},
\be
q = \xi^2\,\Phi^{\prime} \quad .
\label{eq:BE_q_param}
\ee
Such a sphere is unstable to collapse if its
dimensionless radius $\xi$ is larger than
6.451~\citep{Bonnor56}\footnote{Note, as $\xi = 2\pi\,r/\lambda_J$
  this instability criterion is almost equal to Jeans' 
  instability criterion, i.e. $r > \lambda_J$, and the amount of Jeans
  masses within the sphere is $3/\pi^3\,q \sim 1.53$.}.
Given the external pressure $P_{\sm{ext}}$ and the sound speed of such
a spherical cloud the physical radius and total mass are,
respectively:
\bea
R & = & 0.49 \, \frac{c^2}{G^{1/2}} \, P_{\sm{ext}}^{-1/2} \\
M & = & 1.18 \, \frac{c^4}{G^{3/2}} \, P_{\sm{ext}}^{-1/2} \quad .
\eea

%\be
%\frac{\di^2\xi}{\di t^2} =
%  \frac{8}{3\pi^2} \, t_{\sm{ff}}^{-2} \, \phi^{\prime}
%\ee

The ratio of thermal to gravitational energy, $\alpha$, within a
Bonnor-Ebert-Sphere, which varies only slightly with radius, is
completely fixed by its truncation radius $\xi_c$. This can seen by
the following argument. The gravitational potential $\psi$ inside the
sphere is determined by
\be
\frac{\di\psi}{\di\,r} = G\,\frac{M}{r^2} \quad,
\label{eq:BE_potential}
\ee
where the mass $M$ is given by Eq.~(\ref{eq:BE_sphere_mass}). Using
the matching conditions at the edge of the sphere with the
potential in the ambient media, i.e. $\psi \propto r^{-1}$ for $\xi >
\xi_c$, the solution of Eq.~(\ref{eq:BE_potential}) is
\be
\psi = - c^2\,\left[\phi_c + q_c - \phi \right]
\quad \mbox{for} \quad \xi \leq \xi_c \quad,
\label{eq:BE_pot_solution}
\ee
where $\phi_c \equiv \phi(\xi_c)$, $\phi^{\prime}_c \equiv
\phi^{\prime}(\xi_c)$, and $q_c \equiv 
q(\xi_c)$. Therefore, $\alpha = 3/2\,c^2/|\psi|$ is given by
\be
\alpha = \frac{3}{2}\,
  \frac{1}{\phi_c + q_c - \phi} \quad.
\label{eq:BE_alpha}
\ee
With $\xi_c = 6.5$ ($\phi_c = 2.66$, $q_c = 15.85$) $\alpha$ is in the
range of:
\be
 0.081 \le \alpha \le 0.094 \quad.
\label{eq:alpha_vary}
\ee

The rotational to gravitational energy $\beta$ for a rigidly rotating
Bonnor-Ebert-Sphere is given by
\be
\beta = \frac{2}{5} \, \frac{v_{\varphi}^2}{c^2} \,
 \frac{1}{\phi_c + q_c - \phi} 
  =
 \frac{16}{15\pi^2} \, 
 \frac{\xi^2 \, \left(t_{\sm{ff}}\,\Omega\right)^2}
      {\phi_c + q_c - \phi}
\quad,
\label{eq:BE_beta}
\ee
where $v_{\varphi}$ and $\Omega$ are the toroidal velocity and angular
velocity, respectively. For instance, a sphere rotating with
$t_{\sm{ff}}\,\Omega = 0.1$ has values of $\beta$ that span the range
\be
  0 \leq \beta \leq 0.0029 \quad ;
\label{eq:beta_vary}
\ee
and scales roughly proportional to $\xi^2$, while $\beta$ takes its
maximum at the edge of the sphere and is given by $\beta_{\sm{max}} =
0.11\, \left(\phi^{\prime}_c\right)^{-1} \, \left(t_{\sm{ff}} \,
\Omega\right)^2$.

%\subsection{Rotation}
%$\Omega_1 = 2.755\times 10^{-15}\,\mbox{rad}\,\sec^{-1}$
%run A1 $t_{\sm{ff}}\,\Omega = 0.1$
%$\Omega_2 = 5.509\times 10^{-15} \, \mbox{rad} \, \sec^{-1}$
%run A2 $t_{\sm{ff}}\,\Omega = 0.2$

\begin{table}
  \begin{tabular}{|c|c|c|c|c|} \hline
  run & $t_{\sm{ff}} \, [\sec]$ & $\Omega \, [\mbox{rad} \, \sec^{-1}]$ 
      & $t_{\sm{ff}}\,\Omega$ & $\beta_{\sm{max}}$ \\ \hline \hline
  A1  & $3.63\times 10^{13}$  & $2.755\times 10^{-15}$
      & $0.1$ & $2.88\times 10^{-3}$ \\ \hline
  A2  & $3.63\times 10^{13}$  & $5.509\times 10^{-15}$ 
      & $0.2$ & $1.15\times 10^{-2}$ \\ \hline
  A3  & $3.63\times 10^{13}$  & $8.265\times 10^{-15}$ 
      & $0.3$ & $2.59\times 10^{-2}$ \\ \hline
  B68 & $2.12\times 10^{12}$ & $9.430\times 10^{-14}$ 
      & $0.2$ & $1.21\times 10^{-2}$ \\ \hline
  \end{tabular} 
\caption{The initial parameters for the simulations we present in this
  paper. Where $\Omega$, $t_{\sm{ff}}$, and $\beta_{\sm{max}}$ are the
  initial angular velocity, the initial free fall time, and the
  maximum of ratio of the rotational and gravitational energy,
  respectively.} 
\label{tab:runs}
\end{table}

\subsection{Initial conditions: low and high mass cores}

In this paper we present the results of four simulations aiming to
study both low and high mass pre-stellar cores. We start each
simulation with an overcritical Bonnor-Ebert-Sphere which rotates
initially with a constant angular velocity $\Omega$. Three of these
initial models have the same initial core density $\rho_0$ but
different initial angular velocities $\Omega$ (run A1 -- A3) and a
mass of $168\,\Msol$ characteristic of massive star formation
\citep{Chini04}, and one low mass model starting with a core density
according to the observed Bok globule Barnard 68 by~\citet{Alves01}
with a mass of $2.1\,\Msol$ and a medium angular velocity (run
B68). The initial parameters of these runs are summarized in
table~\ref{tab:runs}.

\comment{Observed Bonnor-Ebert-Spheres: 

\citet{Racca02} Coalsack globule 2,
$\xi_{\sm{max}} = 7.0 \pm 0.3$, $\rho_0 = 2.5\times 10^{-19} \, \g \,
\cm^{-3}$, $T = 15 \, \K$, $M = 4.5 \, \Msol$, $P_{\sm{ext}} =
9\times 10^{-11} \, \mbox{Pa}$, $R = 0.1 \, \pc$.

\citet{Harvey01} dark globule B335, $\xi_{\sm{max}} = 12.5 \pm 2.6$,
$T = 10 - 12 \, \K$, $M = 14 \, \Msol$, $R = 0.15 \, \pc$, rotation
by\citet{Frerking87} $\Omega = 1.2\times 10^{-14} \, \mbox{rad} \, \sec^{-1}$ .

\citet{Alves01} Bok globule Barnard 68, $\xi_{\sm{max}} = 6.9
\pm 0.2$, $R = 1.25\times 10^4 \, \AU$ and a core density $\rho_0 =
2.14\times 10^{-18} \, \g \, \cm^{-3}$, $T = 16 \, \K$.}

One particularly interesting object for low mass star formation is the
Bok globule Barnard 68 which was extensively studied
by~\citet{Alves01}. The density profile of this cloud exhibits a close
to perfect Bonnor-Ebert profile with $\xi_c = 6.9 \pm 0.2$
corresponding to a physical radius $R = 1.25\times 10^4 \, \AU$, a
core density of $\rho_0 = 1.0\times 10^{-18} \, \g \, \cm^{-3}$, a
total mass of $M_{\sm{B68}} = 2.1\,\Msol$, and a temperature of $T =
16\,\K$ (we set up this run with an external pressure of
$P_{\sm{ext}}/k_{\sm{B}} = 2.4\times10^5\,\K\,\cm^{-3}$). This
corresponds to model marked by a diamond shown in
Fig.~\ref{fig:BE_scaling}. One can see that this observed system is
nearly immediately scalable from the initial state that we use defined
by the triangle. To simulate the collapse of a realistic molecular
cloud we initialize run B68 with the physical parameters given
above. According to~\citet{Lada03} Barnard 68 might rotate slightly
with a ratio of the rotational to gravitational energy of a few
percent. We initialize the sphere with $t_{\sm{ff}} \, \Omega = 0.2$
corresponding to $\beta_{\sm{max}} \approx 0.01$.

Most of our simulations are designed to study the formation of massive
accretion disks in which massive (O, B) stars may form. The collapse
of massive (up to $120\,\Msol$), initially singular cores has been
simulated by \citet{Yorke02} who included a careful treatment of
radiative transfer effects. Recent observations by \citet{Chini04}
have discovered a massive ($20\,\Msol$) star that is being formed
within a very massive, large accretion disk (at least $100\,\Msol$ of
gas) in the young star forming regin of M 17. We therefore chose the
initial setup for the study of massive star and disk formation
corresponding to runs A1 -- A3 as follows: the cut-off radius is
$\xi_c = 6.5$ corresponding to a physical radius of $R = 1.62\,\pc$
($R = 3.34\times 10^{5}\,\AU$), a unperturbed core density $\rho_0 =
3.35\times 10^{-21}\,\g\,\cm^{-3}$, and sound speed $c =
0.408\,\km\,\sec^{-1}$ ($T = 20\,\K$). This parameters with a $10\%$
overdensity result in a total mass of the gas cloud of $M =168\,\Msol$,
and an external pressure $P_{\sm{ext}}/k_B = 3.2\times 10^3 \, \K \,
\cm^{-3}$. The inital free fall time $t_{\sm{ff}} = \sqrt{3\pi /
32\,G\,\rho_0}$ of our Bonnor-Ebert sphere is $1.1\times
10^6\,\ys$\footnote{Note that the more accurate dynamical time scale
for a Bonnor-Ebert-Sphere is $t_0 = 1/\sqrt{4\pi\,G\,\rho_0} \approx
0.52 \, t_{\sm{ff}}$}.  The edge of the sphere is defined by a density
decrease of the ambient medium by a factor of $100$ whereas the
pressure is continuous at the edge of the sphere. Therefore, the
ambient low density gas is a hundred times warmer ($T_{\sm{amb}} =
2000\,\K$) than the cold gas cloud and the sound speed there is
$c_{\sm{amb}} = 4.08\,\km\,\sec^{-1}$. To make sure that the evolution
of the sphere is not influenced by the simulation box boundaries we
choose the simulation volume to be $(100 \, r_0)^3 \sim (15\,R)^3 \sim
(25\,\pc)^3$.

\bef
\showone{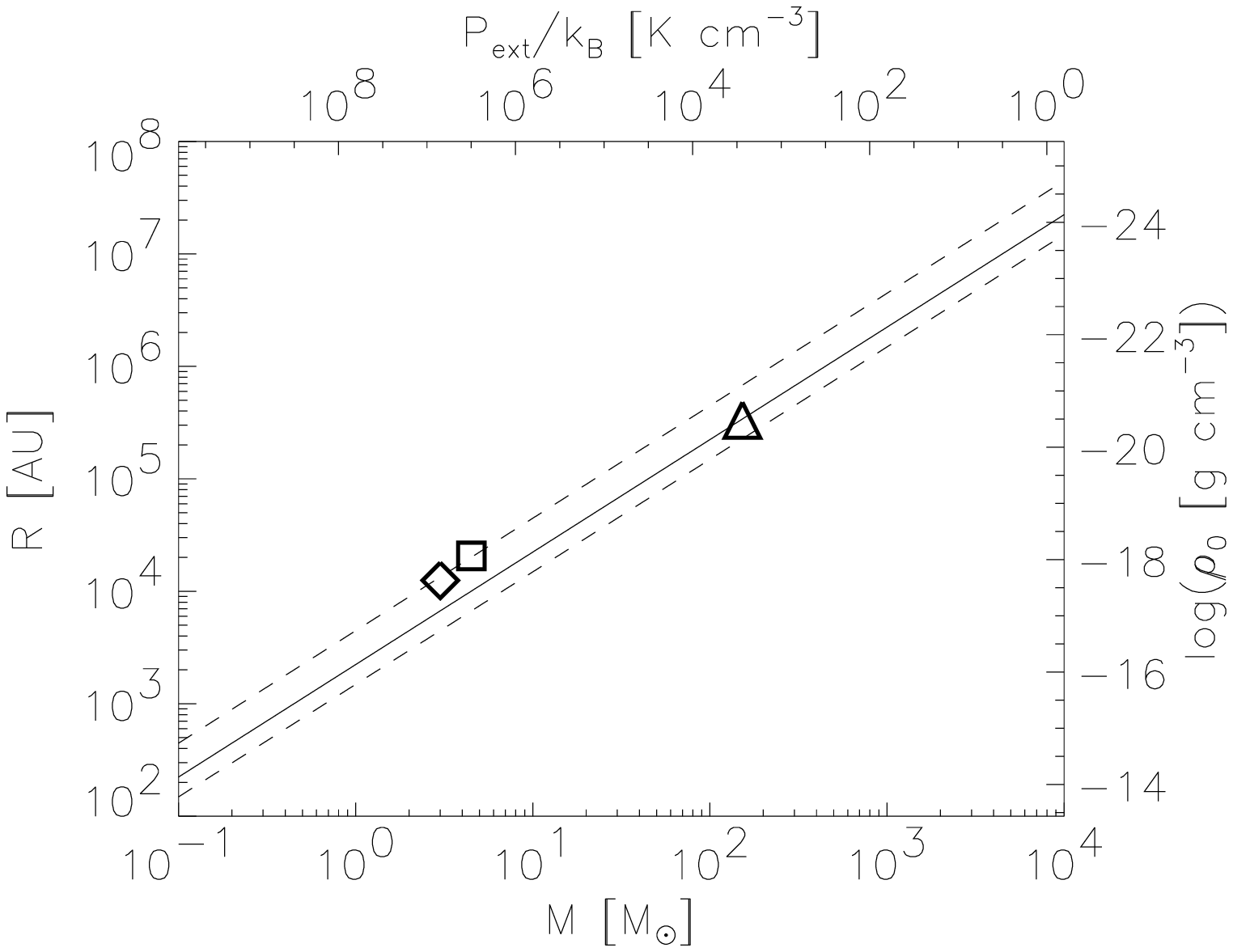}
\caption{Scaling of the physical parameters of critical
  Bonnor-Ebert-Spheres. The solid line shows the parameters for sphere
  with $T = 20 \, \K$, and the upper and lower dashed lines show the
  relations for a sphere with $T = 10 \, \K$ and $T = 30 \, \K$,
  respectively (Note that the axis labeling for $P_{\sm{ext}}$ is only
  valid for $T = 20 \, \K$ and has to be rescaled for different
  temperatures). 
%Note $\rho \propto R^{-1} \, c^2$ and $M \propto R \, c^2$
  The triangle ($\triangle$) refers to our model parameters for A1 --
  A3, the square ($\sq$) shows the Coalsack globule 2 observed
  by~\citet{Racca02}, and the diamond ($\diamond$) marks the
  properties of Barnard 68 (our model B68) as given in
  \citet{Alves01}.}
\label{fig:BE_scaling}
\eef

Fig.~\ref{fig:BE_scaling} shows the scaling relations in the
isothermal regime for the physical mass, radius, core density, and
external pressure for critical Bonnor-Ebert-Spheres. All spheres at a
given temperature with the properties indicated by the shown lines are
equivalent.

Radiative cooling by molecular excitation lines is very efficient in
the low density regime. In fact, molecular clouds will maintain the
same temperature during their collapse until the density rises above
$n \sim 10^{7} \, \cm^{-3}$. We wish to study this transition to less
efficient cooling very carefully and therefore start the collapse of
Bonnor-Ebert-Spheres in this low-density, isothermal phase. Therefore,
we choose the initial density for the runs A1 -- A3 ($n_0 \sim
10^{3}\,\cm^{-3}$) in order to sample the full range of densities
given by the cooling data of~\citet{Neufeld93} and \citet{Neufeld95}
. Their cooling data span a number density of $10^3 \leq n(H_2) \leq
10^{10} \, \cm ^{-3}$.  We emphasize however that these initial
physical conditions can be rescaled only as long as the gas stays at
the same temperature, i.e. contracts isothermally (cf. discussion in
section~\ref{sec:core_formation}).  Therefore, we can apply our
results to observed molecular clouds with different initial physical
parameters.

All runs are set up with a $10\%$ $m = 2$ density perturbation on top
of a slightly enhanced Bonnor-Ebert-Profile, i.e., 
\be \rho = \rho_{\sm{BE}} 
   \left(1.1 + 0.1\, \cos\left(2\,\varphi\right) \right) \quad,
\label{eq:density_perturbation}
\ee 
where $\rho_{\sm{BE}}$ obeys Eq.~(\ref{eq:BE_diff}) and $\varphi$ is
the azimuthal angle (the 10\% overall density enhancement guarantees
the collapse of a critical BE-sphere since it overwhelms the
relatively small amount of rotation). All spheres rotate initially
with a rigid body rotation profile with the amount of rotation given
in table~\ref{tab:runs}.

\subsection{Radiative cooling}

Earlier \citep[for a review see][]{Bodenheimer00} and recent
\citep[e.g.][]{Matsumoto03} simulations of protostellar disk formation
and fragmentation typically use two types of equations of state: (i)
an isothermal equation of state throughout, which assumes that
radiative cooling is always efficient enough to keep the gas at the
same initial temperature, or (ii) an isothermal EOS that is switched
to an adiabatic one at sufficiently high density. We did some
simulations of the latter approach and found problems not only due to
numerical artifacts but also due to physical difficulties because the
sudden change of the equation of state violates energy conservation
(an isothermal EOS corresponds to an infinite heat bath whereas an
adiabatic EOS assumes a finite thermal energy).

To achieve a more realistic, and more physically correct picture of
the formation of protostellar disks, we account in our simulations for
the effect of cooling by collisional excitations of gas molecules.  We
use cooling functions provided by~\citet{Neufeld93, Neufeld95}.  These
authors computed the radiative cooling rates as a function of the gas
temperature, density and optical depth for the temperature and density
range of $10 \leq T \leq 2500 \, \K$ and $10^3 \le n(H_2) \le 10^{10}
\, \cm^{-3}$.  Using steady state molecular abundances for the most
important coolants in molecular clouds (H$_2$, H$_2$O, CO, O$_2$, HCl,
C, and O) they provide a total cooling rate which can be used to
calculate the energy loss due to radiative cooling.  We use their
optical depth parameter for an singular isothermal sphere (which is
appropriate for a sphere with a $1/r^2$ density profile)
\be
{\tilde N}_{\sm{SIS}} = 5.1\times 10^{19} \, 
 \left(\frac{n}{\cm^{-3}}\right)^{1/2} \, 
  \cm^{-2} \quad \mbox{per} \quad \km \, \sec^{-1}
\ee
to read off the total cooling rate from the provided cooling data
base. The cooling rates provided by~\citet{Neufeld93}
and~\citet{Neufeld95} are available for densities $n(H_2) \le 10^{10}
\, \cm^{-3}$. Because the cooling power per $H_2$ molecule in the high
temperature ($T > 100 \, \K$) and high density ($n > 10^{10} \,
\cm^{-3}$) regime is nearly independent of the density, we extrapolate
their data in the high density range assuming that the cooling power
is only a function of temperature. As the maximum temperatures in our
simulations are well below the dissociation temperature of $H_2$
($T_{\sm{diss}} \sim 2000 \, \K$) we do not account for this cooling
process in the present simulations. We also did not consider cooling
processes due to gas-grain interactions which we will implement in
future studies.
However, our results are similar in character to those
of~\citet{Yorke95} who did include dust cooling.

The treatment of the thermal cooling of the gas allows one to
calculate the internal energy density, $\epsilon$, at any position and
time. The pressure, $P$, at each time step is then calculated by: \be
P = \epsilon \left(1 - \gamma \right) \quad, \ee where $\gamma$ is the
adiabatic index. Throughout this study we use a constant $\gamma$ of
$5/3$. Including cooling in our simulations the thermal energy is not
only altered by compressional heating during the collapsing phase but
also by the loss of energy due to radiative emission. We found it is
useful to compare our approach which uses full molecular cooling, with
the earlier adiabatic model studies by calculating an effective EOS,
i.e. $\gamma_{\sm{eff}} \equiv \di P / \di \rho$, at each time step.

\section{Core formation and evolution}
\label{sec:core_formation}

\subsection{Pure isothermal collapse of Bonnor-Ebert-Spheres}
\label{sec:isothermal_general}

\bef
\showtwo{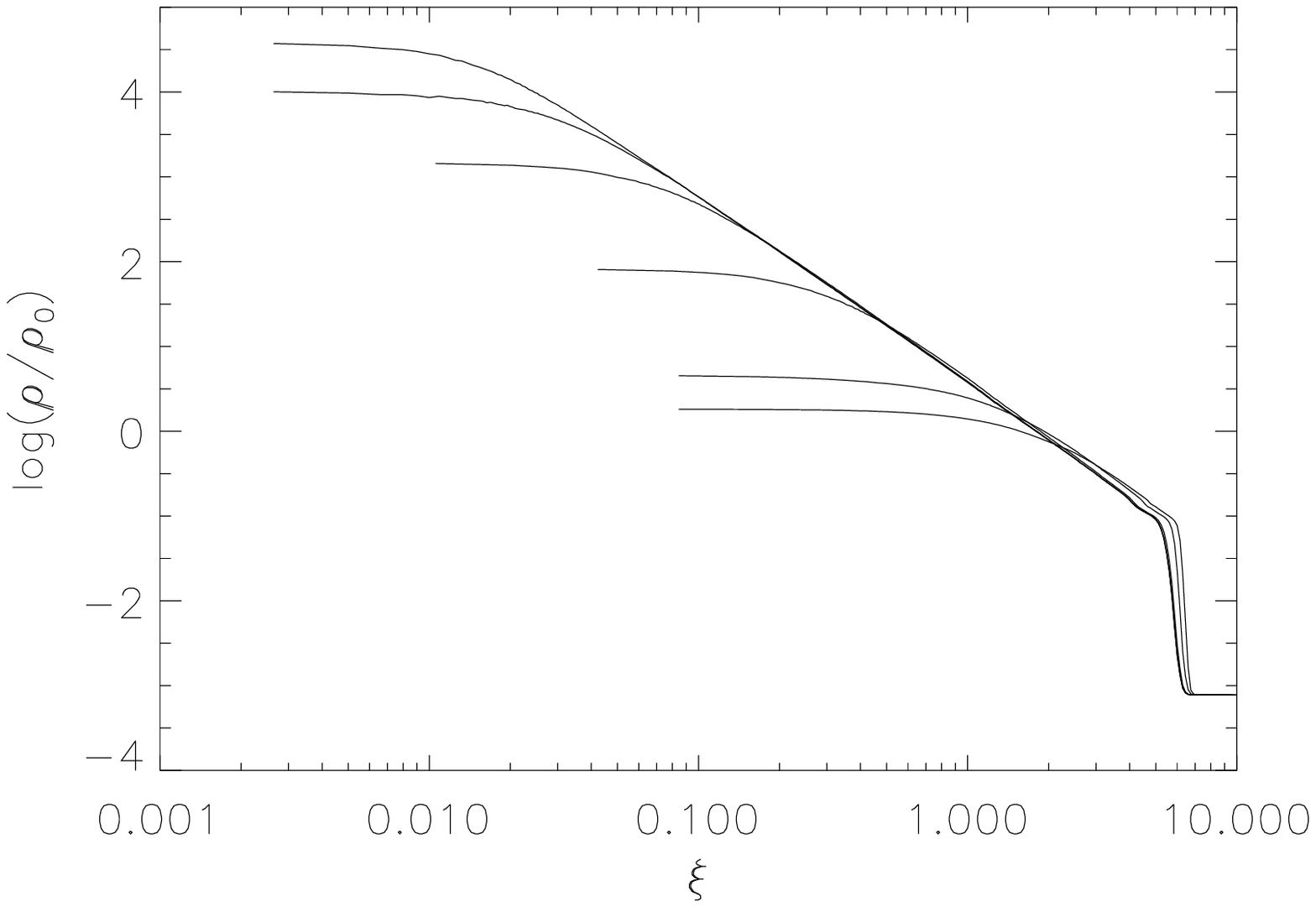} 
	{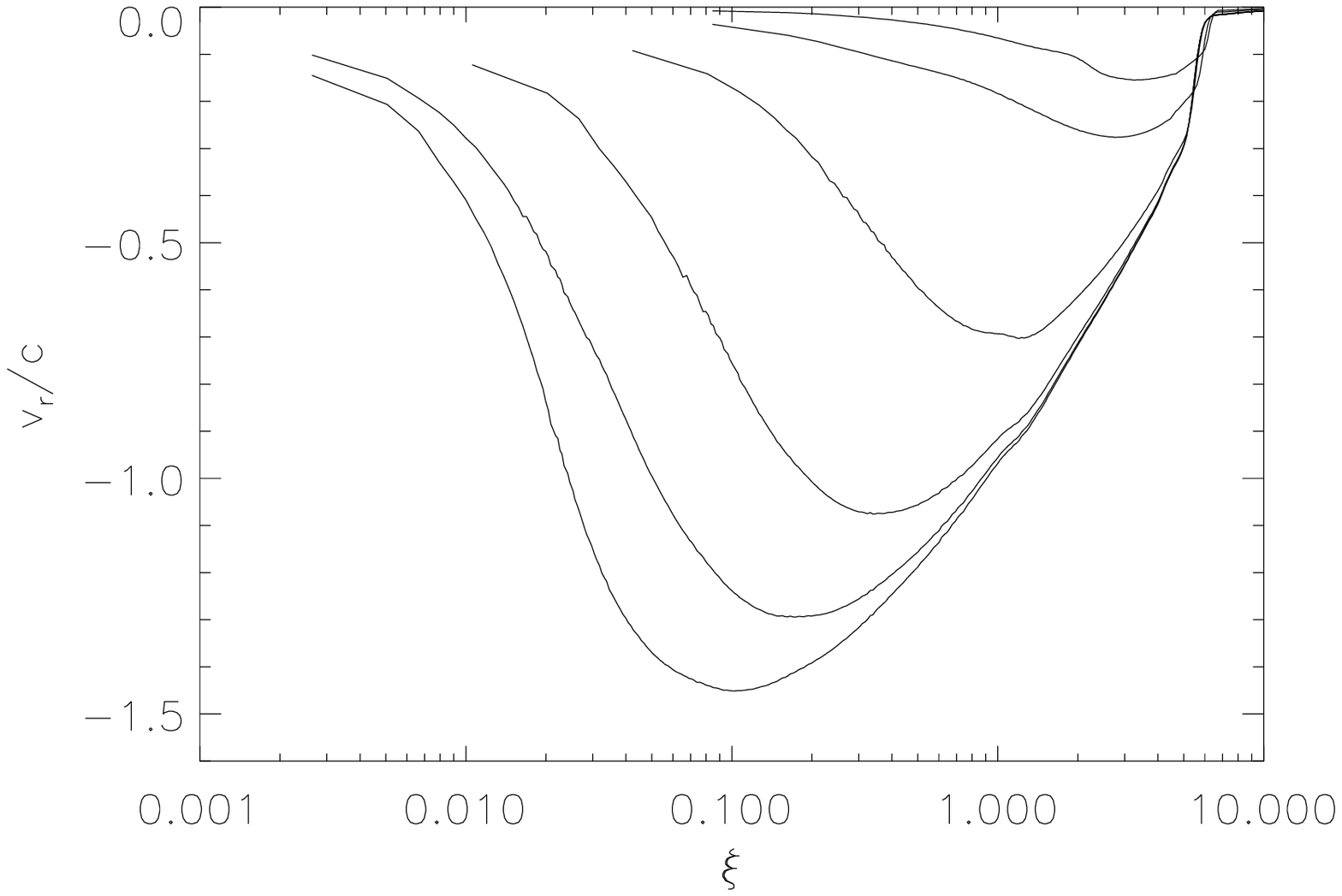}
\caption{Radial profiles of density (first panel) and the radial
  velocity (second panel) at different times in the isothermal
  regime. The sphere starts to collapse from {\em outside-in} leaving
  a $1/r^2$ density profile in the envelope while the flat core region
  shrinks. The maximal infalling velocity appears close to the edge of
  the core and is moving inward with increasing density. The shown
  lines correspond to (from bottom to top of the first panel, and top
  to bottom for the second panel) $t/t_0 = 3.7, 5.6, 7.13, 7.33, 7.36,
  7.37$, where $t_0 = 1/\sqrt{4\pi\,G\,\rho_0} \sim 6.0\times 10^5 \,
  \ys$ and the data are taken from run A1.}
\label{fig:dens_vrad_plt_iso}
\eef

Early theoretical work on star formation \citep[e.g.][]{Hayashi65}
already pointed out that cooling during the initial stage of
collapsing molecular cloud cores is efficient enough to ensure
constant temperatures of the gas over 4 -- 5 orders of magnitude in
density increase. Hence, the isothermal collapse proceeds as a free
falling contraction of the gas. Our simulations -- including molecular
cooling -- resemble this early isothermal phase until $n \sim 10^{7.5}
\, \cm^{-3}$. Fig.~\ref{fig:temp_dens} and Fig.~\ref{fig:gamma_dens}
from our simulation A1 clearly show the evolution in the isothermal
phase at low densities. Because the solutions of collapsing isothermal
Bonnor-Ebert-Spheres are rather general we devote the following
paragraph to review the collapse of a pure isothermal cloud.

A detailed 1D numerical investigation of a collapsing isothermal
Bonnor-Ebert-Sphere was done by~\citet{Foster93}. The results of our
3D simulations in the isothermal regime are very similar to the
results found by~\citet{Foster93}. Fig.~\ref{fig:dens_vrad_plt_iso}
shows the time evolution of the density profile and the radial
velocity profile which have the same trend as the 1D results: after a
sound wave propagates through the sphere at a sound crossing time of
$t_{\sm{sc}} = \xi/\sqrt{4\pi\,G\,\rho_0}$ ($\sim 3.8\times 10^6 \,
\ys$ in the case of runs An and $\sim 2.4\times 10^5 \, \ys$ in the
case of B68) the cloud starts to collapse from {\em
outside-in}. Initially the gravitational acceleration of the
Bonnor-Ebert-Sphere
\be g = \frac{G\,M}{r^2} = c \,
\sqrt{4\pi\,G\,\rho_0} \, \phi^{\prime}
\label{eq:BE_acceleration}
\ee 
is largest at $\xi = 3.0$, ($\phi^{\prime}(3.0) = 0.517$), and
decreases only slightly with radius ($\phi^{\prime}(6.5) = 0.375$).
Soon after the low density material from the outer part of the cloud
is accelerated towards the center the density profile of the sphere
starts to change: a high density flat core with a $1/r^2$ envelope
builds up.  The $1/r^2$ law in the envelope reflects the fact that the
sphere maintains a state that is close to hydrostatic equilibrium in
the envelope.  Now the gravitational acceleration is highest at the
edge of the flat core and decreases with $1/r$. Therefore, material at
the core boundary is constantly accelerated whereas material in the
envelope stays nearly at a constant infalling velocity because the
acceleration becomes very low in this region. As the core density
increases the core shrinks continuously and the $1/r^2$ envelope
profile becomes more dominant. The maximum of the radial velocity
appears around the edge of the flat core region and is moves inward
with time.  Although, the infalling material becomes supersonic, no
shock occurs in this isothermal collapsing phase since the central
gravitational force accelerates the gas in front of the supersonic
material fast enough to prevent a velocity discontinuity. The radial
velocity stays close to zero at the outer edge of the sphere and at
the center of the cloud.

We emphasize that the collapse of an over-critical Bonnor-Ebert-Sphere
differs from the Expansion-Wave solution of~\citet{Shu77} since the
latter is an inside-out collapse of an initially singular isothermal
sphere (SIS). The collapse of a SIS proceeds in a self-similar fashion
leading to a velocity profile that increases with decreasing radial
distance to the center \citep[cf. Fig.~2 of][]{Shu77} whereas the
velocity of a collapsing Bonnor-Ebert-Sphere possesses always a finite
maximum and goes to zero at the center of the sphere.

To get a quantitative estimate of the isothermal collapse described
above we approximate the density profile shortly after the initial
collapse by:
\be
\rho(r) \approx \left\{
\begin{array}{lcr} 
  \rho_{\sm{core}} & : & r \leq r_{\sm{core}} \\
  \rho_{\sm{core}}\,\left(r/r_{\sm{core}}\right)^{-2}
      & : & r \gg r_{\sm{core}}
\end{array} 
\right.
\label{eq:sphere_profile}
\ee 
where $\rho_{\sm{core}}$ is the core density and $r_{\sm{core}}$ is
the radial distance to the core edge. This approximation is valid as
long as the collapse is isothermal. The core becomes non-isothermal
after the density increased above the critical density of $n \sim
10^{7.5} \, \cm^{-3}$ (see section~\ref{sec:isothermal_contraction}).

Using conservation of the mass of the sphere $M$ we can estimate the
core radius for a given core density (with the critical density of
$\rho_{\sm{core}} = 10^{-16} \, \g \, \cm^{-3}$ and a sound velocity
of $c = 0.41 \, \km \, \sec^{-1}$):
\bea
r_{\sm{core}} & \approx & \sqrt{\frac{M}{4\pi\,\rho_{\sm{core}}\,R}} 
= \frac{c}{\sqrt{4\pi\,G\,\rho_{\sm{core}}}} \,
  \left(\xi_c\,\phi^{\prime}_c\right)^{1/2}
\quad,
\label{eq:core_radius}
\\
& \sim & 4.5 \times 10^2 \, \AU \nonumber
\eea
where $R$ is the radius of the sphere and we assumed $r_{\sm{core}}
\ll R$. Using this approximations the total core mass is given by
\bea
M_{\sm{core}} & \approx &
  \frac{4\pi}{3}\,r_{\sm{core}}^3\,\rho_{\sm{core}} 
  = \frac{1}{3}\,M\,\frac{r_{\sm{core}}}{R}
\nonumber \\
 & = & \frac{1}{3}\,\frac{c^3}{\sqrt{4\pi\,G^3\,\rho_{\sm{core}}}} \,
  \left(\xi_c\,\phi^{\prime}_c\right)^{3/2}
\label{eq:core_mass}
\\ 
& \sim & 0.1 \, \Msol \quad . \nonumber
\eea

\subsection{Collapse with molecular cooling: initial isothermal phase}
\label{sec:isothermal_contraction}

\bef
\showone{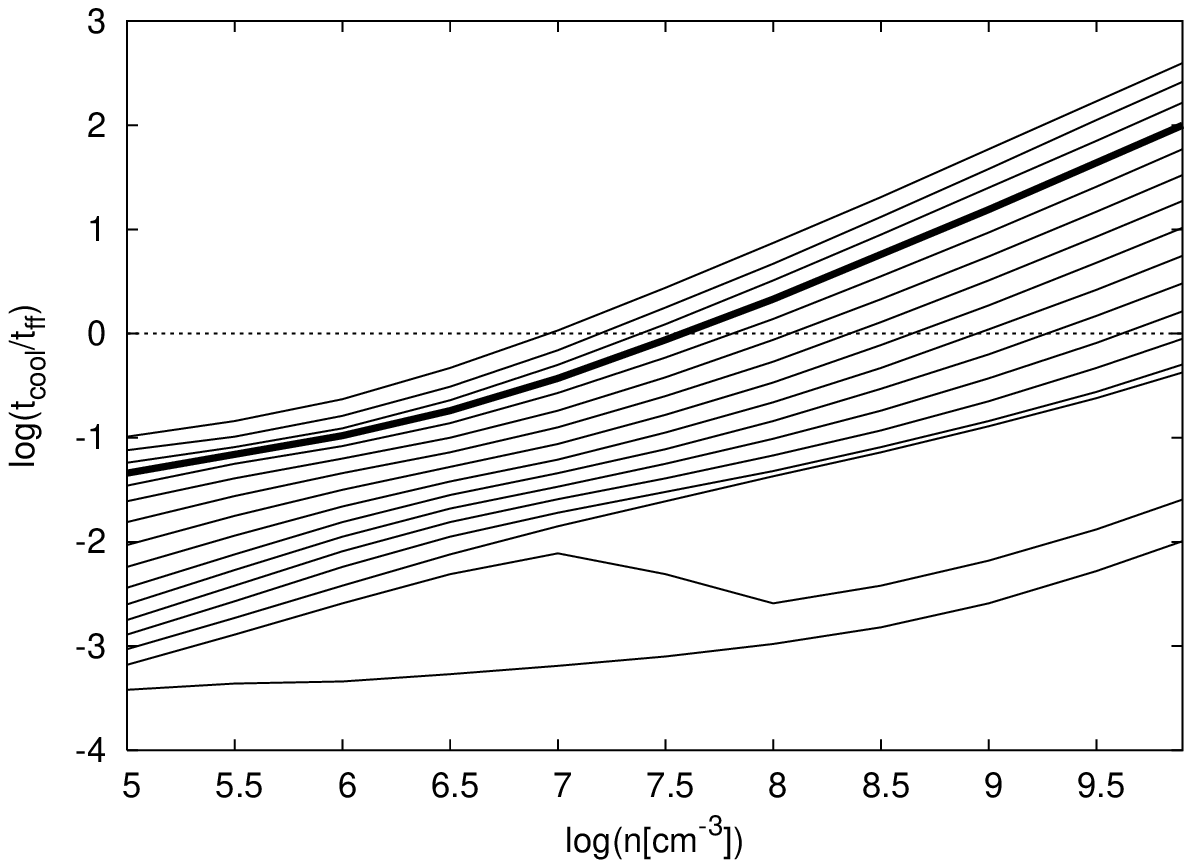}
\caption{Comparison of the local free-fall time $t_{\sm{ff}}$ with the
  cooling time $t_{\sm{cool}}$ in different temperature regimes. For a
  $20\,\K$ cloud (thick line) the cooling time becomes longer than the
  free-fall when its number density exceeds $n \approx
  10^{7.6}\,\cm^{-3}$. The cooling data are taken from
  \citet{Neufeld95} and \citet{Neufeld93}. The temperature range 
  is $T = 10\,\K - 10^{2.5}\,\K$ (from top to bottom) in steps of 0.1
  dex.} 
\label{fig:time_scales}
\eef

Theoretical models of collapsing protostars predict that the formation
of a protostar occurs in several stages
\citep[e.g.][]{Larson69,Larson03}. Already from
Fig.~\ref{fig:time_scales}, which compares the cooling time scale
$t_{\sm{cool}}$ and the local free fall time $t_{\sm{ff}}$, one can
infer that a gravitationally unstable low density cloud will undergo a
isothermal collapse until the cooling by molecular line emission
becomes inefficient when $t_{\sm{cool}} > t_{\sm{ff}}$. For instance,
a $20\,\K$ cloud cannot be efficiently cooled if its core density
exceeds more than $10^{7.6}\,\cm^{-3}$. After the core density reaches
this critical value, the high density region of the gas cloud
contracts almost adiabatically on a much longer time scale. Another
prediction one can infer from Fig.~\ref{fig:time_scales} is the
evolution trajectory of the core in the temperature-density-plane
which follows the track where $t_{\sm{ff}} \sim t_{\sm{cool}}$ in
the inefficient cooling regime. Our simulations are in good agreements
with this prediction.

Initially, due to effective cooling, the unstable sphere maintains its
initial temperature while collapsing on its local free fall
time. Fig.~\ref{fig:temp_dens} shows the evolution trajectory of the
cloud's core in the temperature-density plane from our simulation
A1. The inital contraction stage clearly indicates the regime where
cooling is fast enough to stay at the clouds initial temperature while
core density increases. A quantification of this stage is shown in
Figs.~\ref{fig:pres_dens} and ~\ref{fig:gamma_dens} where we compute
the effective equation of state, i.e. $\gamma_{\sm{eff}} \equiv
\di\log{p} / \di\log{\rho}$. We calculate the trajectory plots by
averaging the variables in the region where $n_{\sm{max}}/2 \le n \le
n_{\sm{max}}$, where $n_{\sm{max}}(t)$ is the maximal density in our
simulation at the time $t$ where $n_{\sm{max}}(t)$ increases
continuously with time.

\bef
\showone{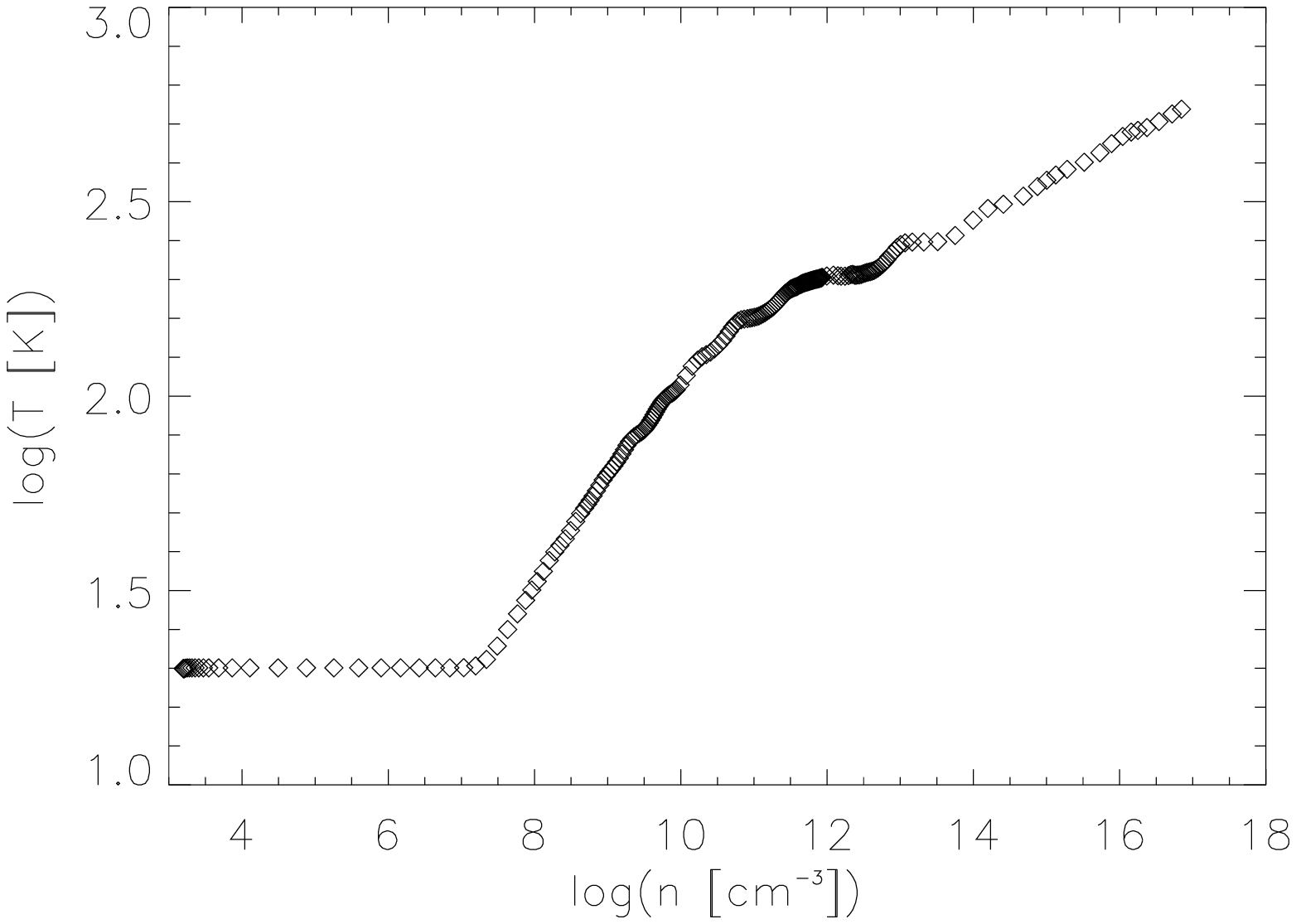}
\caption{Evolution trajectory of the core forming region from our
  simulation in the
  temperature-density plane. Initially, the core undergoes a rapid
  isothermal contraction until the cooling by molecular lines becomes
  inefficient, i.e. $t_{\sm{cool}} > t_{\sm{ff}}$, at a core density
  $n_{\sm{core}} \sim 10^7\,\cm^{-3}$ (from run A1).}
\label{fig:temp_dens}
\eef

\bef
\showone{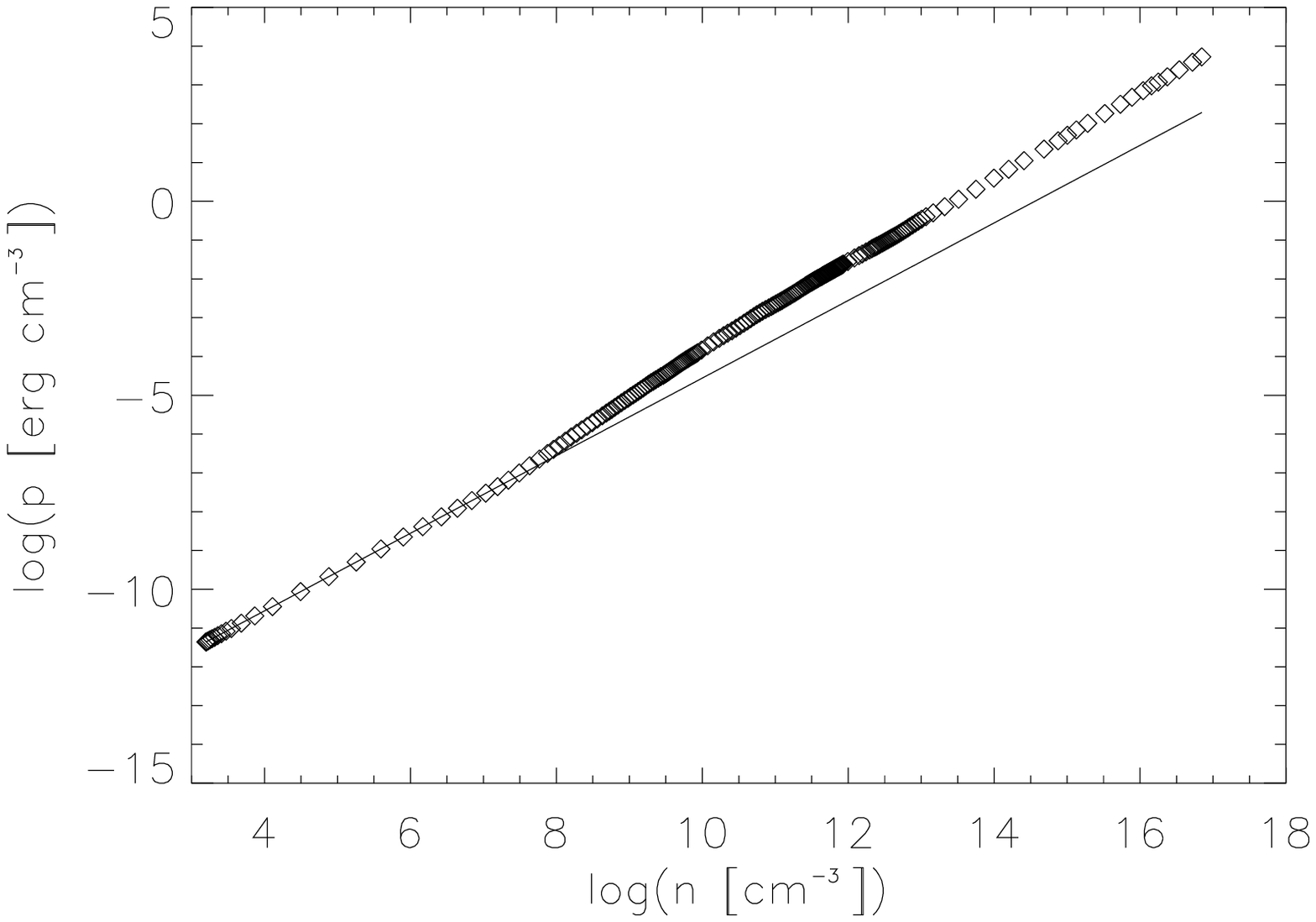}
\caption{Shows the thermal pressure, $P = P(n_{\sm{core}})$, in the
  core forming region.  The inital isothermal collapse (indicated by
  the solid line, $p \propto n$) is followed by a non-isothermal
  contraction with a non-linear pressure response (from run A1).}
\label{fig:pres_dens}
\eef

\bef
\showone{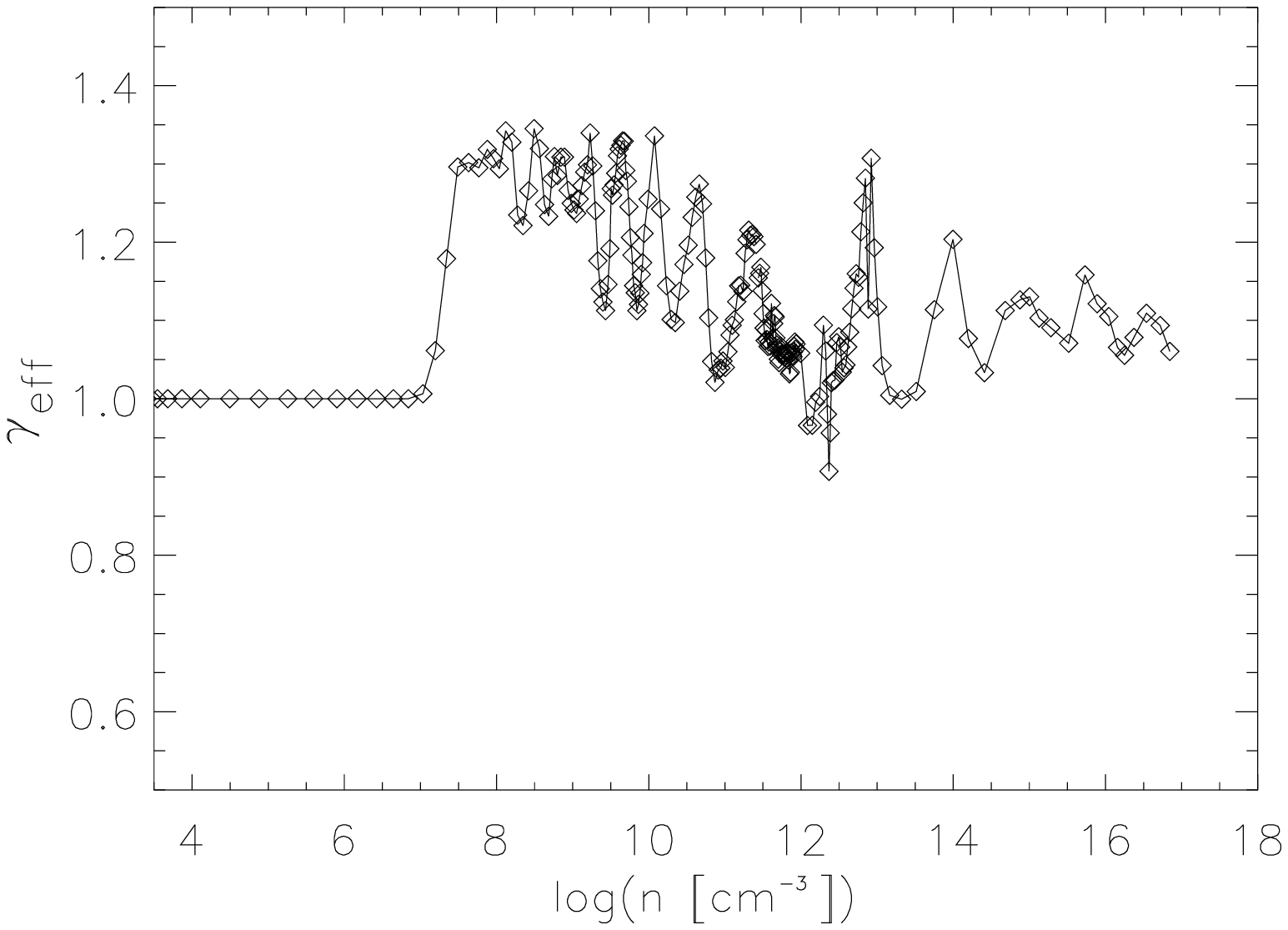}
\caption{The effective equation of state, $\gamma_{\sm{eff}} = \di\log
p/\di\log\rho$,  in the core forming region as a function of the
core density (from run A1).}
\label{fig:gamma_dens}
\eef

From Fig.~\ref{fig:time_scales} and from our simulations
(Fig.~\ref{fig:temp_dens} and Fig.~\ref{fig:gamma_dens}) we get a core
density (i.e. of material interior to radius $r_{\sm{core}}$) at the
end of the isothermal phase of $n_{\sm{core}} \sim 10^{7.5}\,\cm^{-3}$
and $\rho_{\sm{core}} \sim 10^{-16}\,\g\,\cm^{-3}$. Using
Eq.~(\ref{eq:core_radius}) and Eq.~(\ref{eq:core_mass}) gives a core
radius of $r_{\sm{core}} \sim 450\AU$ and a core mass of
$M_{\sm{core}} \sim 0.1\,\Msol$. Note that these values are
independent of the initial mass $M$ and radius $R$ as long as the
inital sphere obeys the Bonnor-Ebert profile given by
Eq.~(\ref{eq:BE_diff}). Other investigations of spherical collapses of
molecular clouds \citep[for a review see][]{Larson03} also predict a
first 'hydrostatic core' when the density reaches $\sim 10^{-10} \, \g
\, \cm^{-3}$ where the core mass is almost independent of the clouds
initial mass or radius. The predicted 'hydrostatic core' mass is about
$0.01\,\Msol$. Our findings give a larger core mass at the end of the
isothermal collapsing phase as cooling becomes inefficient at lower
densities.

Comparing the value of the core mass at the end of the isothermal
phase to the local Jeans mass, $M_J =
\pi^{5/2}/6\,c^3/\sqrt{G^3\,\rho_{\sm{core}}} \approx 0.56\,\Msol$,
(which is independent of the sphere parameters)
gives an estimate of gravitational strength at this point. The fact
that $M_{\sm{core}}/M_J <1$ in the fast contracting isothermal phase
is the reason that the collapsing cloud has not yet broken up into
several fragments. 

With the approximation of the radial density profile
Eq.~(\ref{eq:sphere_profile}) the total mass within a sphere of radius
$r \gg r_{\sm{core}}$ is given by
\bea
M(r) & \approx & \frac{c^2}{G} \, \left(\xi_c\,\phi^{\prime}_c\right)
  \, r
\label{eq:sphere_mass}
\\
     & \approx & 1\, \Msol \, \left(\frac{T}{20\,\K}\right) \,
  \left(\frac{r}{3.3\times 10^{16} \, \cm}\right) 
\quad,
\nonumber
\eea
where we used $\xi_c = 6.5$ for the numerical example. 
Comparing the core mass $M_{\sm{core}}$ to total mass of the sphere,
$M$, we get
\bea
\frac{M_{\sm{core}}}{M} & \approx & \frac{1}{3} \, 
  \left(\frac{\phi^{\prime}_c}{\xi_c}\right)^{1/2} \,
  \left(\frac{\rho_{\sm{core}}}{\rho_0}\right)^{-1/2}
\\
   & \approx & 0.08 \, \left(\frac{\rho_{\sm{core}}}{\rho_0}\right)^{-1/2}
  \quad .
\nonumber
\eea

Using Eq.~(\ref{eq:sphere_mass}) we can compute the mass accretion at
different radii:
\be
\dot{M}(r) \approx - \frac{c^2}{G} \, \left(\xi_c\,\phi^{\prime}_c\right) \,
  v_r  \quad,
\label{eq:mdot_iso_base}
\ee 
where $v_r$ is the radial infall velocity. The gravitational
acceleration outside the core region is $g \approx c^2 \, \left(\xi_c
\, \phi^{\prime}_c\right) /r$ and the radial velocity $v_r \approx c^2
\, \left(\xi_c \, \phi^{\prime}_c\right) \, \left(4\pi \, G \,
\rho_{\sm{core}}\right)^{-1/2} / r$ assuming that the velocity changes
on the dynamical time of the core region~\footnote{This assumption
predicts a $r^{-1}$ dependence of the radial velocity outside the core
region. Our simulations show only a weak dependence of the velocity on
the radius. Using the {\em local} dynamical time gives a velocity
which is independent of $r$. From our simulations we conclude that the
increase of the infall velocity occurs on a time scale somewhat in
between the dynamical time and the local free fall time. Using either
time scale give the same velocity and mass accretion at the core
edge.}. Together with the size of the core region
Eq.~(\ref{eq:core_radius}) we get a good estimate of the mass
accretion onto the core: 
\bea 
\dot{M}_{\sm{core}} & \approx & \frac{c^3}{G}
   \, \left(\xi_c\,\phi^{\prime}_c\right)^{3/2} 
\label{eq:mass_accretion_iso} 
\\
  & \sim & 6.1\times 10^{-5} \,
      \Msol/\yr \, \left(\frac{T}{20\,\K}\right)^{3/2}
\nonumber
\eea
Eq.~(\ref{eq:mass_accretion_iso}) shows that the mass accretion of a
collapsing Bonnor-Ebert-Sphere is independent of its initial mass and
radius. This result is applicable to non-merging, i.e. low mass, cloud
cores and similar to the case of a collapsing SIS~\citep{Shu77}.

\bef
\showone{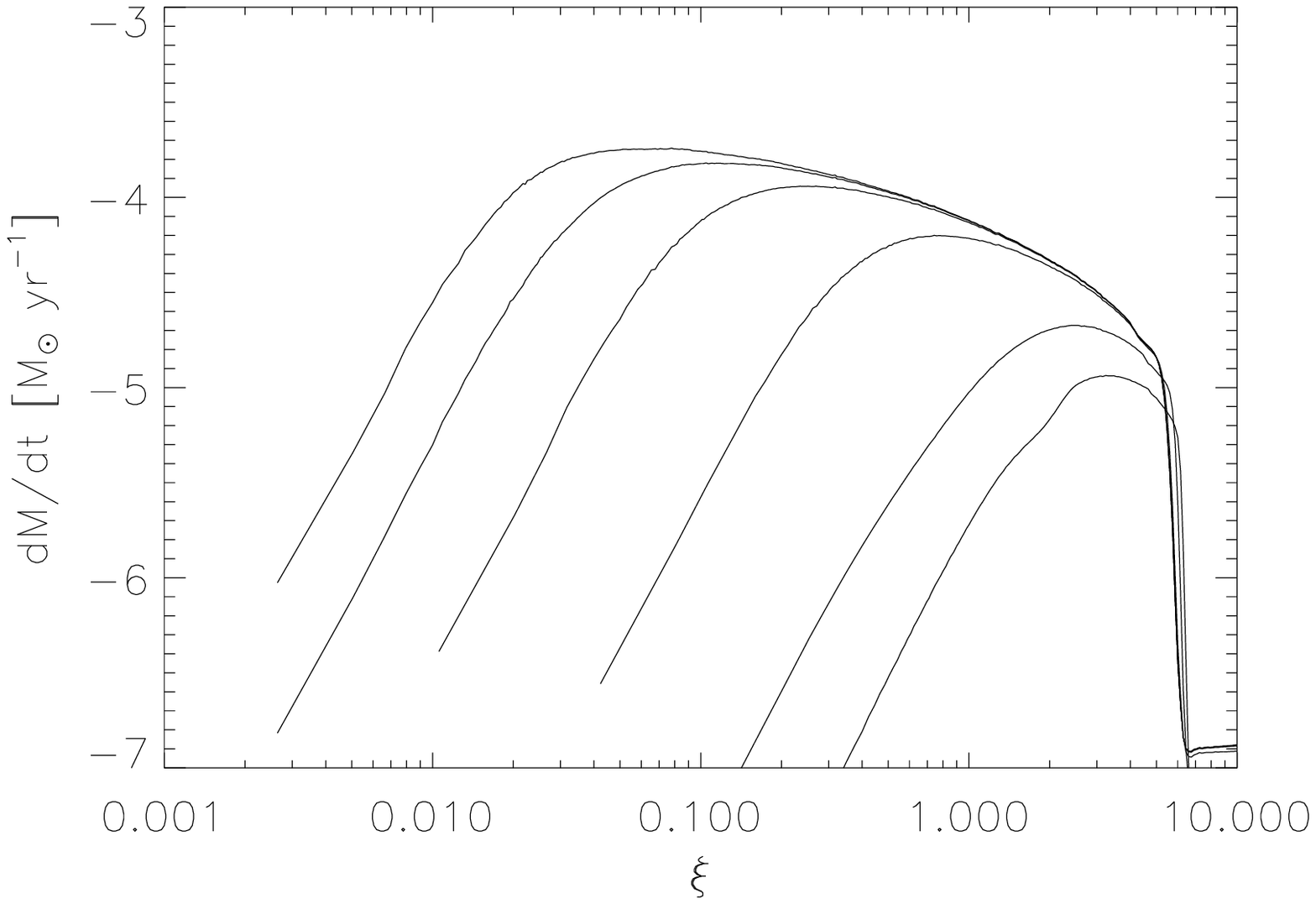}
\caption{Spherical averaged radial profiles of the mass accretion in
  the isothermal regime corresponding to
  Fig.~\ref{fig:dens_vrad_plt_iso}. The maximum mass accretion occurs
  roughly at the core radius $r_{\sm{core}}$ which moves steadily
  towards the center (from run A1).}
\label{fig:mdot_plt_iso}
\eef

Although Eq.~(\ref{eq:mass_accretion_iso}) suggests that the mass
accretion is constant in time we observe a slowly increasing $\dot{M}$
which is due to the steady increase of the radial velocity.
At a given radius $r \gg r_{\sm{core}}$ in the envelope the mass
accretion approaches a constant value. The mass accretion has its
maximum at the at the core radius and drops quickly towards the core
center as the radial velocity sharply at radii $r < r_{\sm{core}}$
Fig.~\ref{fig:mdot_plt_iso} shows the evolution of the mass accretion
in the isothermal regime.

\subsection{Collapse with cooling: post-isothermal phase}

When the central density exceeds the critical density $n_{\sm{crit}}
\approx 10^{7.5}\,\cm^{-3}$ thermal energy produced by gravitational
contraction is no longer radiated efficiently and the core starts to
heat up. The resulting increase of the thermal pressure slows down the
contraction of the central gas region and the effective equation of
state becomes stiffer. This is reflected in the increase of the
effective equation of state, $\gamma_{\sm{eff}}$.  Our simulations
(cf. Fig.~\ref{fig:gamma_dens}) show a steep increase of
$\gamma_{\sm{eff}}$ from $1.0$ to $\approx 1.3$ after the core density
rises above the critical cooling density. Due to the complex cooling
process $\gamma_{\sm{eff}}$ does not stay at a fixed value but varies
with time with a trend to decrease in the regime $10^8 \la
n_{\sm{core}} \la 10^{12} \, \cm^{-3}$. The strong variability of
$\gamma$ in Fig.~\ref{fig:gamma_dens} reflects also the multiple shock
occurrence during the collapse (cf. Fig.~\ref{fig:temp_sim_01}). Note
also that the data points in this plot do not correspond to the time
steps in the simulation but to the larger time steps at which we got
output data files. Sampling the data points on a higher time
resolution might result in a smoother graph.

\bef
\showone{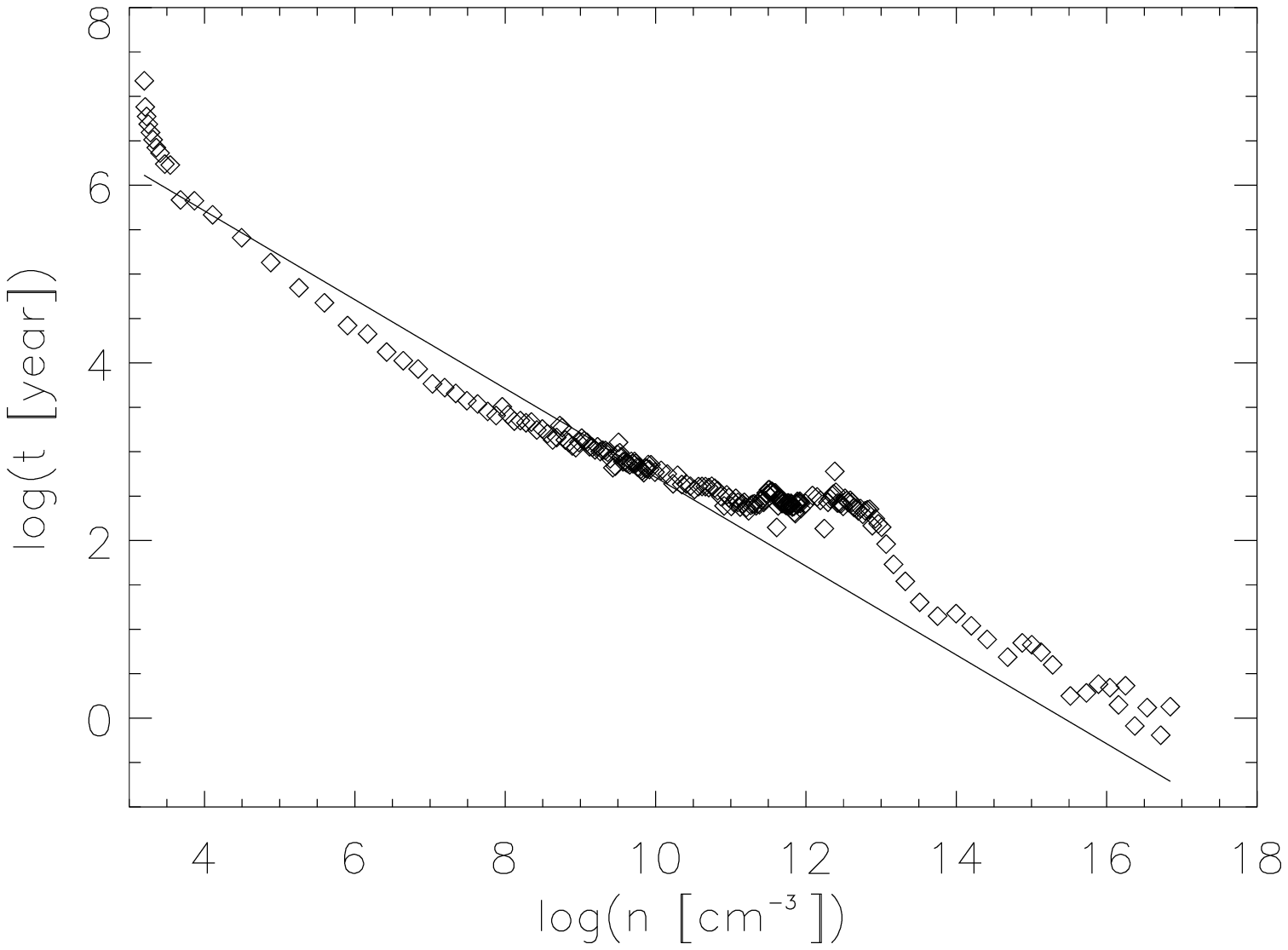}
\caption{Evolution time scale, $t_{\sm{evol}} \equiv
  n_{\sm{core}}/\dot{n}_{\sm{core}}$, as a function of the core
  density $n_{\sm{core}}$. The solid line shows the local free-fall
  time, i.e. $t \propto n_{\sm{core}}^{-1/2}$ (from run A1).}
\label{fig:time_dens}
\eef

In the post-isothermal phase, the density increase in the center of the
gas cloud is dictated by the cooling time $t_{\sm{cool}}$ as opposed
to the dynamical time $t_{\sm{ff}}$. Fig.~\ref{fig:time_dens} shows
the evolution time scale of the core region, i.e. $n_{\sm{core}} /
\dot{n}_{\sm{core}}$ as a function of the density. After the core
density reached $\sim 10^8 \, \cm^{-3}$ the collapse proceeds on a
time scale that is determined by cooling process. At densities of $\ga
10^{10} \, \cm^{-3}$ the core density increases on a nearly constant
time scale much larger than the local free fall time marking a phase
of 'hydrostatic equilibrium' \citep[cf. also][]{Larson69}. At later
times the temperature of the core (at $n \ga 10^{13} \, \cm^{-3}$ and
$T \ga 250 \, \K$) rises to a point when cooling by molecular
excitations becomes more efficient (due to the abundant production of
H$_2$O and its multitude of excitation levels) and the collapse
proceeds again almost on the local free fall time.

The slower contraction of the warmer core region, as compared to the
isothermal case, causes the supersonic infalling material to shock at
the edge of the warm core. These first shocks occur preferentially
above and below the disk plane where the gas is not rotationally
supported at a typical distance of $\sim 650\,\AU$ (for the runs A$n$)
from the center. Temperatures at this point reach $70 - 80\,\K$ and
the temperature profile is discontinuous rising from $30\,\K$ to
$70\,\K$ between the outer pre-shock region and the shocked region
within a few tens of $\AU$ (cf. Fig.~\ref{fig:temp_z_line}). The
velocities at this point reach values of $1.5\,\km\,\sec^{-1}$ ($\Ma
\approx 2.5$). As infalling gas piles up at the shock boundary, a
density discontinuity also develops in the post-shock region which
separates a low density envelope and a high density pre-protostellar
disk. This post-shock region slowly moves towards the center of the
gas cloud and its temperature increases further while supersonic gas
from the outside envelope hits this shock boundary.

\bef
\showtwo{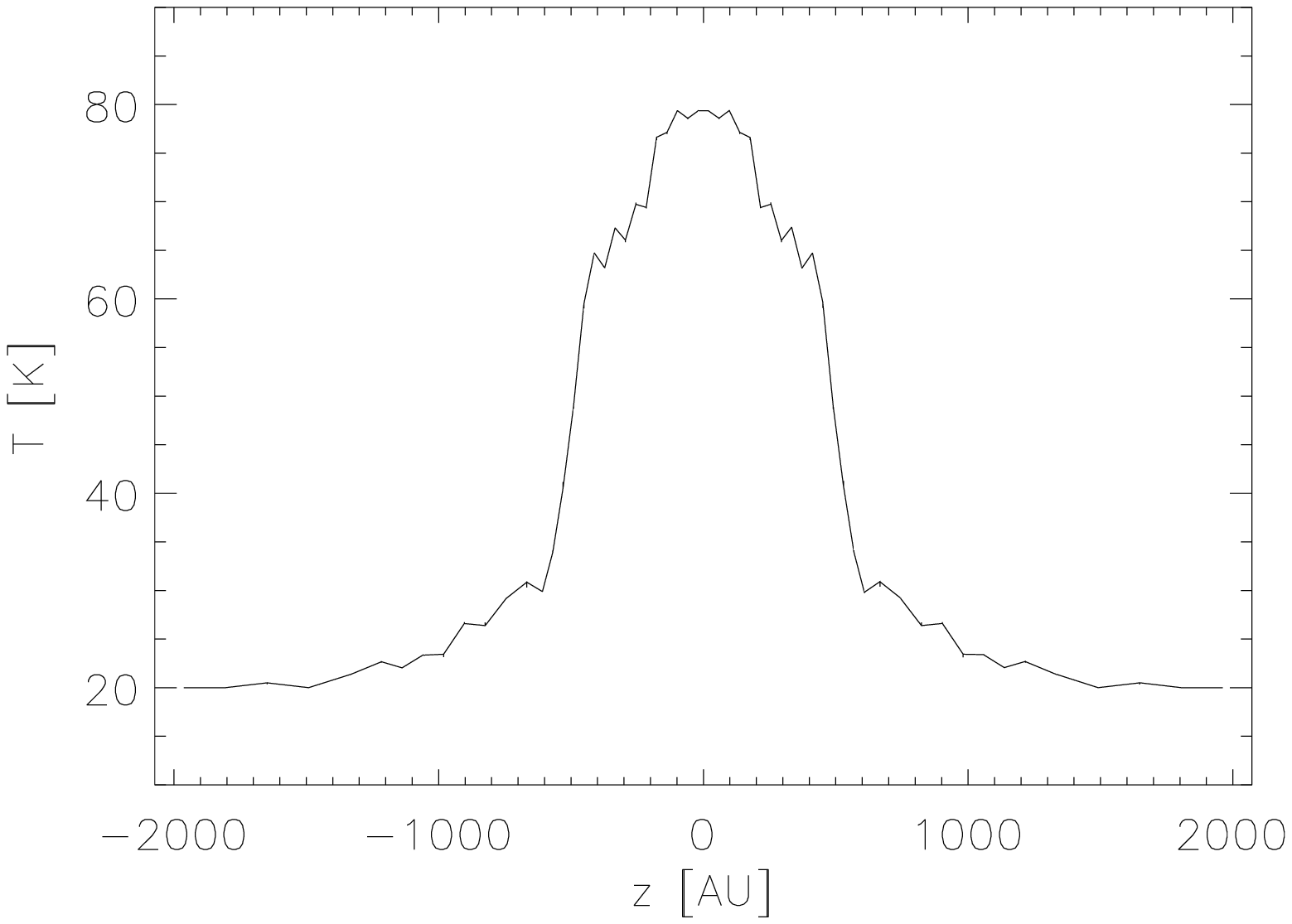}
	{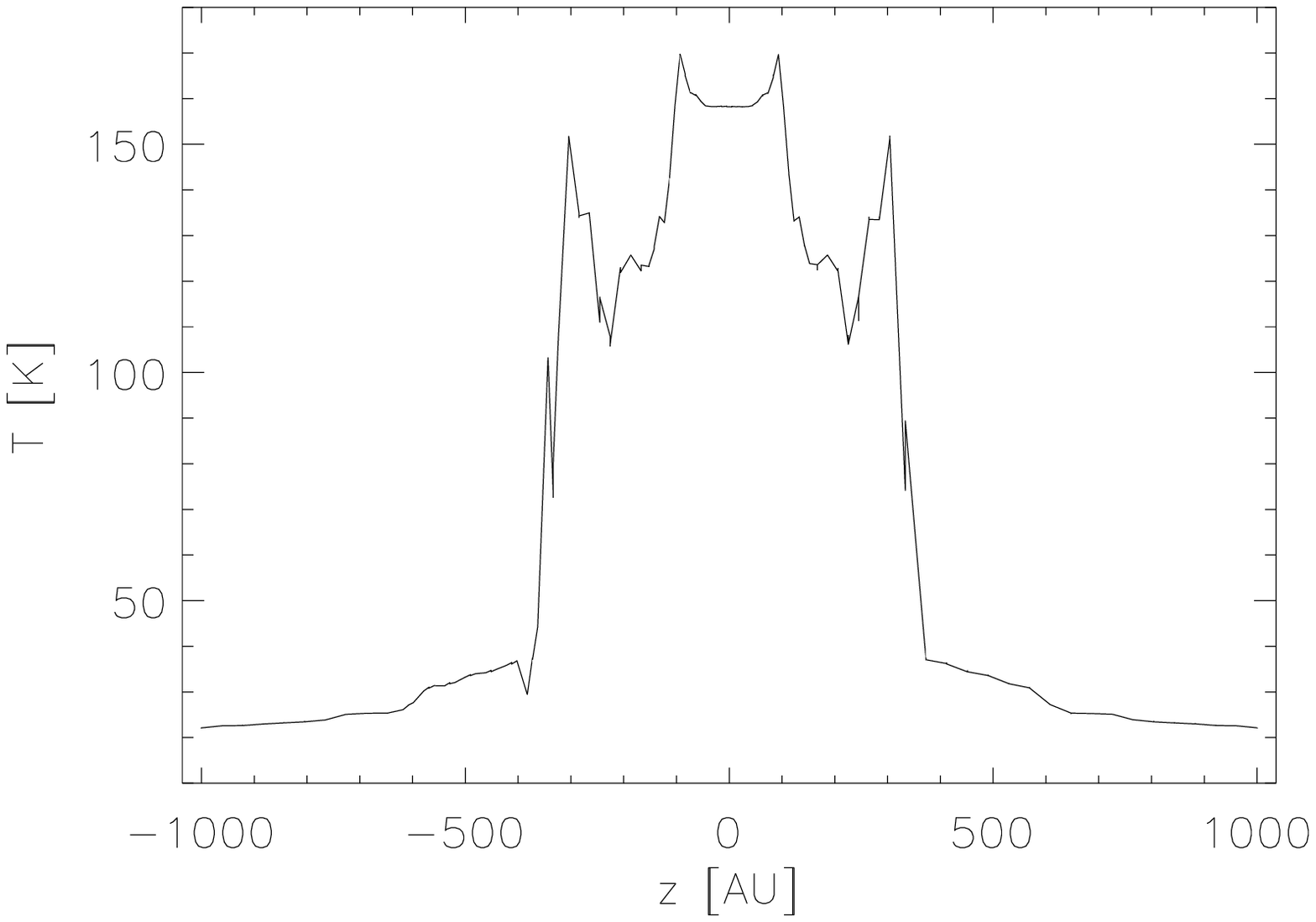}
\caption{Temperature line profiles along the $z$-axis through the
  center of the gas cloud for run A1. The first panel shows the line
  profile after the first shock develops when the central density
  reached $2\times 10^9 \, \cm^{-3}$. The second panel shows the
  temperature profile at a later stage when $n_{\sm{core}} \sim
  10^{11}\,\cm^{-3}$. At this time a second shock in the inner core
  region develops when supersonic material falls onto the
  protoplanetary disk (from run A1).}
\label{fig:temp_z_line}
\eef

Gas inside this first shock region is subsonic and therefore not shock
heated leading to a lower temperature region behind the shock. As the
core density and the core temperature continue to increase, gas inside
the pre-protostar region becomes supersonic and a secondary shock
develops at $\sim 100\,\AU$ above and below the disk plane (cf. second
panel of Fig.~\ref{fig:temp_z_line}). Typical temperatures at the
secondary inner shock boundaries are $\sim 150\,\K$ whereas the core
stays cooler. This stage (the core density is now $\sim
10^{11}\,\cm^{-3}$) marks the first development of the protostellar
disk.

Similar to the outer first shock, the inner shock fronts move slowly
towards the disk plane while the local Mach number rises to $\Ma \sim
4$ corresponding to a shock velocity of $3\,\km\,\sec^{-1}$. Again,
the collapsing core region is fed only by gas inside the inner shock
fronts whereas gas falling from outside onto the shocks is stalled at
the shock boundaries and heats up the gas in these shock regions.

\bef
\showtwo{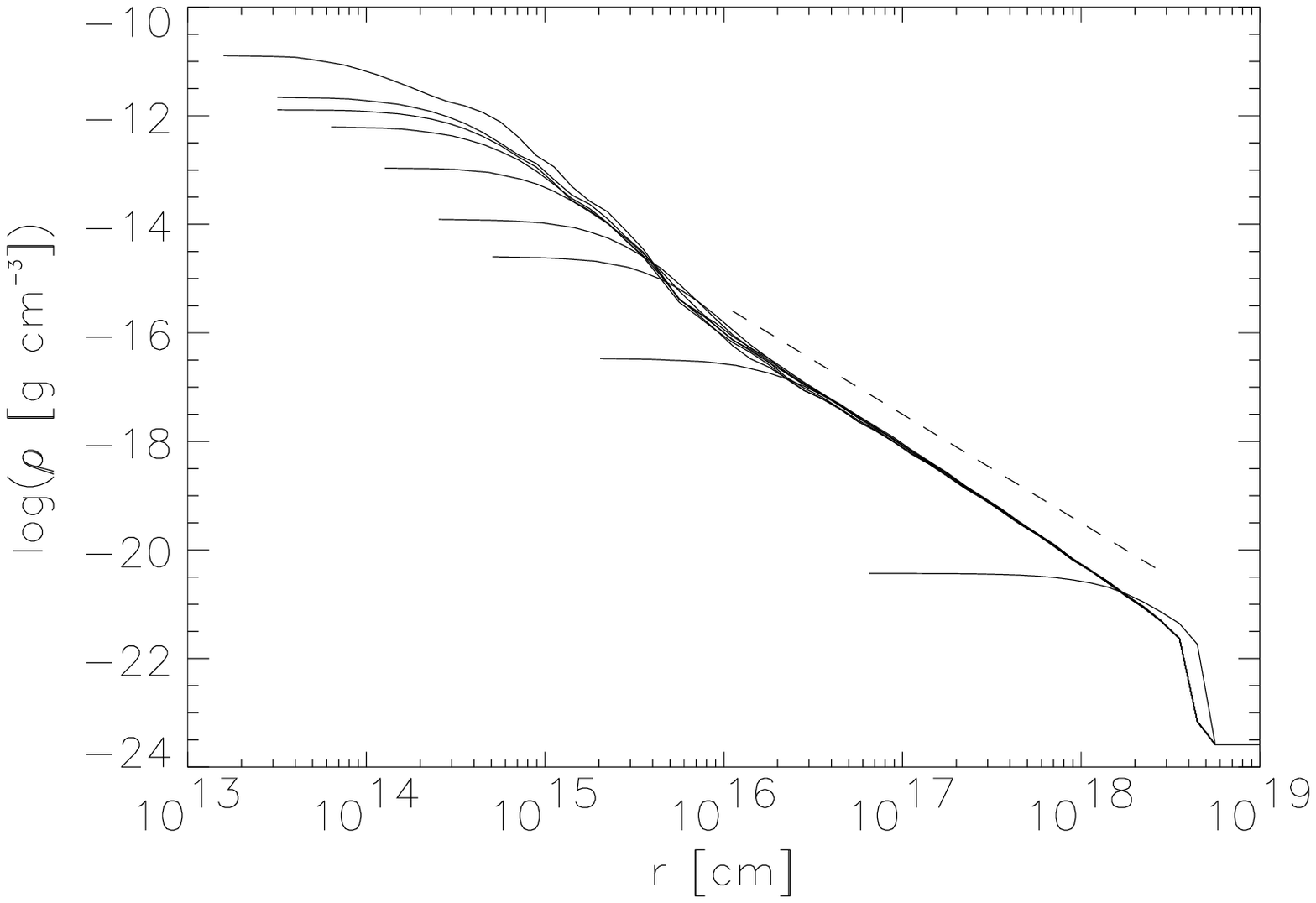}
	{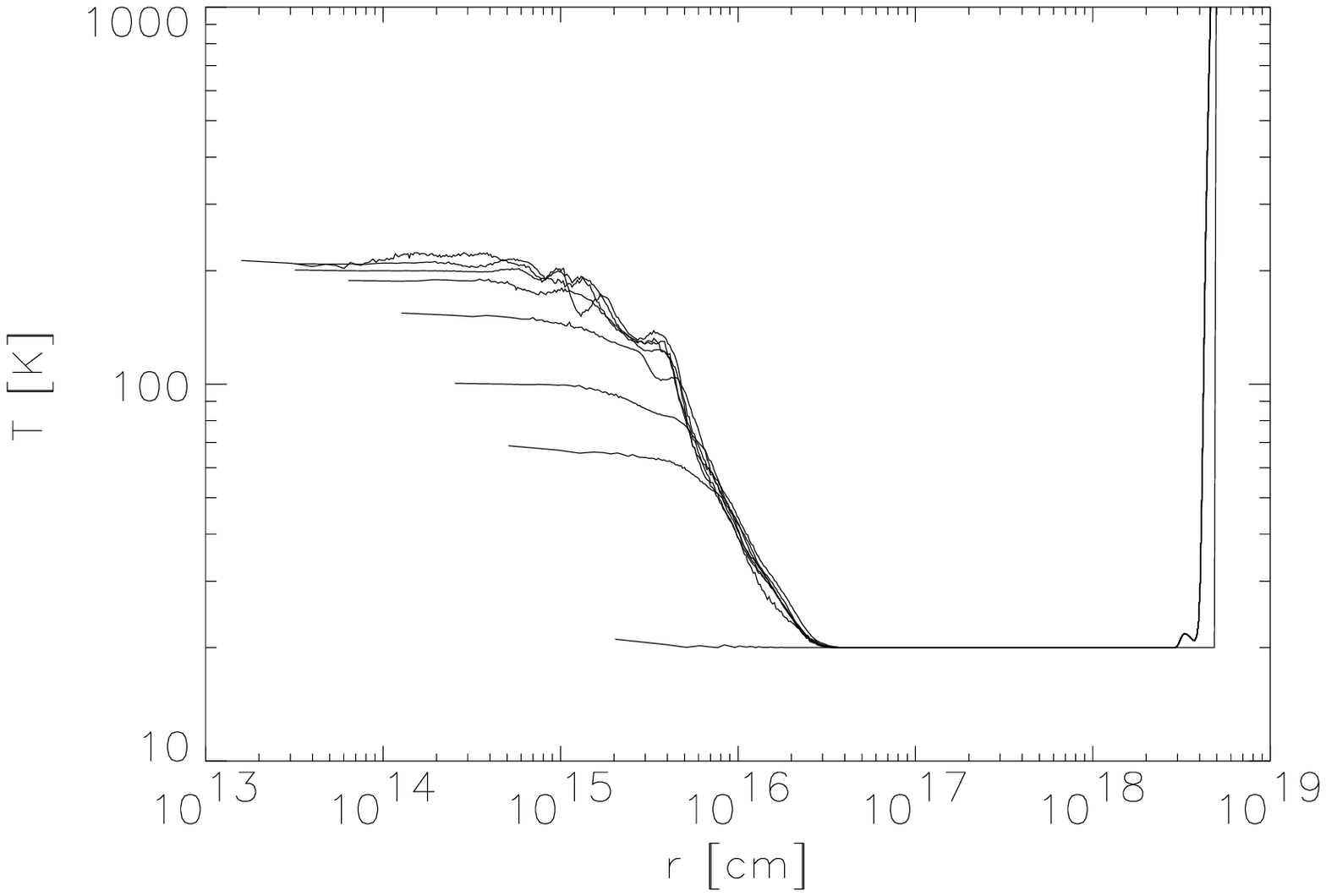}
\caption{Radial profiles of the density (upper panel) and temperature
  (lower panel) at different times. As the simulation resolution
  increases with time due to the Jeans refinement criterion the shown
  line profiles also increase in resolution. The dashed short line
  shows a $r^{-2}$ density profile. The data are compiled from run A1.}
\label{fig:dens_plt}
\eef

\bef
\showone{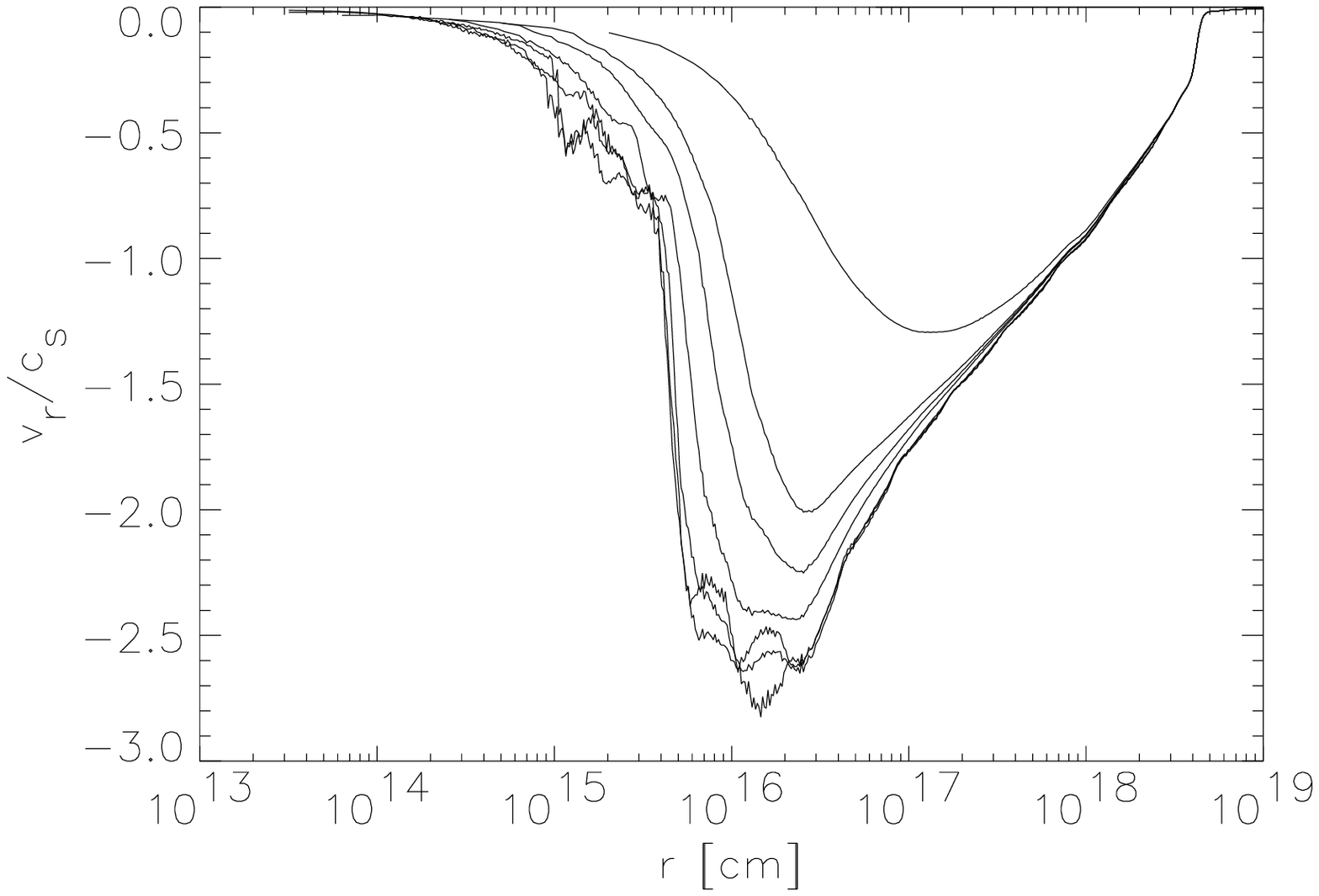}
\caption{Radial profiles of the radial velocity at different times
  (this profiles correspond the those shown in
  Fig.~\ref{fig:dens_plt}). During the first stages of the spherical
  collapse the radial velocity increases steadily with time until it
  reaches the maximum value of $\sim 3 \, c$ and the peak velocity is
  moving towards the inner part of the cloud. After the temperature in
  core region begins to increase the maximum of the radial Mach number
  stays at $r \sim 10^{16} \, \cm$ (from run A1.}
\label{fig:vrad_plt}
\eef

\bef
\showtwo{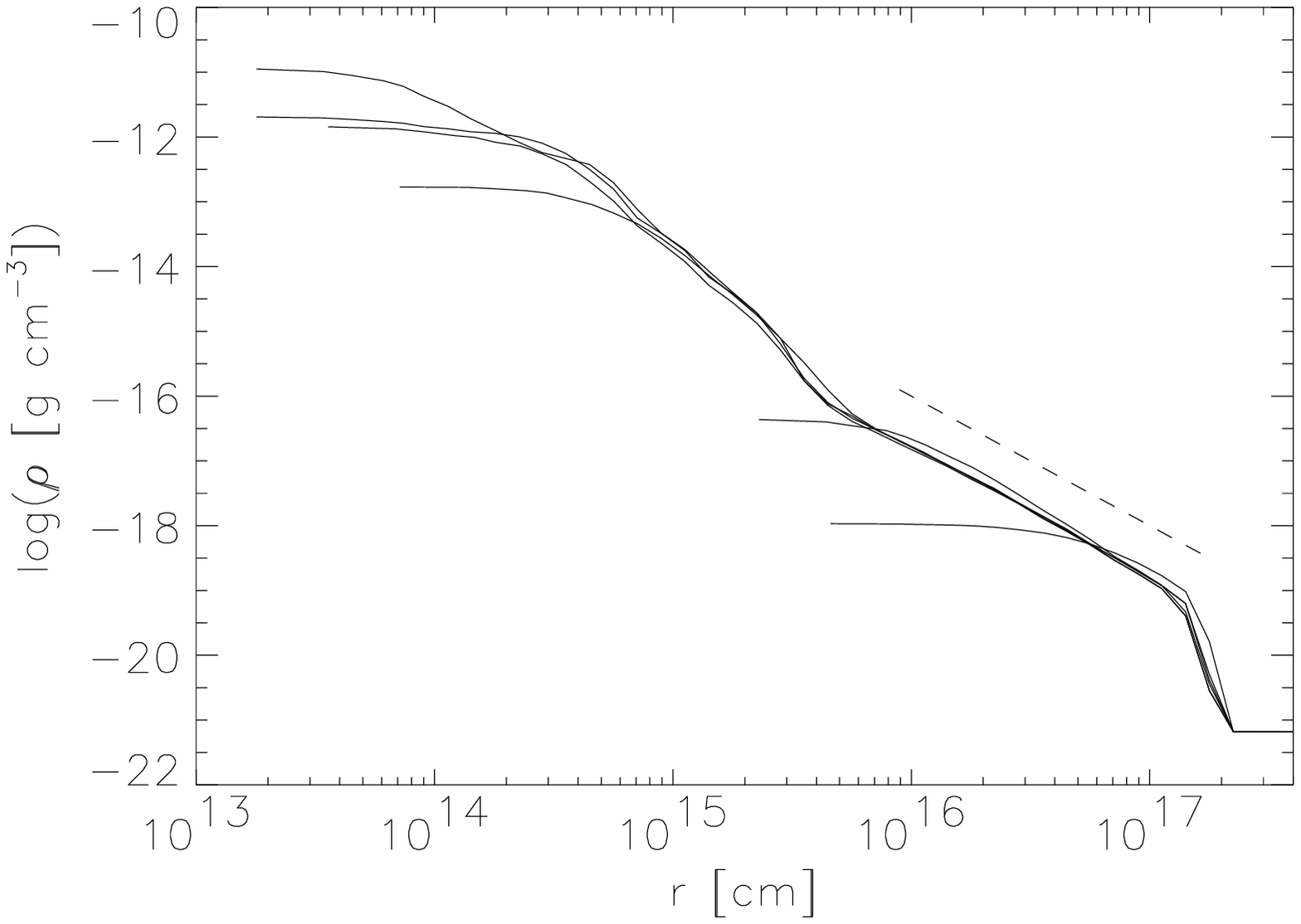}
	{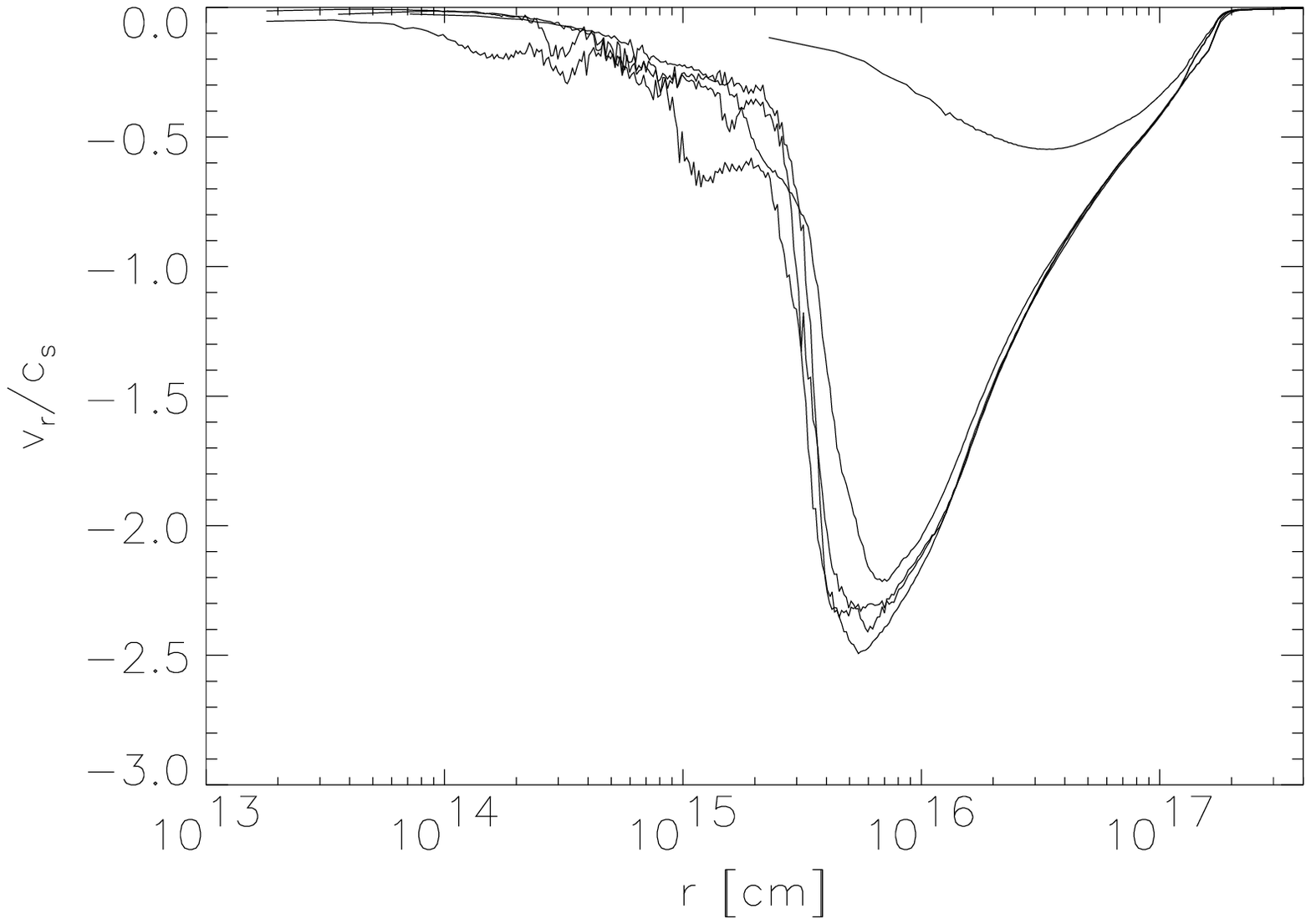}
\caption{Similar to figures Fig.~\ref{fig:dens_plt} and
  Fig.~\ref{fig:vrad_plt} except that the profiles are for run B68. The
  dashed line in the density panel shows a $r^{-2}$ profile.}
\label{fig:dens_plt_B68}
\eef

Fig.~\ref{fig:dens_plt} shows the time evolution of the density
profile and temperature profile, respectively, in the non-isothermal
regime from simulation A1. The second line from the bottom in the
density plot marks the last stage of the isothermal collapse phase.
In the non-isothermal phase the density profile deviates from the
$r^{-2}$ profile because of the slower contraction of the core during
this regime. The density falls of more steeply than $r^{-2}$ with
increasing radius in the heated up region and approaches a $r^{-2}$
profile in the isothermal regime at $r \ga 10^{-16} \, \cm$. The
transition region appears at a radius of $\sim 7\times 10^{15} \, \cm$
which is in good agreement with the theoretical prediction of
Eq.~(\ref{eq:core_radius}).  The temperate profile rises steeply with
decreasing radius, roughly as $r^{-2}$, whereas the core region has a
flat temperature profile. The profiles $\rho(r)$ are radially bin-ed
spherical averages, i.e. $\rho(r) = \int \di\Omega\,\rho(\vx)$, where
$\di\Omega = \di\cos\theta\,\di\varphi$ is the differential solid
angle.

Fig.~\ref{fig:vrad_plt} shows the evolution of radial infall velocity
as a function of radius from run A1. The peak velocity increases with
time and 'moves' towards smaller radii (the first line from the top
refers to the second line from the bottom of Fig.~\ref{fig:dens_plt}
which marks the end of the isothermal regime). Due to the shock which
develops at the edge of the core the velocity profile becomes steeper
with time whereas the peak velocity approaches a constant value of
$\sim 3 \, c$ in the spherical collapse phase. The velocity at the
center drops always to zero.

In Fig.~\ref{fig:dens_plt_B68} we show the evolution of the 1D density
profiles for run B68. Because the initial core density $\rho_0$ in
this case is higher than for the runs A1 -- A3
($\rho_{\sm{B68}}/\rho_{\sm{A1}} \sim 300$) the isothermal region with
$\rho \propto r^{-2}$ is smaller but the kink in the profile which
separates the non-isothermal core region from the isothermal envelope
appears at the same physical radius of $\sim 7\times 10^{15} \,
\cm$. This result is again in agreement with the theoretical
prediction. Finally, we note that a similar double shock structure has
been seen in 2D collapse simulations by~\citet{Yorke95} that feature
dust as the coolant.

\section{Disk formation}
\label{sec:disk_formation}

As described in the previous section, the appearance of the secondary
shock establishes the initial state for the formation of the protostellar
disk. The protostellar disk builds up and achieves a column density of
$\Sigma \approx 10^{1.5}\, \g \, \cm^{-2}$ and a disk height $h \sim
200\,\AU$. As the shock fronts move towards the equatorial plane, the
density and temperature increase on a time scale slower than the local
free fall time. Fig.~\ref{fig:dens_height_surf} (run A1) and
Fig.~\ref{fig:dens_surf_B68} (run B68) show the time evolution of the
column density which we compute as follows:
\be
  \Sigma(R) \equiv \int\,\di z\,\rho(x,y,z) \quad .
\label{eq:column_dens}
\ee 
where $R = \sqrt{x^2+y^2}$ is cylindrical radius and we evaluate
the integral~(\ref{eq:column_dens}) throughout the entire simulation
box. Similar to the radial density profile the column density profile
develops a envelope and a flat core region. The non-isothermal core is
separated from the envelope by a steep increase in density which leads
to a kink in the radial density profile. In general, we find that the
column density falls off roughly as $r^{-2}$. The disk height
decreases while the disk gets denser and reaches a value of $\sim
10\,\AU$ when the column density becomes $\sim 10^3\,\g\,\cm^{-2}$.

\comment{more steeply ($\sim R^{-2}$) than predicted
by~\citet{Hayashi81} ($\Sigma \propto R^{-3/2}$) and}

In the case of run B68 the ring structure is clearly visible in latest
density profile.

\bef
\showtwo{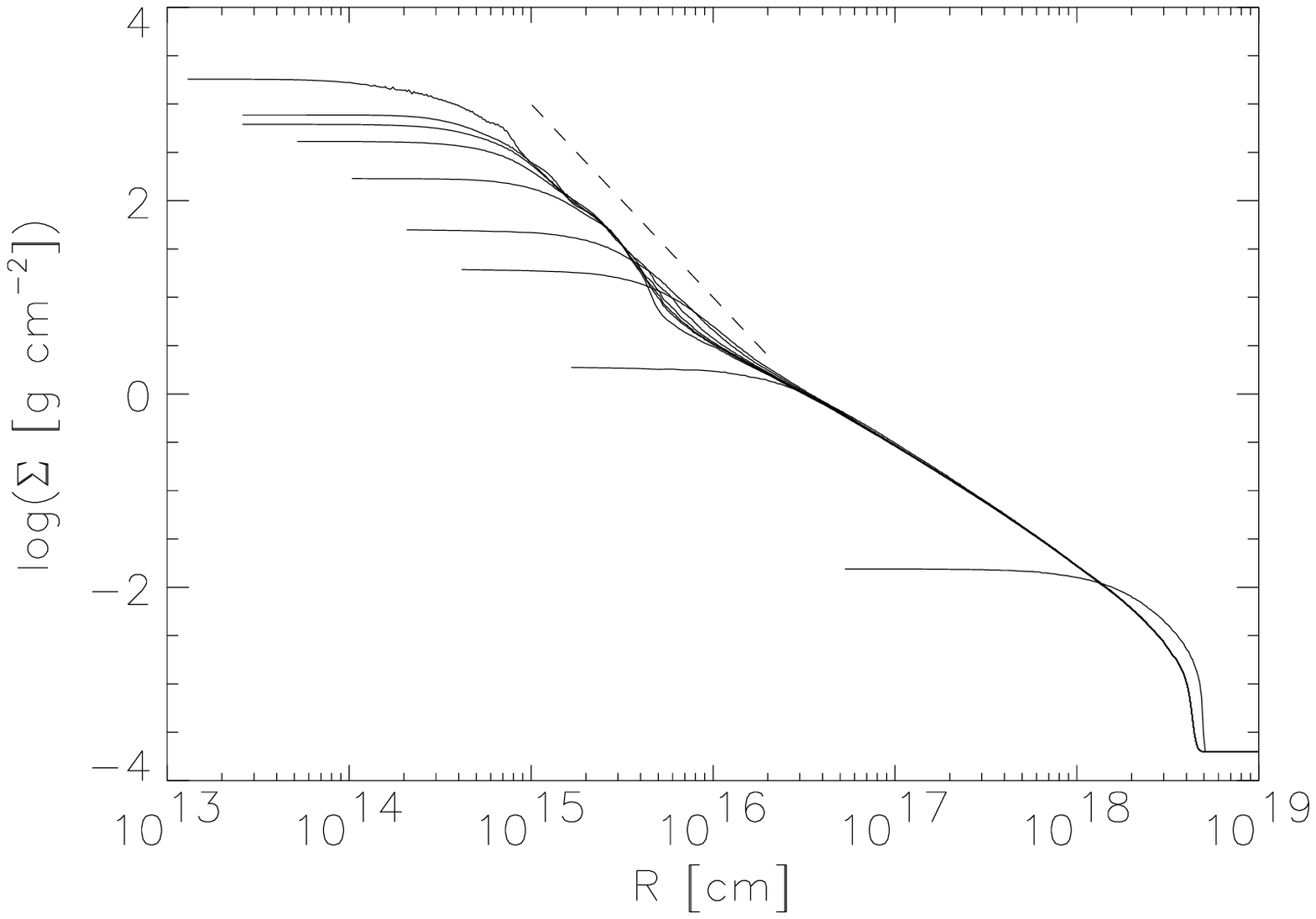}
	{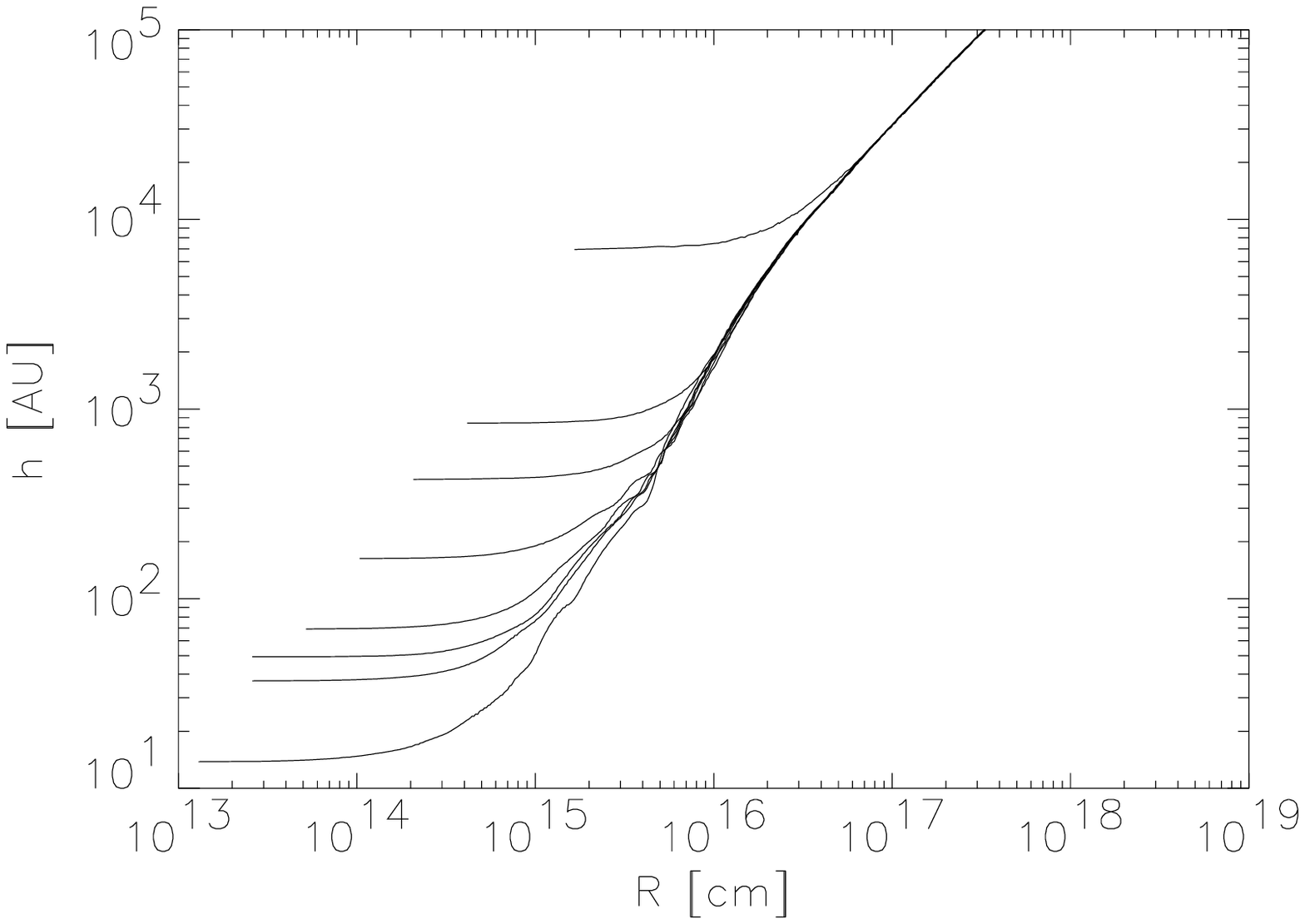}
\caption{Radial profiles of the column density $\Sigma$ and disk height
  $h$ at different times. For comparison, the dashed line shows a
  $R^{-2}$ profile. The shown profiles correspond to the density
  shown in Fig.~\ref{fig:dens_plt} (run A1).}
\label{fig:dens_height_surf}
\eef

\bef 
\showthree{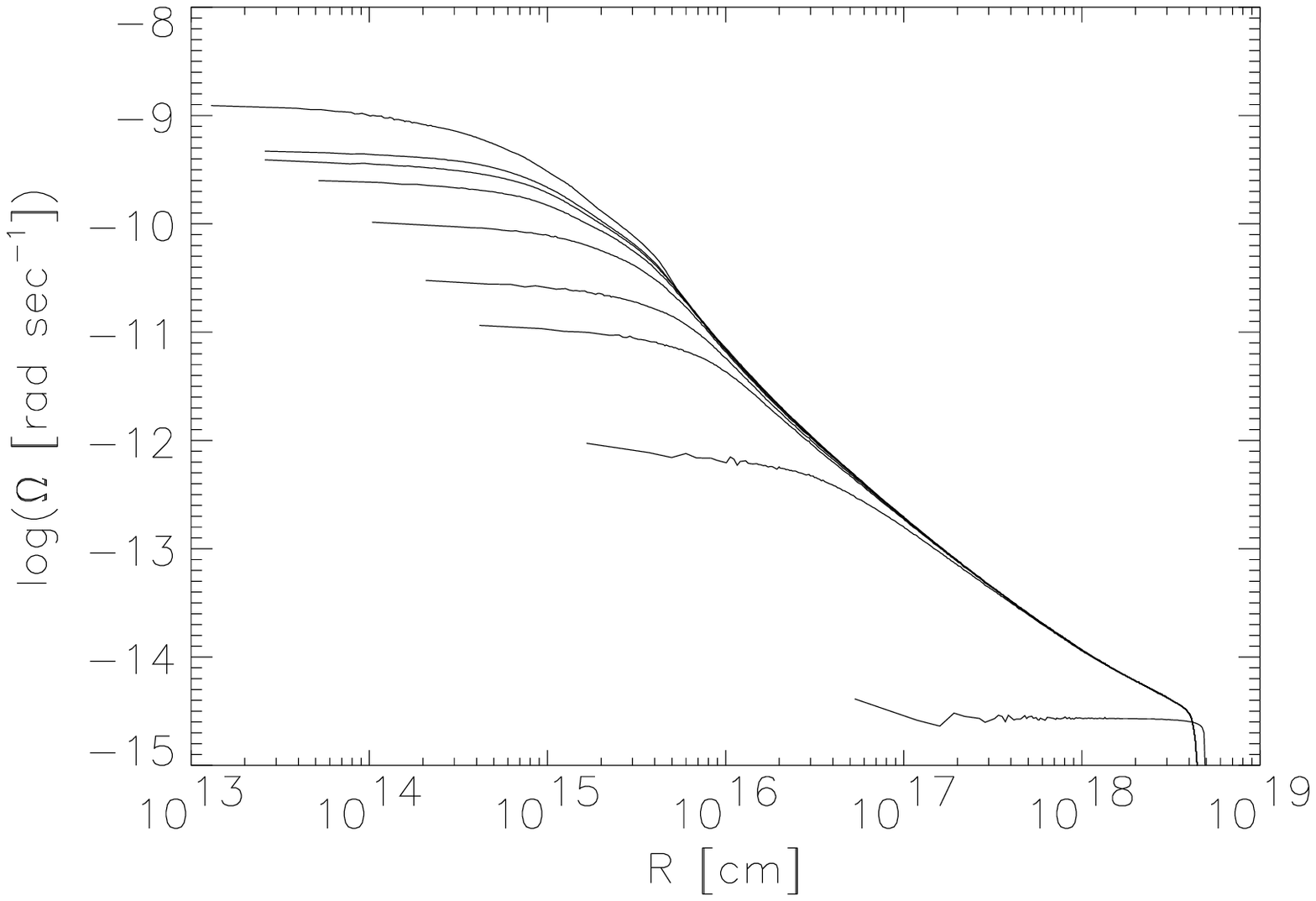} 
	  {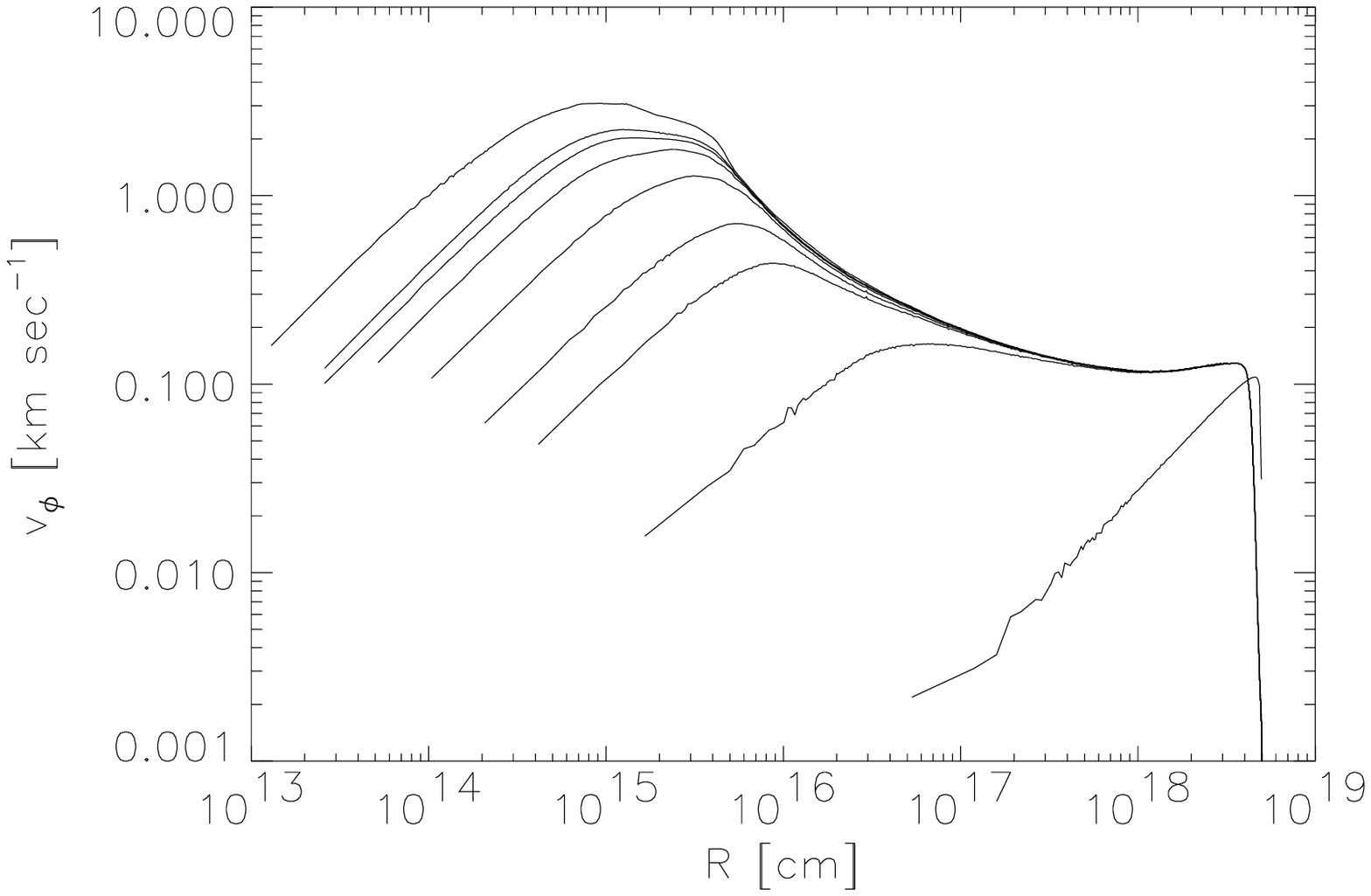}
	  {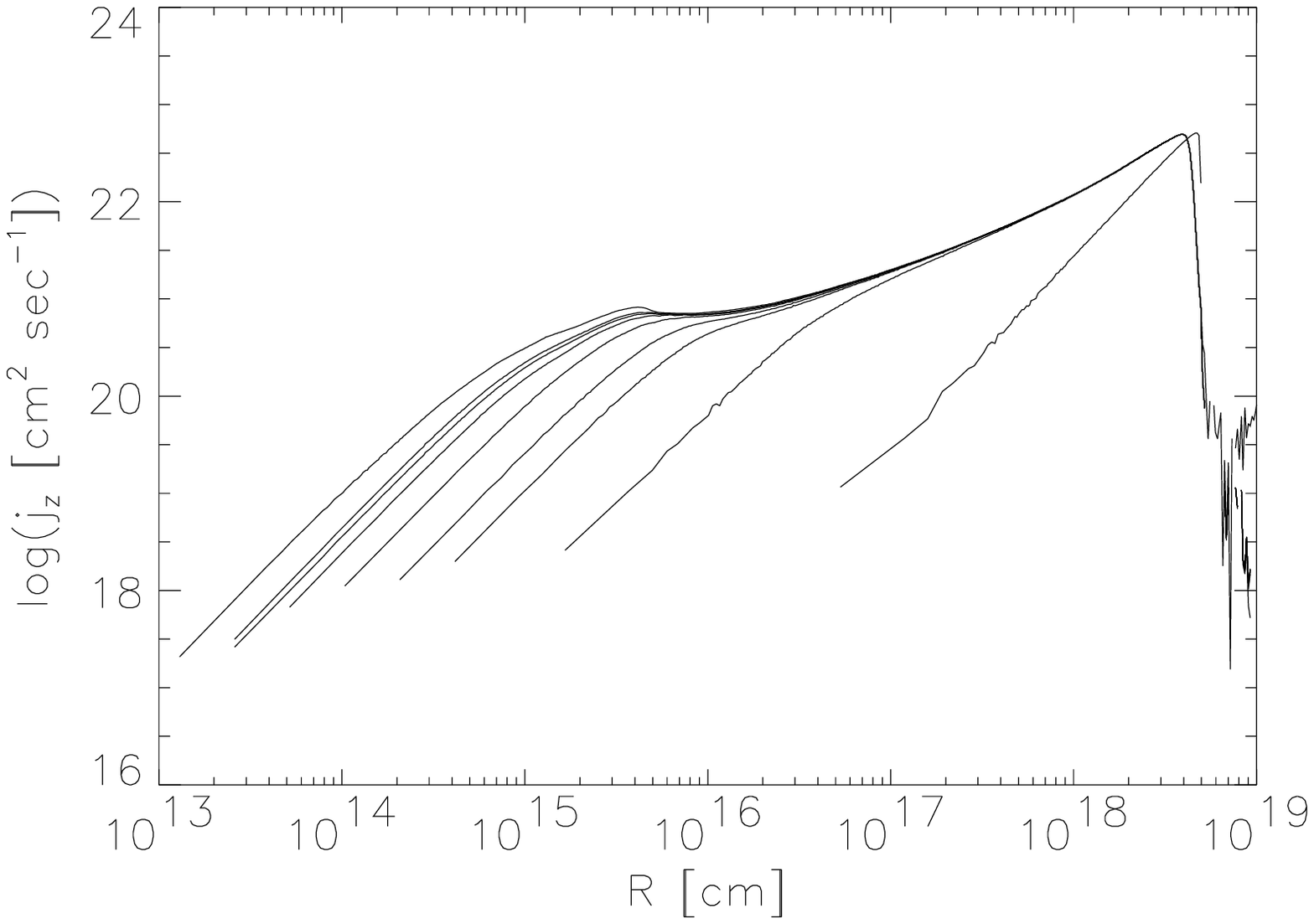}
\caption{Radial profiles of the angular velocity $\Omega$, the
  toroidal velocity $v_{\varphi}$, and the specific angular momentum
  $j_z$ at different times. The shown profiles correspond to the
  density shown in Fig.~\ref{fig:dens_height_surf} (run A1).}
\label{fig:omega_vphi_surf}
\eef

\bef
\showtwo{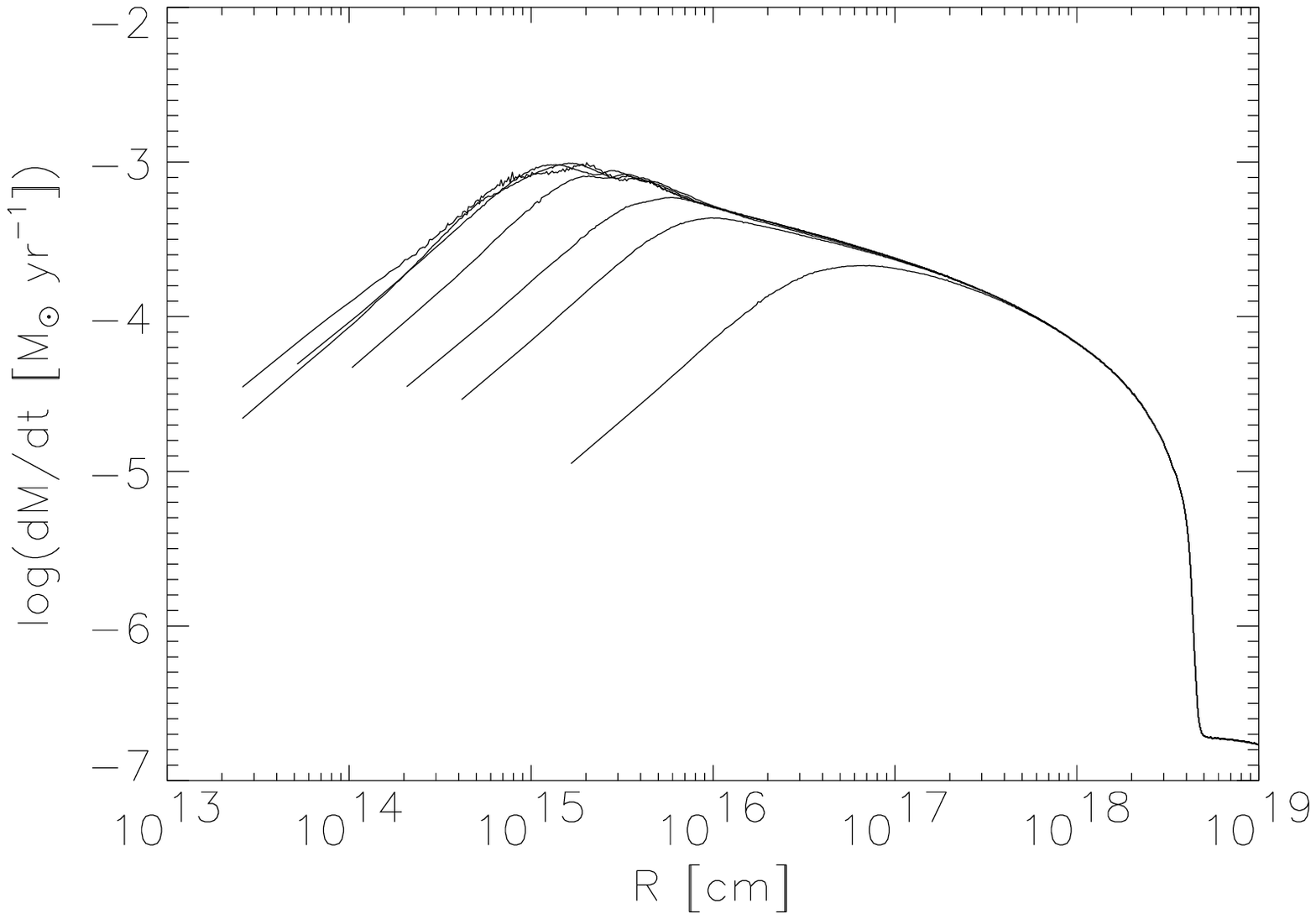}
	{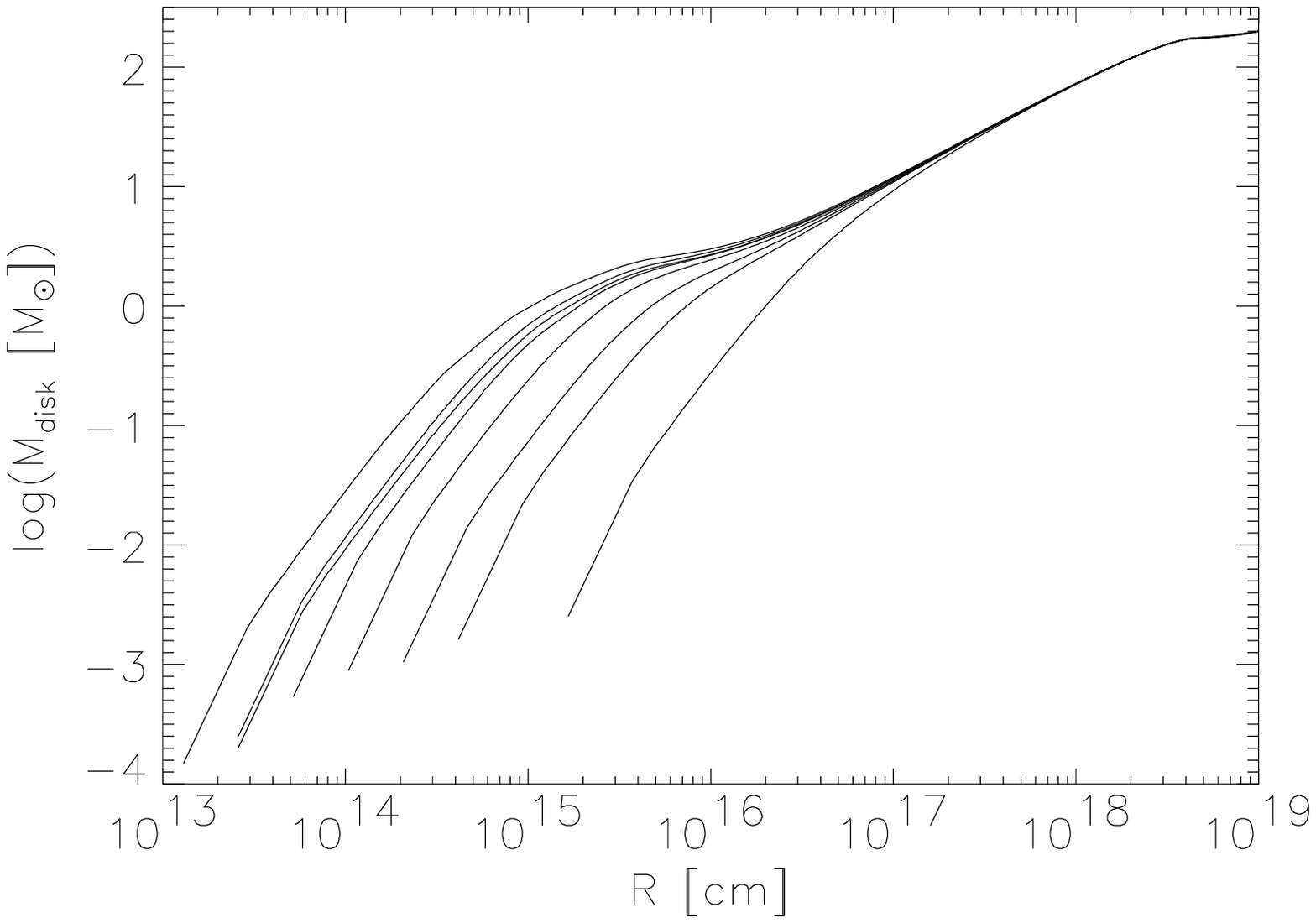}
\caption{Radial profiles of the mass accretion $\dot{M}$ and the
  cumulative disk mass $M_{\sm{disk}}$ at different times. The shown
  profiles correspond to the density shown in Fig.~\ref{fig:dens_plt}
  (run A1).}
\label{fig:mdot_mass_surf}
\eef

\bef
\showtwo{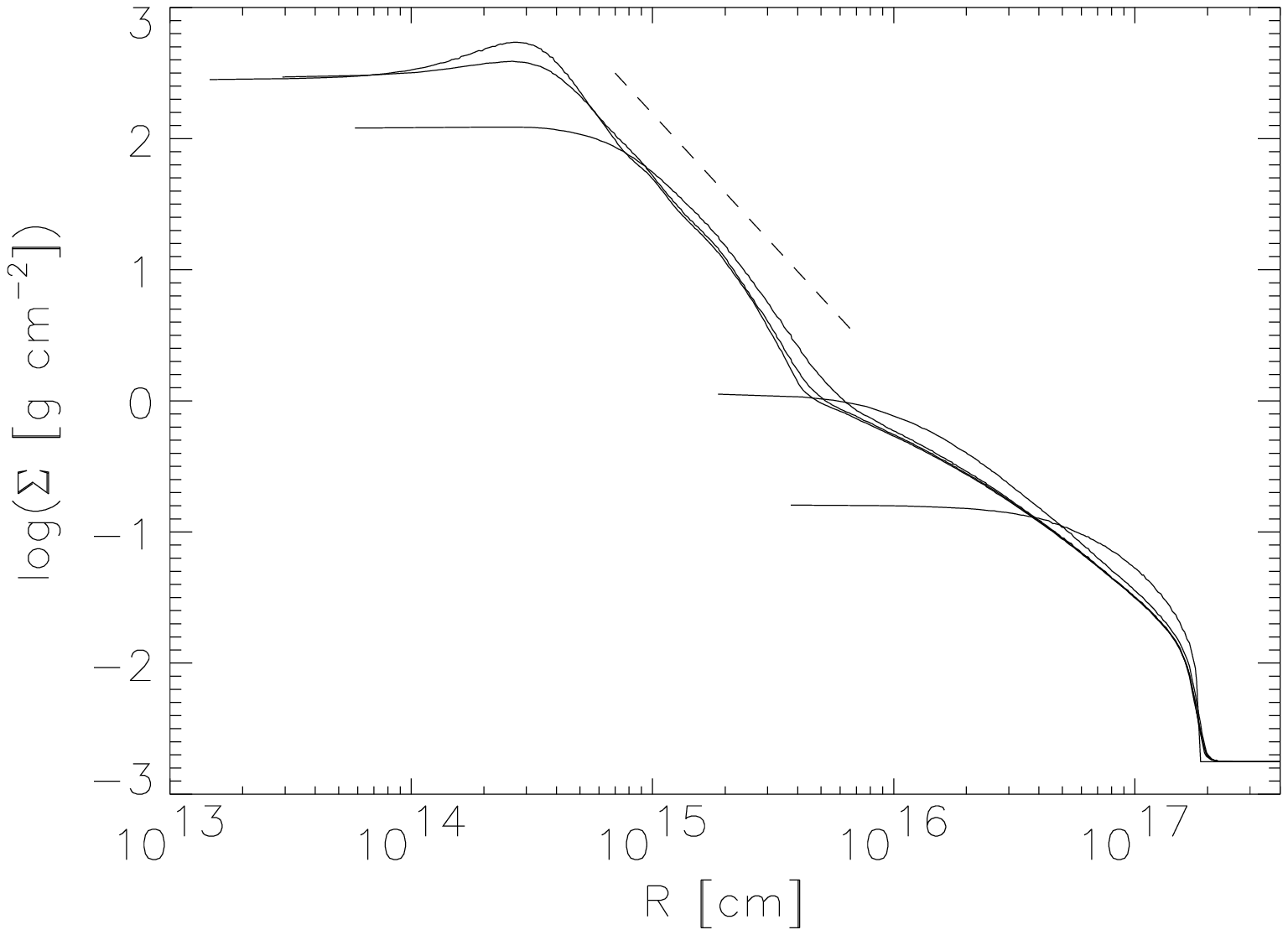}
	{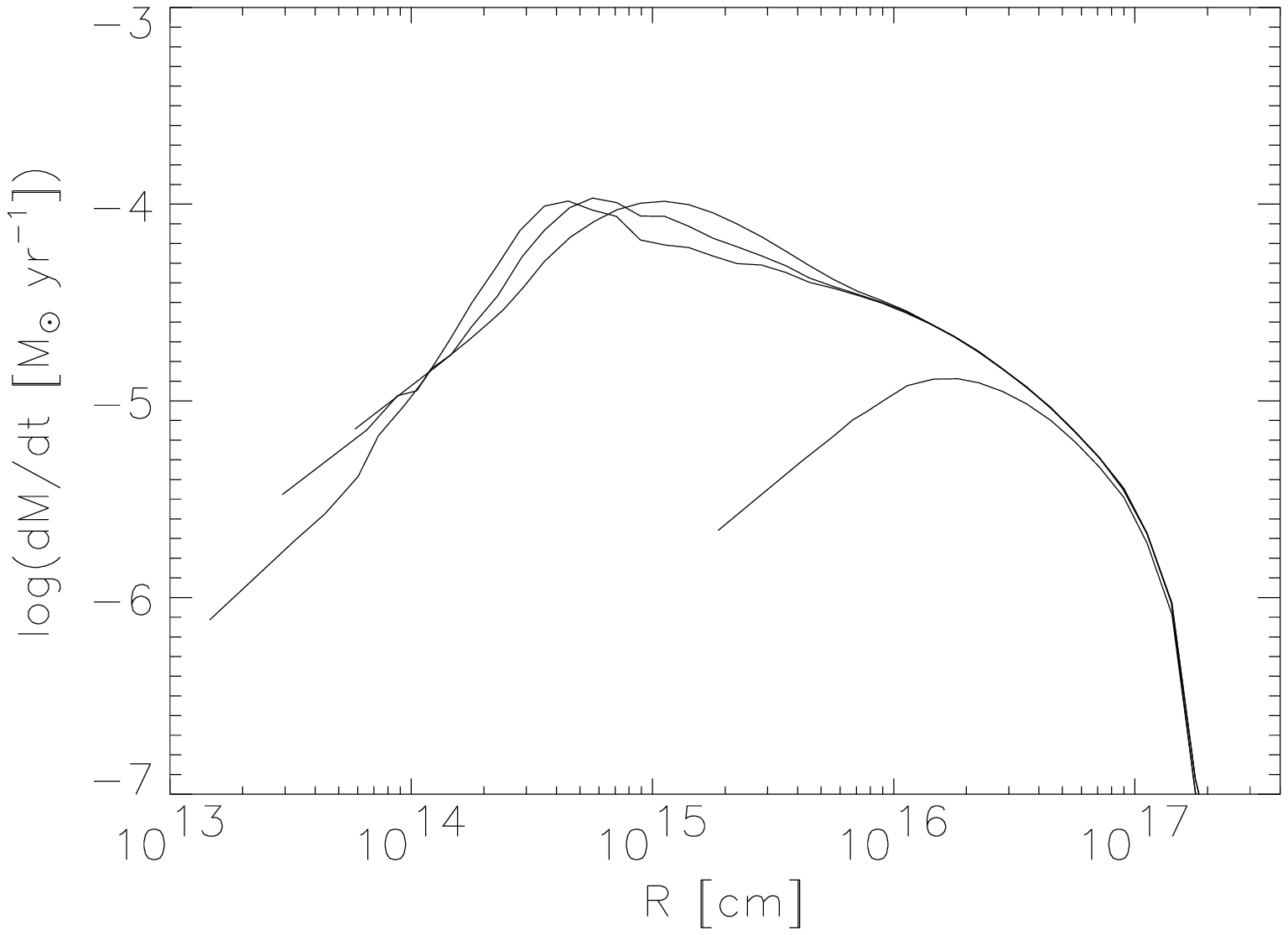}
\caption{Similar figures than Fig.~\ref{fig:mdot_mass_surf} and
  Fig.~\ref{fig:dens_height_surf} for run B68. The ring structure is
  clearly visible in the surface density at late stages. The dotted
  line shows a $R^{-2}$ profile.}
\label{fig:dens_surf_B68}
\eef

Once a disk-like object forms, one can define the disk height $h$ by
\be
h(R) = \frac{\Sigma(R)}{\rho(R)} \quad.
\label{eq:disk_height}
\ee
Fig.~\ref{fig:dens_height_surf} shows the evolution of $h$ for run A1
with time. The initial thick protostellar disk flattens to solar nebular
disk with a disk height of a few tens of $\AU$.

At the time when the protostellar disk first forms, it rotates close
to a solid body with a angular velocity of $\Omega \sim
10^{-11}\,\mbox{rad}\,\sec^{-1}$ whereas the outer envelope follows a
rotation law according to its density distribution, i.e $\Omega
\propto r^{-1}$. Fig.~\ref{fig:omega_vphi_surf} shows the evolution of
the angular velocity for run A1: while the disk is slowly accreting
gas to a core density of $\rho \sim 10^{-11} \, \g \, \cm^{-3}$ its
core spins up to $\Omega \sim 10^{-9} \, \mbox{rad} \,
\sec^{-1}$. These numbers show that the core region is rotationally
supported against gravitational collapse, i.e. $\Omega \sim \sqrt{4\pi
\, G \, \rho /3}$.

We show the mass accretion rate of the disk, $\dot{M}$, in
Fig.~\ref{fig:mdot_mass_surf} (run A1) and
Fig.~\ref{fig:dens_surf_B68} (run B68). This quantity is defined by
the standard vertically averaged continuity equation:
\be
  \frac{\di M}{\di t} = - 2\pi\,v_r\,R\,\Sigma \quad,
\ee
where $v_r$ is the radial velocity in the disk. In the case of run A1
after the protostellar disk forms the mass accretion at the outer edge
of the disk is $\sim 5\times 10^{-4} \, \Msol / \yr$ and increases
slightly with time to $\sim 10^{-3} \, \Msol / \yr$ while the disk
radius decreases. For run B68 the mass accretion is highest at the
ring's outer radius and reaches a maximal value of $\sim 3\times
10^{-4} \, \Msol / \yr$ there. Most of the infalling material is
deposited at this outer rim of the disk and the mass accretion
decreases towards the disk center as $\dot{M} \propto r$.  The
infalling gas outside of the disk reaches a constant terminal velocity
of the order of the sound speed (cf. the radial velocity plot of
Fig.~\ref{fig:vrad_plt}) which results in a constant mass accretion in
time in this outer region.  In the second panel of
Fig. ~\ref{fig:mdot_mass_surf} we compile the evolution of the total
mass in the disk, $M_{\sm{disk}}(R) = 2\pi \int \di R \, R \, \Sigma$,
for run A1 which reaches a few solar masses.

\section{Disk evolution: Bars, Rings, and fragmentation}
\label{sec:fragmentation}

Although we started all runs with a $10\%$ $m = 2$ density
perturbation, we did not observe any signs of fragmentation or
non-central collapse in the isothermal regime. This is due to low mass
(compared to the Jeans mass) in the collapsing core region
(cf. Sec.~\ref{sec:isothermal_contraction}). The first features which
appear are a bar-type or ring-type structure. They are first evident
after that stage of disk formation during which the non-isothermal
core slowly accretes the surrounding gas.
 
It is known that thin disks are unstable to fragmentation if its
Toomre $Q$-parameter~\citep{Toomre64}
\be
Q = \frac{c\,\kappa}{\pi\,G\,\Sigma} \quad,
\label{eq:Toomre_Q}
\ee
becomes smaller than unity, where $\kappa$ is the epicyclic frequency:
\be
\kappa = \left(4\,\Omega^2 + R\,\frac{\di\Omega^2}{\di R} \right)^{1/2}
\quad.
\label{eq:kappa}
\ee 

The first unstable modes which are expected to grow and possibly
persist for some dynamical times are the $m=0$ ring-mode and $m=2$
bar-mode. We find that a ring-type structure develops in the case with
$t_{\sm{ff}} \, \Omega = 0.2$ (runs A1 and B68) with a dimensionless
radius of $\xi_{\sm{ring}} \sim 6\times 10^{-3}$. For run A2 this
corresponds to a physical radius of $\sim 300\, \AU$ and for run B68
to $\sim 18 \, \AU$. In both cases the ring-structure persists only a
few dynamical times and either fragments into a binary system (run A2)
or collapses to a bar (run B68).  In the case with $t_{\sm{ff}} \,
\Omega = 0.1$ (run A1) and $t_{\sm{ff}} \, \Omega = 0.3$ (run A3) no
such ring structures develop in the simulations. Instead a bar-like
structure with spiral arms forms without a intermediate formation of a
ring structure. A ring structure in the Coalsack Globule 2 was
recently discovered by \citet{Lada04}. Our simulations may pertain to
this system; in particular, since this cloud can be well described
with a sub-critical Bonnor-Ebert profile.

We show 2D images of the density, temperature, and $Q$-parameter from
the runs A$n$ and B68 at different times in
Figs.~\ref{fig:dens_sim_01} -- \ref{fig:qparm_sim_B68} . The bar that
forms in the slow rotating run A1 shows no signs of fragmentation
although the $Q$-parameter drops below unity when the bar density is
very high, $> 10^{-8} \, \g \, \cm^{-3}$. Slight instabilities in this
high density and temperature ($600 - 800 \, \K$) regime cannot grow
quickly enough to lead to fragmentation as the inefficient cooling
prevents a fast collapse of overdensities. In the case of faster
rotation (run A2), instabilities form earlier resulting in a smaller
$Q$-value than in run A1. These instabilities are large enough to lead
to the fragmentation of the ring-type structure. In this situation, we
observe the break-up into two fragments where each of them has a disk
with spiral structure. These disks are surrounded with a circumstellar
disk, and are connected by a tidal stream.  The separation between the
members of this binary system is $\sim 260 \, \AU$ which is roughly
twice the size of the spiral structures itself. Each of these
protostellar systems has a total mass of $\sim 1.3 \, \Msol$. Run A3
with $t_{\sm{ff}} \, \Omega = 0.3$ (Figs.~\ref{fig:dens_sim_03} --
\ref{fig:qparm_sim_03}) does not form a ring structure but collapses
to a very twisted spiral arms and shows no sign of fragmentation. The
simulation of Barnard 68 (run B68, Figs.~\ref{fig:dens_sim_B68} --
\ref{fig:qparm_sim_B68}) with $t_{\sm{ff}} \, \Omega = 0.2$ results in
a shortly lived ring structure which collapses to a bar. This bar has
a size of $\sim 130 \, \AU$ and a total mass of $\sim 0.1 \, \Msol$.

The double-shock structure of these disks is easily seen in the
density plots, as well as the temperature structure, and velocity
fields in the x-z plane cuts.

The complex temperature structure due to cooling and the resulting
pressure response in the non-isothermal regime prevents a simple
analysis to decide which kind of structure one can expect from a
rotating collapsing cloud. From our simulations we conclude that
during the formation of the pre-stellar disk a ring structure appears
at a radius of 
\be 
  \xi = \xi_{\sm{ring}} = \xi_{\sm{ring}}(t_{\sm{ff}} \, \Omega) \quad,
\label{eq:ring_radius}
\ee 
whenever the cloud rotates fast enough. In order to decide whether
such possibly formed ring will fragment into several pieces, one has to
compare this radius with the size of the non-isothermal core where the
enhanced thermal pressure stabilizes the core region against
fragmentation.  In section~\ref{sec:isothermal_general} we showed that
the core radius of a collapsing Bonnor-Ebert-Sphere is (almost)
independent of the initial cloud parameters and only determined by the
density where $t_{\sm{ff}} \sim t_{\sm{cool}}$
(cf. Eq.~\ref{eq:core_radius}).  For a ring to fragment its size must
not be much smaller than this warm core size which gives the condition
\be 
\xi_{\sm{ring}} \sim
  \left(\frac{\rho_0}{\rho_{\sm{core}}}\right)^{1/2} \, \left(\xi_c \,
  \phi_c^{\prime}\right)^{1/2} \quad.
\label{eq:ring_frag_cond}
\ee

We find that ring structures which do not obey the
condition~(\ref{eq:ring_frag_cond}) collapse to a single bar rather
than fragmenting into two or more pieces.
Figs.~\ref{fig:dens_sim_B68} to~\ref{fig:temp_sim_B68} show the time
evolution of the density and temperature of the simulation B68 where a
firstly formed ring condense to a bar. Note also that the core region
rotates with $\Omega(r) \approx \mbox{const.}$ whereas the disk
rotates differentially outside the core region
(cf. Fig.~\ref{fig:omega_vphi_surf}) where shearing effects enhance
instabilities. Another mechanism which facilitates the fragmentation
of a ring structure is the appearance of a shock at the core edge. A
ring structure of similar size as the core can accrete more material
than smaller rings increasing the possibility of fragmentation.

In the case where $t_{\sm{ff}} \, \Omega = 0.1$ (run A1), we find that
a 2-armed spiral structure develops when $\rho_{\sm{core}} \sim
10^{-11} \, \g \, \cm^{-3}$ and $T \sim 200 \, \K$. If the initial
sphere rotates with $t_{\sm{ff}} \, \Omega = 0.2$ (run A2) a ring
structure forms earlier when the density and temperature are
respectively $\sim 10^{-14} \, \g \, \cm^{-3}$ and $\sim 85 \,
\K$. This confirms the general picture that slowly rotating clouds
collapse to bars and thin filaments whereas faster rotating objects
form a ring-type structure. Recent numerical investigation of rotating
Bonnor-Ebert-Spheres by~\citet{Matsumoto03} showed that spheres with
$t_{\sm{ff}} \, \Omega \approx 0.2$ collapse to a ring which fragments
into several pieces.

The theory of bars in disk-like structures \citep[see][]{Binney87}
predicts that the bar velocity $\Omega_{\sm{b}}$ must be larger than
the so called pattern velocity: 
\be 
\Omega_{\sm{p}} \equiv \Omega - \frac{1}{2} \, \kappa 
\ee 
of the disk. We show the evolution of the pattern velocity as a
function of disk radius in Fig.~\ref{fig:Omega_p}. As the bar size,
$R_{\sm{b}}$ is limited within the corotation radius $R_{\sm{co}}$,
where the corotation radius is given by $\Omega(R_{\sm{co}}) =
\Omega_{\sm{b}}$, and the disk velocity falls off at the outer core
the bar velocity is limited by:
\be
\Omega_{\sm{p}} < \Omega_{\sm{b}} < \Omega \quad .
\label{eq:bar_condition}
\ee

The disk velocity in the core region is roughly independent of the
radius $r$ which results in a small pattern velocity, e.g.
$\Omega_{\sm{p}} \ll \Omega$. Due to the different time scales of the
core evolution $\sim t_{\sm{ff}}$ and the rotation period $\sim
\Omega^{-1}$ where $t_{\sm{ff}} \, \Omega \ll 1$ we were not able to
follow a full rotation of the condensed bars. Nevertheless, we
estimate the average bar speed for our different simulations (see
table~\ref{tab:summary}) as,
few~$\times 10^{-10}$ (run A1), few~$\times 10^{-12}$ (run A3), and
$\sim 10^{-10} \, \mbox{rad} \, \sec^{-1}$ (run B68) which is in
agreement with condition~(\ref{eq:bar_condition}). Note that the run
with the fastest initial rotation (run A3, $t_{\sm{ff}} \, \Omega =
0.3$) results in the slowest bar velocity. This might be due to the
fact the the bar of run A3 is much larger than the bar of run A1
($t_{\sm{ff}} \, \Omega = 0.1$) and its pronounced spiral arms
released sufficient angular momentum to the surrounding disk.

\begin{table}
  \begin{tabular}{|c|c|c|c|c|} \hline
  run & structure & $\Sigma \propto R^{-n}$
%      & $\rho_{\sm{env}}$ 
      & $\Omega_{\sm{b}}\,  [\mbox{rad} \, \sec^{-1}]$ 
      & $M_*\,[\Msol$] \\ \hline \hline
  A1  & bar  & $1.9$ 
%      & $2.05$
      & $5\times 10^{-11}$ 
      & 2.5 \\ \hline
  A2  & binary system  & $2.0$  
%      & $2.1$
      & -- 
      & $1.3$ each\\ \hline
  A3  & bar \& spirals & $1.9$ 
%      & $2.05$
      & $6\times 10^{-12}$ 
      & 2.2 \\ \hline
  B68 & ring/bar & $2.0$ 
%       & $2.0$
      & $\sim 10^{-10}$ 
      & $0.1$ \\ \hline
  \end{tabular} 
\caption{Summary of the simulation results. Where the disk profiles
  are measured in the non-isothermal regime, 
% $\rho_{\sm{env}}$ is the density distribution in the envelope, 
  $\Omega_{\sm{b}}$ is the bar velocity, and $M_*$ is the
  central mass of the object(s).}
\label{tab:summary}
\end{table}

Finally, we note that the column density profiles of the disks formed
in our simulations are rather steep, obeying
\be
\Sigma \propto R^{-1.95\pm0.05} \quad.
\label{eq:disk_profiles}
\ee
This is steeper than Hayashi models \citep{Hayashi81} but in good
agreement with disk models inferred from the measurement by
\citet{Kuchner04} of planets in extrasolar systems, who found
$\Sigma \propto R^{-2.0\pm0.5}$.

\befone
\showone{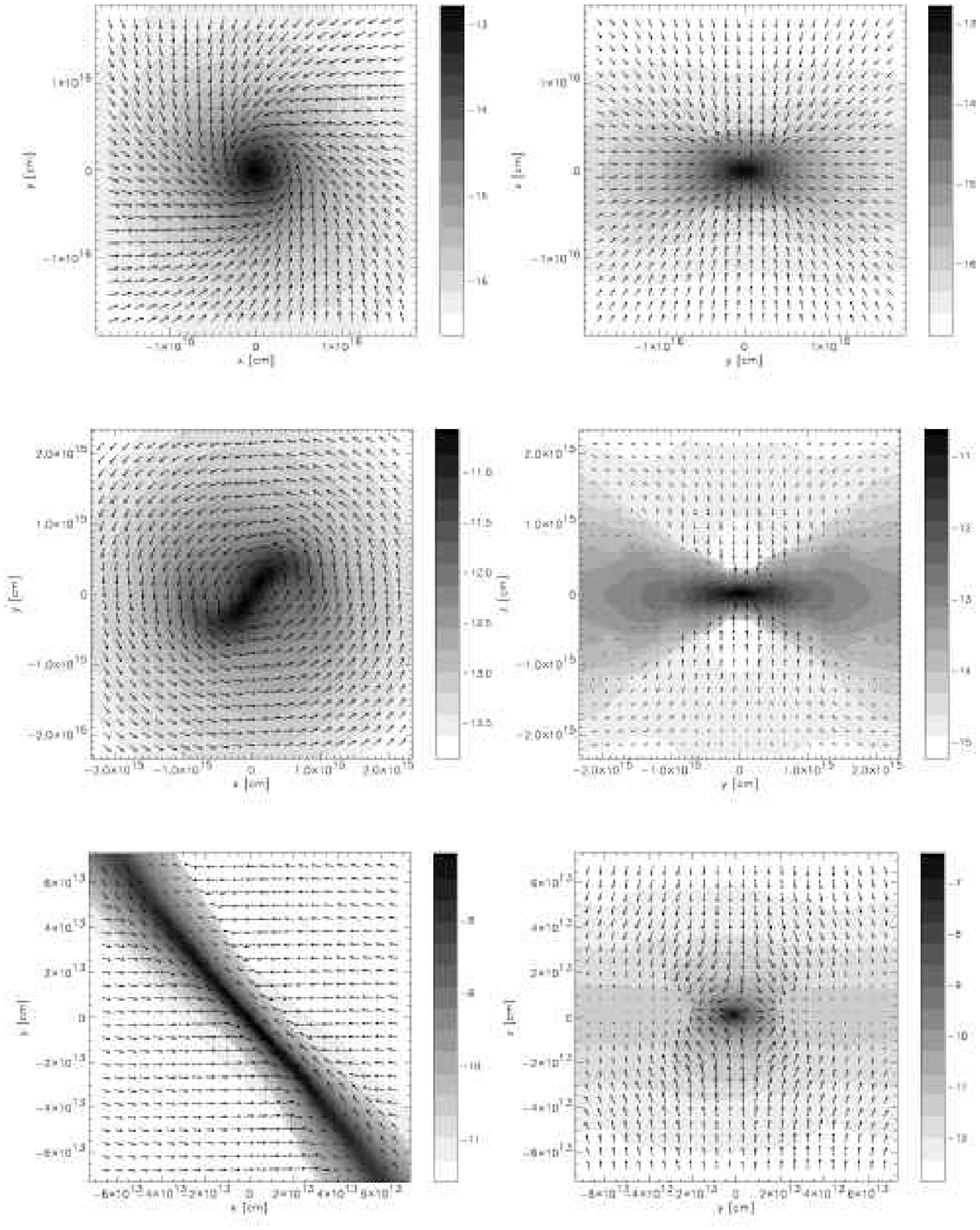}
\caption{Shows 2D slices of the mass density (logarithmic scale in
  $[\g \, \cm^{-3}]$) and velocity field at different times for run
  A1. After the protostellar disk forms a spiral feature builds up and
  collapses to a thin filament. The left panels and right panels show
  the evolution in the disk plane and perpendicular to the disk plane,
  respectively. Note that the highest resolution areas correspond
  (from top to bottom) to 256, 512, and 1024 pixels in each
  dimension.}
\label{fig:dens_sim_01}
\eefone

\befone
\showone{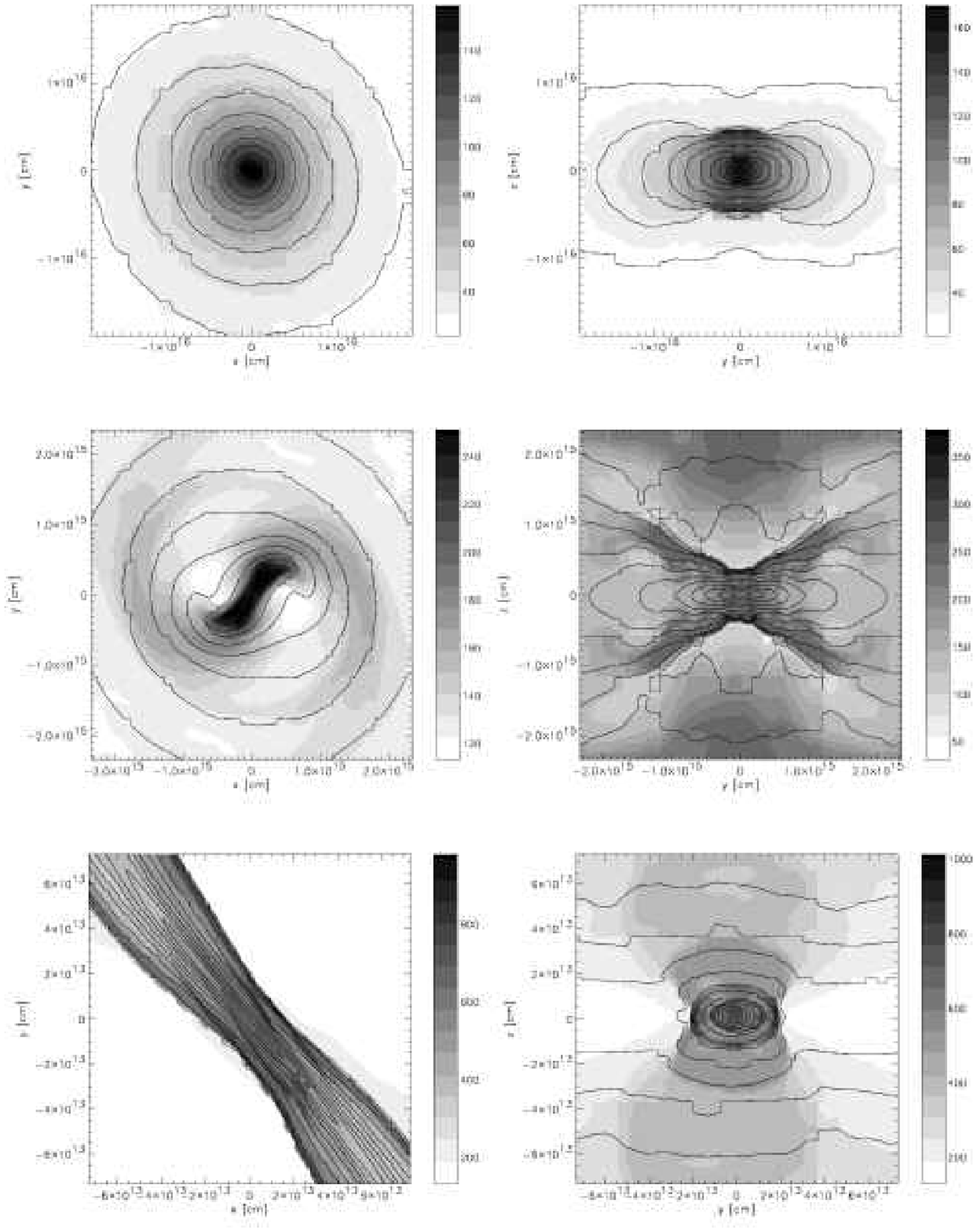} 
\caption{Evolution of the temperature field (gray scale in $[\K]$) and
density (contour lines). The panels correspond to these of
Fig.~\ref{fig:dens_sim_01} (run A1).}
\label{fig:temp_sim_01}
\eefone

\befone
\showone{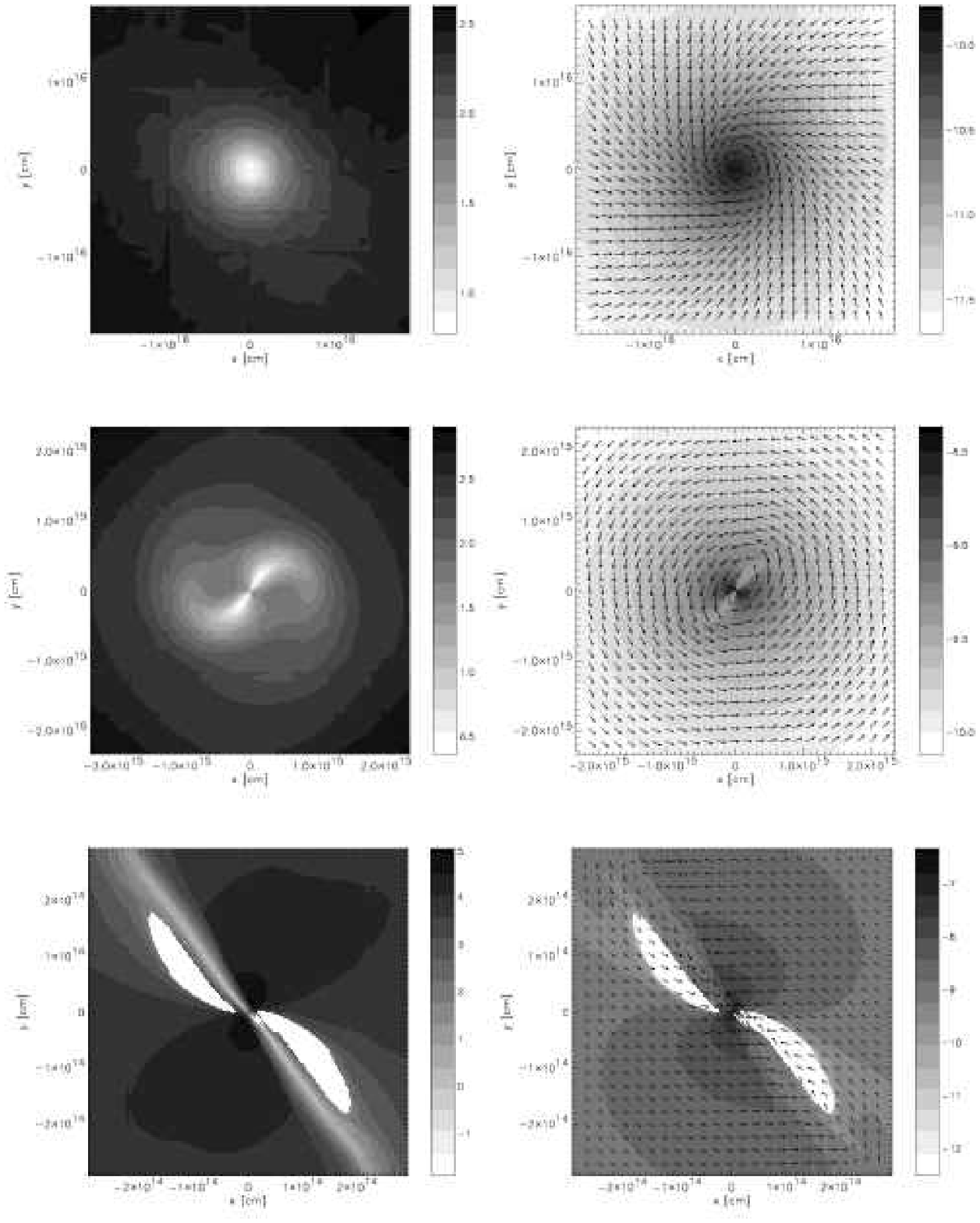} 
\caption{Shows the evolution of the Toomre $Q$-parameter
  (left panel, logarithmic scale) and the angular velocity $\Omega$
  (right panel, logarithmic scale in $[\mbox{rad}\,\sec^{-1}]$. The panels
  correspond to these of Fig.~\ref{fig:dens_sim_01} (run A1).}
\label{fig:qparm_sim_01}
\eefone

\befone
\showone{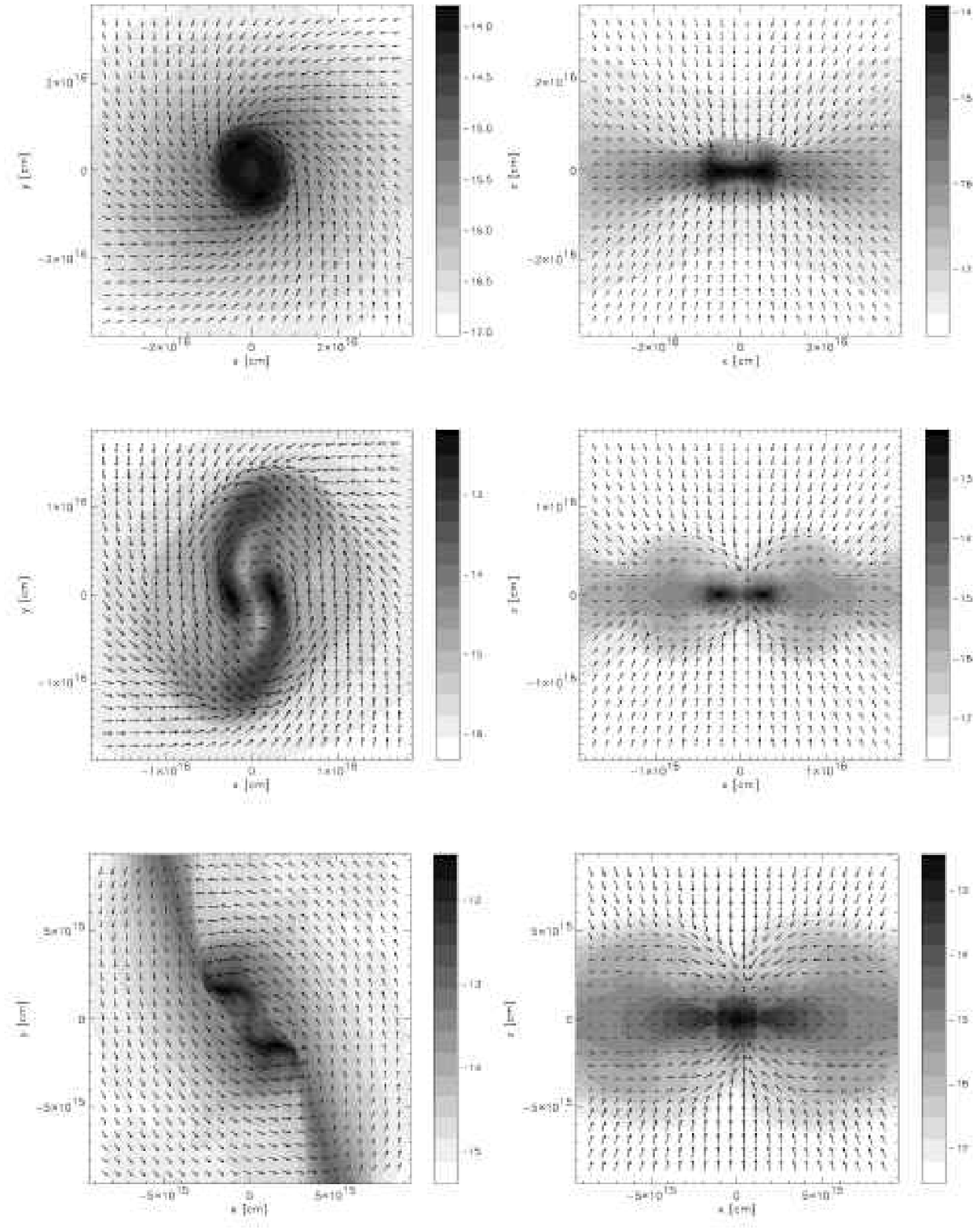} 
\caption{Shows 2D slices of the mass density (logarithmic scale in
  $[\g \, \cm^{-3}]$) and velocity field at
  different times for run A2.} 
\label{fig:dens_sim_02}
\eefone

\befone
\showone{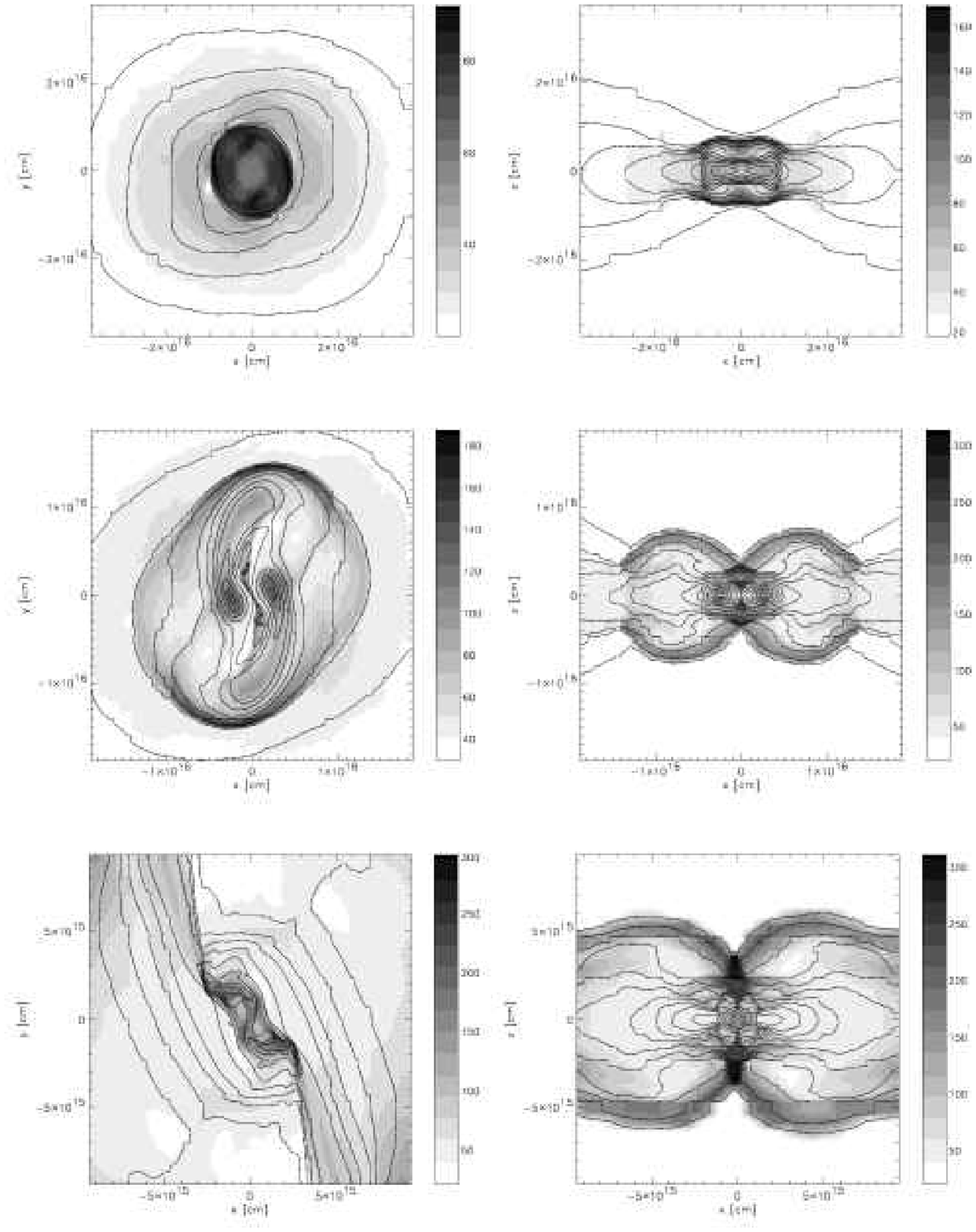} 
\caption{Evolution of the temperature field (gray scale in $[\K]$) and
density (contour lines). The panels correspond to these of
Fig.~\ref{fig:dens_sim_02} (run A2).}
\label{fig:temp_sim_02}
\eefone

\befone
\showone{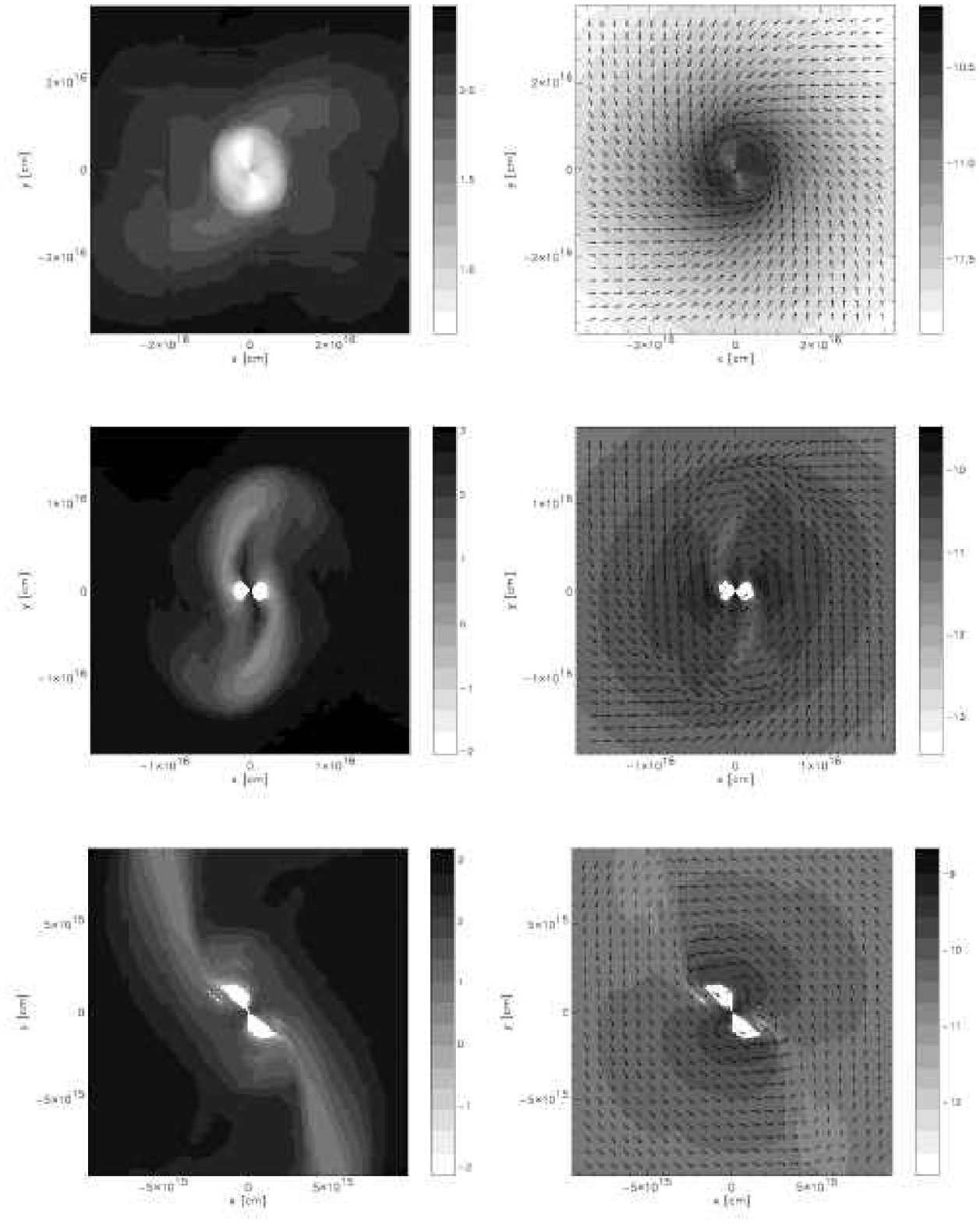} 
\caption{Shows the evolution of the Toomre $Q$-parameter
  (left panel, logarithmic scale) and the angular velocity $\Omega$
  (right panel, logarithmic scale in $[\mbox{rad}\,\sec^{-1}]$. The panels
  correspond to these of Fig.~\ref{fig:dens_sim_02} (run A2).}
\label{fig:qparm_sim_02}
\eefone

\befone
\showone{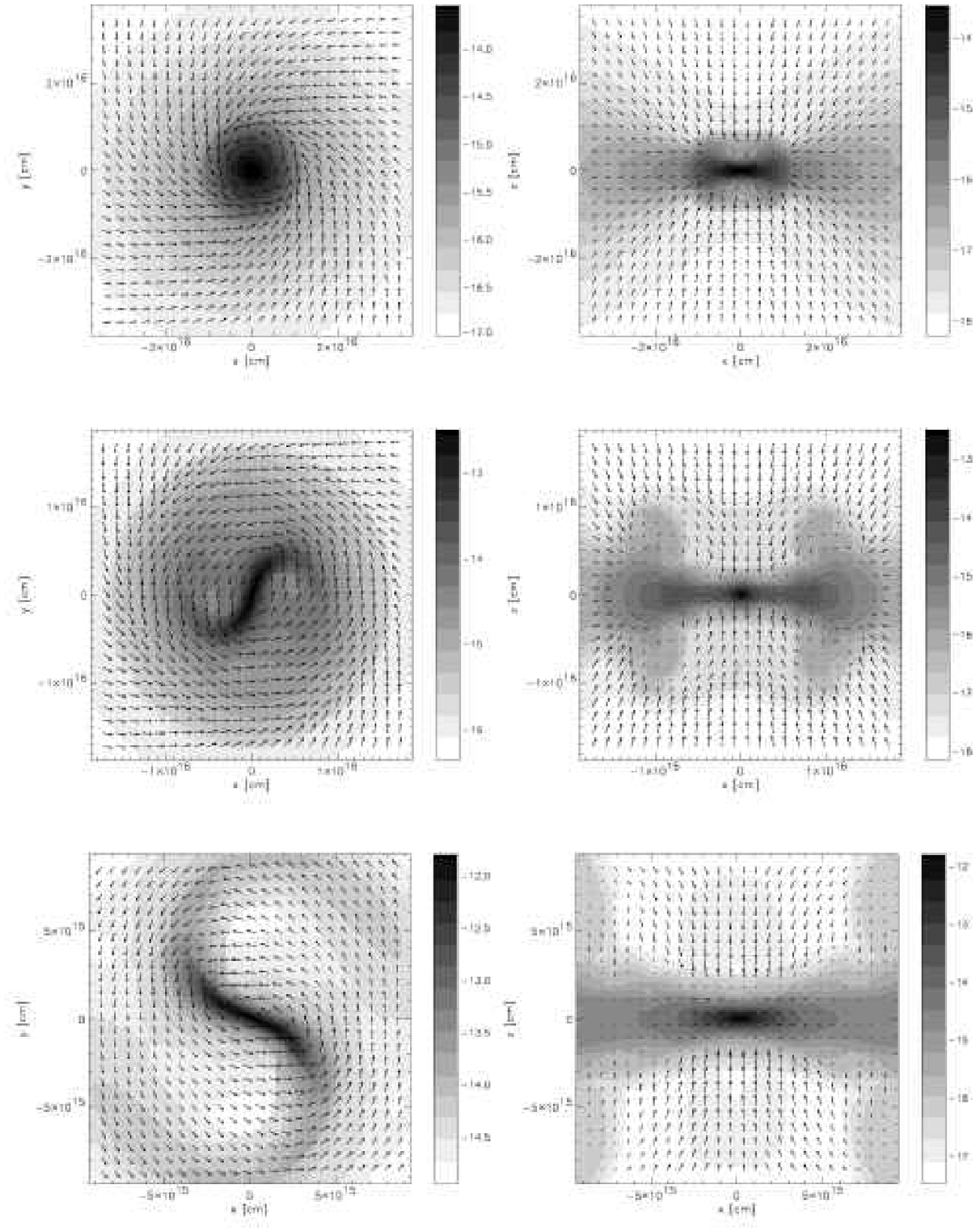} 
\caption{Shows 2D slices of the mass density (logarithmic scale in
  $[\g \, \cm^{-3}]$) and velocity field at
  different times for run A3.} 
\label{fig:dens_sim_03}
\eefone

\befone
\showone{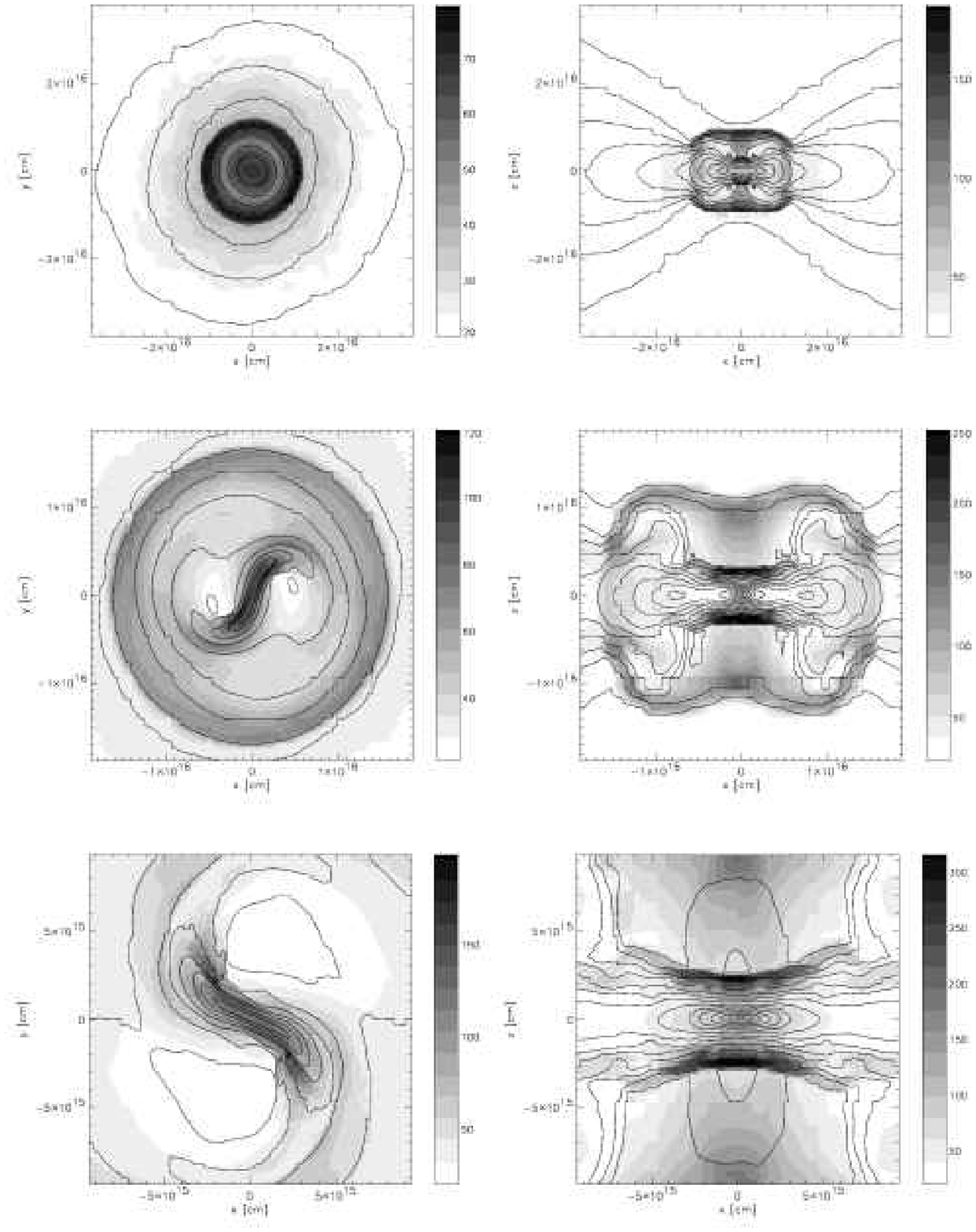} 
\caption{Evolution of the temperature field (gray scale in $[\K]$) and
density (contour lines). The panels correspond to these of
Fig.~\ref{fig:dens_sim_03} (run A3).}
\label{fig:temp_sim_03}
\eefone

\befone
\showone{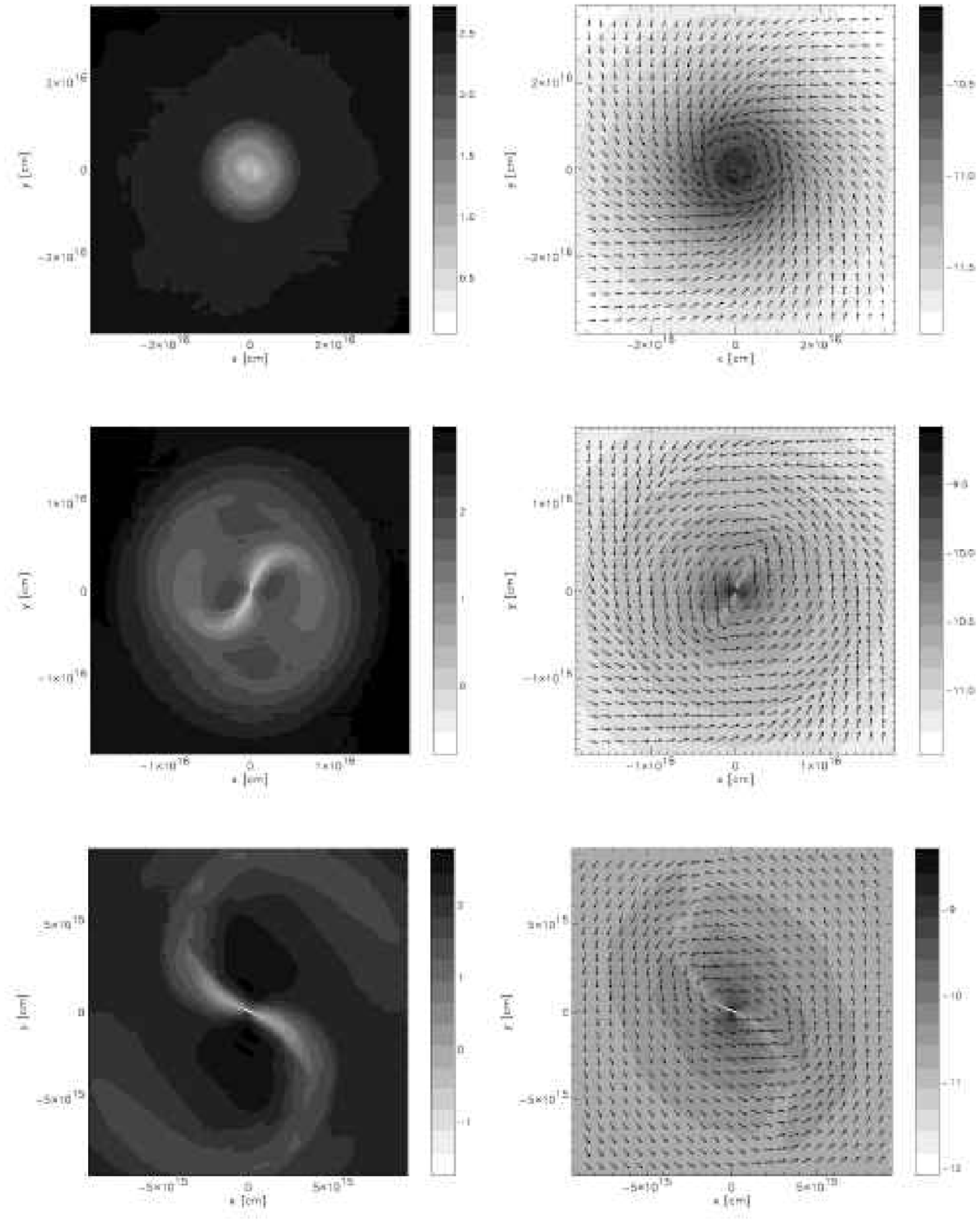} 
\caption{Shows the evolution of the Toomre $Q$-parameter
  (left panel, logarithmic scale) and the angular velocity $\Omega$
  (right panel, logarithmic scale in $[\mbox{rad}\,\sec^{-1}]$. The panels
  correspond to these of Fig.~\ref{fig:dens_sim_03} (run A3).}
\label{fig:qparm_sim_03}
\eefone

\befone
\showone{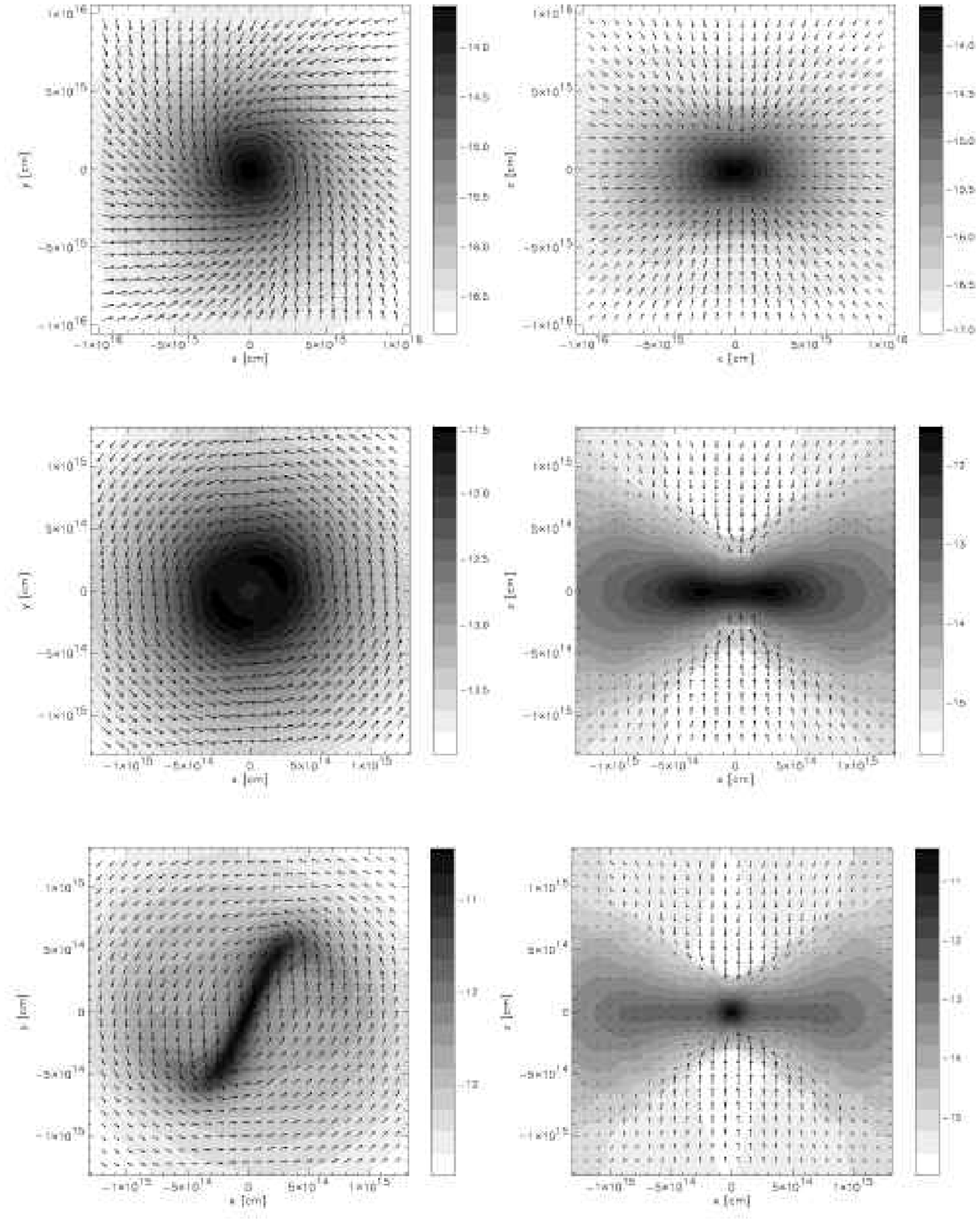} 
\caption{Shows 2D slices of the mass density (logarithmic scale in
  $[\g \, \cm^{-3}]$) and velocity field from the simulation B68 at
  different times: from top to bottom: $t = 5.35 \, , \, 5.40 \, , \,
5.41 \, t_{\sm{ff}}$ corresponding $3.597 \, , \, 3.628 \, , \, 3.633
  \, \times 10^{5} \, \ys$.}
\label{fig:dens_sim_B68}
\eefone

\befone
\showone{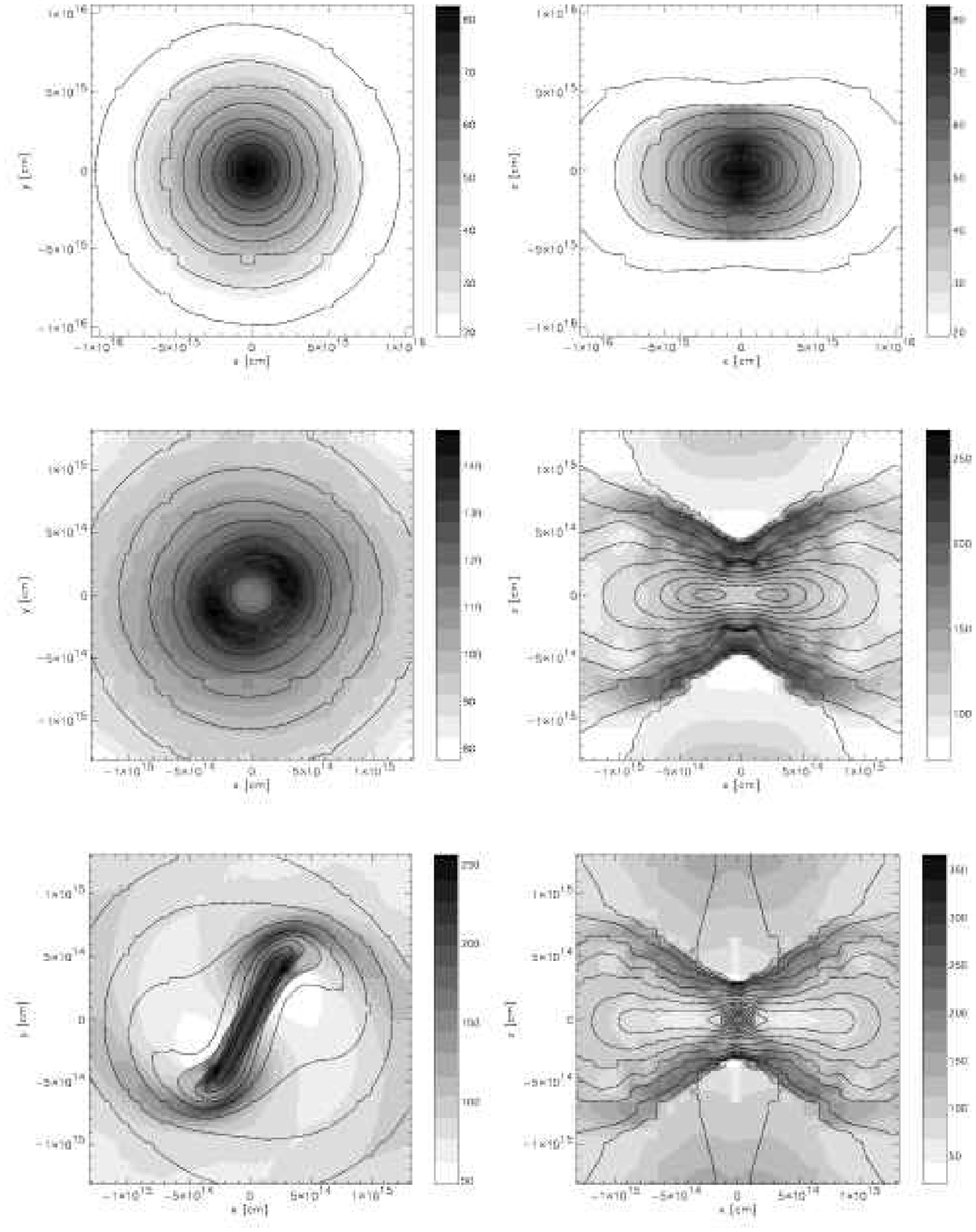} 
\caption{Evolution of the temperature field (gray scale in $[\K]$) and
  density (contour lines). The panels correspond to these of
  Fig.~\ref{fig:dens_sim_B68} (run B68).}
\label{fig:temp_sim_B68}
\eefone

\befone
\showone{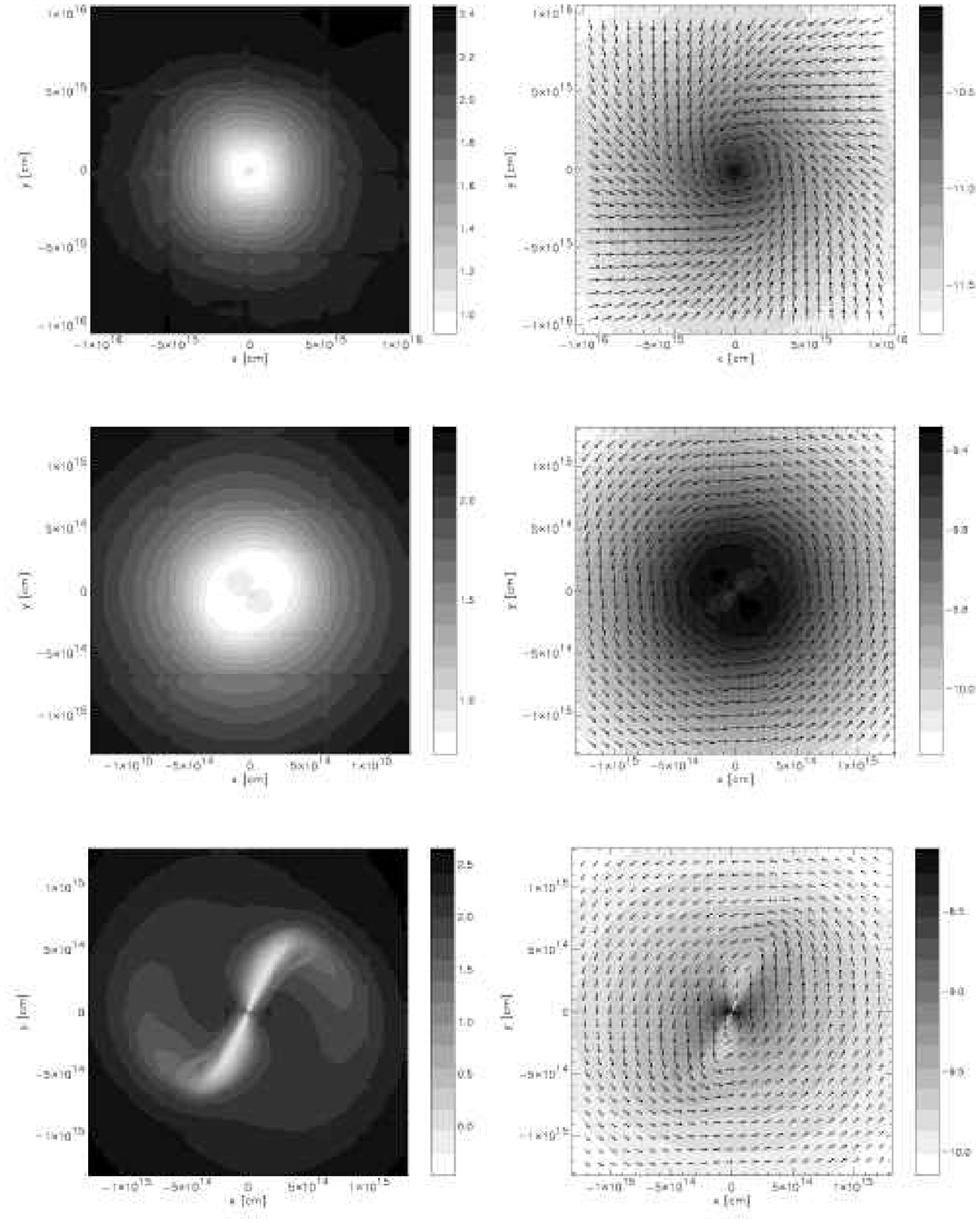} 
\caption{Shows the evolution of the Toomre $Q$-parameter
  (left panel, logarithmic scale) and the angular velocity $\Omega$
  (right panel, logarithmic scale in $[\mbox{rad}\,\sec^{-1}]$. The panels
  correspond to these of Fig.~\ref{fig:dens_sim_B68} (run B68).}
\label{fig:qparm_sim_B68}
\eefone

\bef
\showone{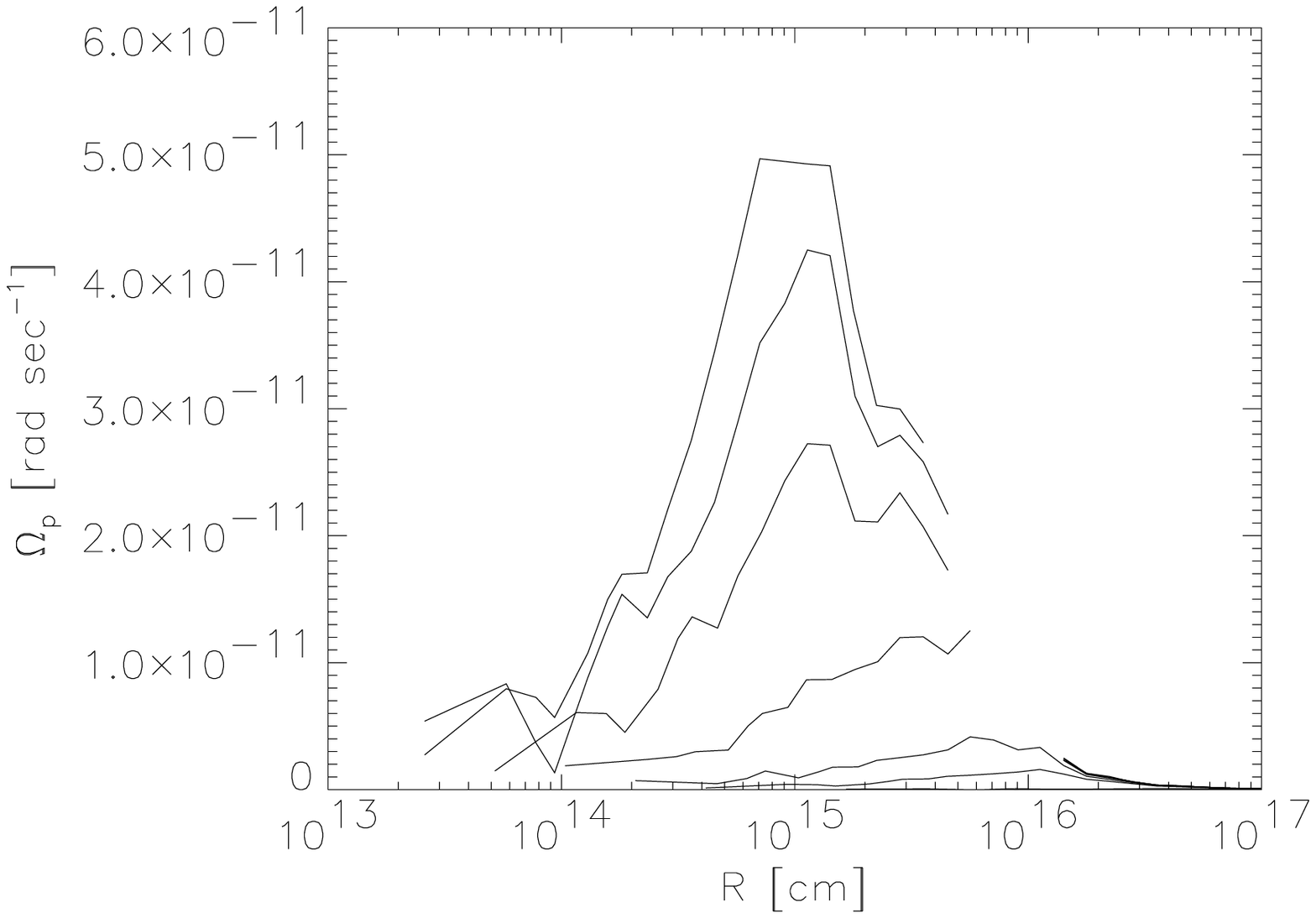}
\caption{Evolution of the pattern velocity $\Omega_p$ as a function of
radius $R$. The data are taken from run A1.}
\label{fig:Omega_p}
\eef

\section{Summary and discussion}
\label{sec:summary}

In this work we studied the collapse of rotating marginal stable
Bonnor-Ebert-Spheres using 3D hydrodynamical simulations based on AMR
technique. Compared to former numerical studies of collapsing
molecular clouds, we included the effect of radiative cooling by
molecular line emissions. This more realistic approach shows that the
effective equation of state is a complex function of density and time
during the collapsing phase and can not be approximated by a time
independent EOS. Initially the gas cloud collapses isothermally on a
free fall time until the core density reaches the critical point where
cooling becomes less efficient (i.e. $t_{\sm{cool}} > t_{\sm{ff}}$) at
$n \sim 10^{7.5} \, \cm^{-3}$. During this isothermal phase the cloud
collapses from outside-in where the density profile approaches to a
(steadily increasing) flat core region with a $r^{-2}$ envelope. The
infall velocity peaks at the edge of the core where it reaches $\Ma
\sim 3$ and drops to zero at the center of the gas cloud.
Subsequently the temperature in core region starts to rise and an
increasing pressure support stabilizes the gas clump preventing
immediate fragmentation. As cold gas from the envelope falls onto the
warm core region shock fronts build up, separating the cold envelope
from the heated up core. All our simulations show a similar double
shock structure: an early outer shock and a later inner shock. Both
shock fronts move slowly toward the core center where the inner shock
region sets the conditions for the first appearance of the
protostellar disk. The appearance of these shocks determine also the
accretion rate onto the core (resp. bar) as material outside the shock
region is stalled at the shock boundaries.  We find peak accretion
rates for the low and high mass systems of $3\times 10^{-4} \,
\Msol/\yr$ and $\sim 10^{-3} \, \Msol/\yr$, respectively. In this work
we did not keep track of the composition of the gas during the
collapse of the molecular cloud, but its is known that the chemistry
might be altered within the shock region \citep[e.g.][]{Jorgensen04}.

Depending on the initial rotation and density a ring mode or a bar
develops in the disk plane. Only one of our four simulations result
into the fragmentation of the protostellar disk. In this particular
case, a ring structure developed first which then fragments into a
binary system in which each system contains a bar structure. The
preferred appearance of bars in our simulations supports theoretical
predictions that the most likely growing instability mode in disks is
the $m=2$ bar mode.

Due to the limitations of our numerical scheme we are not able to
follow the bar evolution for several rotation periods as the time
scales of the bar rotation and core evolution diverge
(i.e. $t_{\sm{rot}} \gg t_{\sm{ff}}$) shortly after the bar
forms. Further fragmentation of the bars might happen during the phase
of hydrogen dissociation ($T > 2000 \, \K$) where the effective
equation of state drops below $4/3$ or when tidal interactions of the
spiral amrs become stronger. 

A comparison of our A2 simuations -- which produces a distinct ring
that fragments into a binary -- with the recent observation of a
massive disk in M 17 \citep{Chini04} is interesting. The observations
reveal an inner torus (within a $100\,\Msol$ disk) whose radius is
$\sim 3.8\times 10^{16}\,\cm$. The radius of the dense inital ring
that forms in our simulation A2 is of the same order ($\sim
10^{16}\,\cm$, see Fig.~\ref{fig:dens_sim_02}) which is in good
agreement with these observations.

\comment{ It is also possible that our simulations results
(particularly our low density runs A$n$) are the pre-stage of a single
high mass star similar to the one observed by~\citet{Chini04} in the
Omega nebula (M 17). These authors observed a $20\,\Msol$ star
embedded in a $100\,\Msol$ disk which central density has a elongated
shape and might be due to a bar or a binary system.}

\section*{Acknowledgement}

The authors would like to thank Tom Abel and James Wadsley for useful
and inspiring discussions. We are grateful to David Neufeld and
his collaborators for providing us with their cooling data. Thanks
also to the referee, Andreas Burkert, for useful comments. The FLASH
code was in part developed by the DOE-supported Alliances Center for
Astrophysical Thermonuclear Flashes (ASCI) at the University of
Chicago. Our simulations were carried out on a 128 CPU AlphaServer SC,
which is the McMaster University node of the SHARCNET HPC Consortium.
R.B. is supported by the SHARCNET Postdoctoral Fellowship programm,
and R.E.P. is supported by the Natural Sciences and Engineering
Research Council of Canada.

\bibliographystyle{mn} 
\bibliography{astro}

\end{document}